\documentclass[preprint,authoryear]{elsarticle}

\usepackage{amsmath}
\usepackage{amsfonts}
\usepackage{epsfig}
\usepackage{graphics}
\usepackage[utf8]{inputenc}
\usepackage{url}
\usepackage{siunitx}
\usepackage{nomencl}
\usepackage[modulo]{lineno}
\usepackage{hyperref}
\usepackage{gensymb}
\usepackage{booktabs}
\usepackage{xcolor}

\newcommand{\dd}[2]{\frac{\partial #1}{\partial #2}}
\newcommand{\ddi}[2]{\partial #1 / \partial #2}
\def\bs{\boldsymbol}
\def\bu{\bs{u}}
\def\umag{U}
\def\div{\mathrm{div}\ }
\def\ke{k-$\epsilon$\ }
\def\LAD{\mathrm{LAD}}

\newcommand{\subfigimg}[3][,]{%
  \setbox1=\hbox{\includegraphics[#1]{#3}}
  \leavevmode\rlap{\usebox1}
  \rlap{\hspace*{-15pt}\raisebox{\dimexpr\ht1-2\baselineskip}{\footnotesize #2}}
  \phantom{\usebox1}
}
\newcolumntype{P}[1]{>{\centering\arraybackslash}p{#1}}

\makeatletter
\def\ps@pprintTitle{%
 \let\@oddhead\@empty
 \let\@evenhead\@empty
 \def\@oddfoot{}%
 \let\@evenfoot\@oddfoot}
\makeatother

\begin{document}

\begin{frontmatter}
\title{RANS solver for microscale pollution dispersion problems in areas with vegetation: Development and validation}

\author[addr1]{Viktor Šíp\corref{cor1}}
\ead{viktor.sip@fs.cvut.cz}
\author[addr1]{Luděk Beneš}
\ead{ludek.benes@fs.cvut.cz}

\address[addr1]{Department of Technical Mathematics, Faculty of Mechanical Engineering, Czech Technical University in Prague. Karlovo náměstí 13, 121 35 Prague 2, Czech Republic.}
\cortext[cor1]{Corresponding author}

\begin{abstract}
  We present a description and validation of a finite volume solver aimed at solving the problems of microscale urban flows where vegetation is present.
  The solver is based on the five equation system of Reynolds-averaged Navier-Stokes equations for atmospheric boundary layer flows, which are complemented by the \ke turbulence model.
  The vegetation is modelled as a porous zone, and the effects of the vegetation are included in the momentum and turbulence equations.
  A detailed dry deposition model is incorporated in the pollutant transport equation, allowing the investigation of the filtering properties of urban vegetation.
  The solver is validated on four test cases to assess the components of the model: the flow and pollutant dispersion around the 2D hill, the temporal evolution of the rising thermal bubble, the flow through and around the forest canopy, and a hedgerow filtering the particle-laden flow.
  Generally good agreement with the measured values or previously computed numerical solution is observed, although some deficiencies of the model are identified. These are related to the chosen turbulence model and to the uncertainties of the vegetation properties.
\end{abstract}

\end{frontmatter}

\section{Introduction}

Air pollution in urban areas poses a significant health risks to the inhabitants, and our capability to predict the wind flow and pollutant dispersion in these areas is thus crucial in mitigating the negative effects of the continuing urbanization.
Computational modelling of urban flows is however challenging. Urban areas are characterized by complex geometries, and the air flow is typically fully turbulent.
Furthermore, the thermal effects play a significant role, which cannot be neglected when trying to understand the causes and effects of the urban heat island.
And importantly, the flow is heavily influenced by urban vegetation, which may block or deflect the air flow, as well as affect the turbulence levels. It also has a capacity to filter the aerosol particles from the polluted air. Its role in the pollution dispersion is thus of increasing interest of researchers \citep{LitschkeKuttler08,Janhall15}.

Both Reynolds-averaged Navier-Stokes (RANS) and Large Eddy Simulation (LES) approaches are used for urban scale CFD simulations \citep{Blocken15}, however, RANS approach is the usual choice when computational costs are of consideration. Thermal effects are often neglected, as it makes possible to use the incompressible flow model often present in the publicly available commercial and open-source CFD solvers \citep{BlockenEtAl12,VranckxEtAl15}.

Range of related tree canopy models were developed for the \ke turbulence model \citep{SvenssonHaeggkvist90,Green92,LiuEtAl96,KatulEtAl04}, and although no specific one is universally accepted, variants of the model were successfully employed for urban flow problems with vegetation \citep{SteffensEtAl12,KenjeresKuile13}.

Aerosol particles are removed from the polluted air by the dry deposition process inside the vegetation.
\citet{PetroffEtAl08a} list four main mechanical processes playing role in the dry deposition: Brownian diffusion, interception, inertial impaction and sedimentation. This rich background of the dry deposition process is rarely fully reflected in the microscale CFD models.
In some small scale dispersion studies the deposition was not considered at all, such as in \citep{BuccolieriEtAl11}, where the authors cited the negligible filtering potential of the vegetation as the main reason. Other used a constant rate of pollutant deposition \citep{VranckxEtAl15}, or various models based on the underlying processes of different level of detail \citep{TiwaryEtAl05,Bruse07,SteffensEtAl12}. However, there is currently no generally accepted and extensively validated dry deposition model for the microscale vegetation flow problems.

In this paper, we present the description and validation of a finite volume solver for microscale urban flows.
The solver is based on RANS equations suitable for modelling of the atmospheric boundary layer flows, including the thermally driven flows. The vegetation effects are included in the \ke turbulence model, and a detailed, physically based dry deposition model is incorporated in the pollutant transport equation.
A preliminary version of the solver was used for designing an optimal near-road barrier \citep{SipBenes16-enumath}.

The solver is validated on four test cases aimed to test different aspects of the model: the air flow and pollutant dispersion over a 2D hill, a rising thermal bubble, the flow through and above the forest canopy, and a hedgerow filtering the particle-laden flow.

\section{Model description}

\subsection{Governing equations}
\label{sec:governing-equations}

\subsubsection{Fluid flow}
\label{sec:fluid-flow}

The fluid flow is described by the Reynolds-averaged Navier-Stokes (RANS) equations. We assume that the flow may be modelled as incompressible. Furthermore, we employ the Boussinesq approximation, which states that the variation of density from its background state may be neglected everywhere except in the gravity term, and instead of the equation for total energy we prefer to use the equation for potential temperature.
Let pressure, density and potential temperature be decomposed into their background components in hydrostatic balance dependent only on the vertical coordinate, denoted with subscript $_0$, and their fluctuations, denoted with superscript~$^*$: $p = p_0 + p^*$, $\rho = \rho_0 + \rho^*$, and $\theta = \theta_0 + \theta^*$. Then the RANS equations read
\begin{align}
  \div \bu & = 0, \label{eq:eqs-1} \\
  \dd{\bu}{t}  + \div (\bu \otimes \bu)
  & = - \frac{1}{\rho_{\mathrm{ref}}} \nabla p^* + \div ((\nu + \nu_T) \nabla \bu) + \bs{f}_g + \bs{S}_u,  \label{eq:eqs-2} \\
  \dd{\theta}{t} + \div (\theta \bu)
  & = \div \left( \left(k_L/\rho_\mathrm{ref} c_p + \nu_T/Pr_T \right) \nabla \theta \right) + \frac{q}{c_p}, \label{eq:eqs-3}
\end{align}
where $\bs{u} = (u_1, u_2, u_3)$ is the velocity vector, $\rho_\mathrm{ref}$ is the background density at some reference point, typically the lowest point in the domain, $\nu$ and $\nu_T$ are the laminar and turbulent kinematic viscosity, $\bs{f}_g = (0, 0, \frac{\theta^*}{\theta_0} g)^T$ and $g$ is the gravitational acceleration magnitude. The term $\bs{S}_u$ stands for the momentum loss in the vegetation, and is described in Sec. \ref{sec:model-canopy}.
In the potential temperature equation, $k_L$ is the heat conduction coefficient of the air, and $c_p$ is the specific heat at constant pressure. The turbulent Prandtl number $Pr_T$ is here taken equal to 0.9.
Finally, $q$ is the density of heat sources per unit mass.

\subsubsection{Turbulence model}
\label{sec:turbulence-model}

The turbulence is modelled by a standard \ke model proposed by \citet{LaunderSpalding74}, which describes the behaviour of the turbulence kinetic energy (TKE) $k$ and its dissipation $\epsilon$. We use it in the following form:
\begin{align}
  \dd{(\rho k)}{t}
  + \div (\rho k \bu)
  = &
  \div \left(\left( \mu + \frac{\mu_T}{\sigma_k} \right) \nabla k \right)
  + P_k
  - \rho \epsilon + \rho S_k
  \label{eq:tke}, \\
  \dd{(\rho \epsilon)}{t}
  + \div (\rho \epsilon \bu)
  = &
  \div \left(\left( \mu + \frac{\mu_T}{\sigma_\epsilon} \right) \nabla \epsilon \right)
  + C_{\epsilon_1} \frac{\epsilon}{k} P_k
  - C_{\epsilon_2} \rho \frac{\epsilon^2}{k}
  + \rho S_\epsilon
  \label{eq:eps},
\end{align}
where $\mu$ and $\mu_T$ are the laminar and turbulent dynamic viscosity,
and the production of the turbulent kinetic energy has the form
\begin{equation}
  P_k = \tau_{ij}^R \dd{u_i}{x_j},
\end{equation}
with $\tau_{ij}^R = \mu_T \left(\dd{u_i}{x_j} + \dd{u_j}{x_i} \right) - \frac{2}{3} \rho k \delta_{ij}$. The terms $S_k$ and $S_\epsilon$ model the vegetation effects, and are described in detail in Sec. \ref{sec:model-canopy}.
The turbulent viscosity is coupled to the modelled variables through the relation
\begin{equation}
\mu_T = C_\mu \rho \frac{k^2}{\epsilon}.
\end{equation}

The effects of buoyancy are not included in the turbulence model. We have ignored these effects under the justification of our focus on the geometrically complex urban areas, where the turbulence generation by shear overshadows the buoyancy effects.

Constants of the standard \ke model are alternated, so that the model can sustain a horizontally homogenous flow over a flat topography without any obstacles when suitable boundary conditions (see Sec. \ref{sec:bound-conds}) are used. If that is to be, the constants of the model and the von Kármán constant $\kappa$ have to be tied together by the expression \citep{RichardsHoxey93}
\begin{equation}
  \sigma_\epsilon = \frac{\kappa^2}{( C_{\epsilon_2} - C_{\epsilon_1} ) \sqrt{C_\mu}}.
  \label{eq:keps-constants-abl}
\end{equation}
In this work, we generally use the following set of constants satisfying this relation:  $C_{\epsilon_1} = 1.44$, $C_{\epsilon_2} = 1.92$, $C_\mu = 0.09$, $\sigma_k = 1.0$, $\sigma_\epsilon = 1.167$, although also a different choice of $C_\mu$ is tested for vegetation flows in Sec. \ref{sec:forest-canopy-flow}.

\subsubsection{Pollutant transport}
\label{sec:pollutant-transport}

We assume that the pollutant is present in such small concentrations that it does not change the properties of the air. In that case, its transport can be modelled using the passive scalar equation,
\begin{equation}
  \frac{\partial c}{\partial t}
  + \div (c \bu)
  =
  \div \left(\frac{\nu_T}{Sc_T} \nabla c \right)
  + S_g + S_c.
  \label{eq:passive-scalar-eq}
\end{equation}
where $c$ is the mass concentration of the pollutant, and $Sc_T$ is the turbulent Schmidt number, value of which is chosen on a case by case basis.
The molecular diffusion of the pollutant is neglected here, since the turbulent diffusion is typically much larger in the atmospheric boundary layer.
The term $S_g$ represents the gravitational settling of particles, and $S_c$ stands for the source (and sink) term, which includes the sink caused by the dry deposition on the vegetation described in Sec. \ref{sec:dry-deposition-model}.

The gravitational settling term has the form
\begin{equation}
  S_g = - \div (c \bs{u_s}),
  \label{eq:grav-settling}
\end{equation}
where $\bs{u_s} = (0, 0, -u_s)$ is the gravitational settling velocity vector oriented towards the ground. The settling velocity of the particle of the diameter $d_p$ and density $\rho_p$ is given by the Stokes' equation,
\begin{equation}
  u_s = (d_p^2 \rho_p g C_C)/(18 \mu),
  \label{eq:settling-vel}
\end{equation}
where
\begin{equation}
  C_C = 1 + 2\frac{\lambda}{d_p} \left(1.257 + 0.4 \exp \left(-1.1 \frac{d_p}{2\lambda} \right) \right)
\end{equation}
is the Cunningham correction factor and $\lambda = \SI{0.066}{\um}$ is the mean free path of the particle in the air \citep{SeinfeldPandis06}.

\subsection{Boundary conditions for ABL flows}
\label{sec:bound-conds}

For typical ABL flow simulations, we use the inlet boundary conditions and wall functions given by \citet{RichardsHoxey93}. We will give the formulas for velocity, turbulence kinetic energy, and its dissipation here to avoid needless repetition in further text. Detailed description of the boundary conditions is given for each test case separately.

At the inlet, the wind velocity is given by the log wind profile,
\begin{equation}
  u(z) = \frac{u_*}{\kappa} \ln \left( \frac{z + z_0}{z_0} \right),
\end{equation}
where $u_*$ is the friction velocity, $\kappa$ is the von Kármán constant and $z_0$ is the surface roughness length. The turbulence kinetic energy and its dissipation are set to
\begin{equation}
  k (z) = \frac{u_*^2}{\sqrt{C_\mu}},\quad \epsilon(z) = \frac{u_*^3}{\kappa (z + z_0)}.
  \label{eq:inlet-k-eps}
\end{equation}

At the ground, wall functions are used.
Normal component of velocity vector is set to zero, and the wall shear stress acting on the near-ground cell with the centre at height $z_p$ is prescribed as
\begin{equation}
  \tau_w^R = \frac{\kappa C_\mu^{0.25} k^{0.5} U}{\ln \left(\frac{z_p + z_0}{z_0} \right)}.
\end{equation}
Zero value of turbulence kinetic energy is prescribed at the ground, and its production in the near-wall cells is given as \citep{ParenteEtAl11}
\begin{equation}
  P_{k,w}    = \frac{(\tau_w^R)^2}{\rho \kappa C_\mu^{0.25} k^{0.5} (z_p + z_0)}.
\end{equation}
The equation for TKE dissipation is not solved in the near-wall cells, and its value is instead calculated as
\begin{equation}
  \epsilon_w = \frac{C_\mu^{0.75} k^{1.5} }{\kappa (z_p + z_0)}.
\end{equation}

\subsection{Vegetation models}
\label{sec:vegetation-models}

We model the vegetation as a porous zone, described by its \textit{leaf area density} (LAD) profile.
Leaf area density (given in \si{\m\squared\per\m\cubed}) is defined as a total one-sided leaf area per unit volume.
In the problems presented here the vegetation is horizontally homogeneous, and its LAD profile therefore varies only with the vertical coordinate.

\subsubsection{Canopy flow model}
\label{sec:model-canopy}

\citet{WilsonShaw77} summarized the effects of the vegetation on the air flow in four points:
\begin{enumerate}
\item It extracts the momentum from the mean flow due to the aerodynamic drag of the vegetation elements.
\item The extracted energy is converted to the turbulence kinetic energy in the wakes formed behind the obstructions.
\item The energy of the large-scale turbulent motions is transformed into smaller scale turbulent motions, enhancing the turbulent dissipation in the canopy.
\item The turbulence kinetic energy production is increased due to the heat transfer between the plant surface and the air.
\end{enumerate}
The last phenomenom is often neglected as having little effect, however, all other mechanisms should be reflected in the vegetation model.
We adopted the model described by \citet{KatulEtAl04}, which specifies the additional terms in the momentum and turbulence equations as follows.
The momentum sink  in Eq. (\ref{eq:eqs-2}) caused by the form drag is given as
\begin{equation}
  \bs{S}_u = - (C_d \LAD\ \umag) \bu,
\end{equation}
where $U$ is the velocity magnitude, and $C_d$ is the drag coefficient. Typical values are $0.1 \leq C_d \leq 0.5$ \citep{KatulEtAl04,EndalewEtAl09}. Viscous drag is considered negligible relative to the form drag, and is not included in the momentum sink.
The source term in the TKE equation (\ref{eq:tke}) reads as
\begin{equation}
  S_k = C_d \LAD (\beta_p \umag^3  - \beta_d \umag k).
\end{equation}
The positive part of the term represents the energy converted from the mean flow kinetic energy to the turbulence kinetic energy, and parameter $\beta_p$ is the fraction of the converted energy. The negative part reflects the short-circuiting of the Kolgomorov cascade. Finally, the term in the dissipation equation (\ref{eq:eps}) is
\begin{equation}
  S_\epsilon = C_{\epsilon_4} \frac{\epsilon}{k} S_k.
  \label{eq:veg-eps}
\end{equation}
Its form was derived using the dimensional analysis.
Constants of the model are set to $\beta_p = 1.0, \beta_d = 5.1, C_{\epsilon_4} = 0.9$.

\subsubsection{Dry deposition model}
\label{sec:dry-deposition-model}

The dry deposition is a complex process depending on the physical and chemical properties of the aerosol, micrometeorological conditions, or vegetation surface properties. \citet{PetroffEtAl08a} identified four main mechanisms of the dry deposition:
\begin{itemize}
\item \textit{Brownian diffusion}, affecting predominantly small particles with diameter ${d_p < \SI{0.1}{\um}}$.
\item \textit{Interception}, occurring when a particle following the streamline passes too close to the obstacle and gets captured on it.
\item \textit{Impaction}, i.e. a collision of the particle which does not follow the streamline with the obstacle due to the inertia of the particle. Impaction is further differentiated into the \textit{inertial} and the \textit{turbulent impaction} by its cause.
\item \textit{Sedimentation}, which stands for the collision of the particle with the obstacle due to the downward motion of the particle caused by the gravitational force. Sedimentation is the dominant process for particles sizes $d_p > \SI{10}{\um}$.
\end{itemize}

\begin{figure}[ht]
  \centering
  \begin{tabular}{@{}p{0.35\linewidth}@{\qquad}p{0.35\linewidth}@{}}
    \subfigimg[width=0.95\linewidth]{A)}{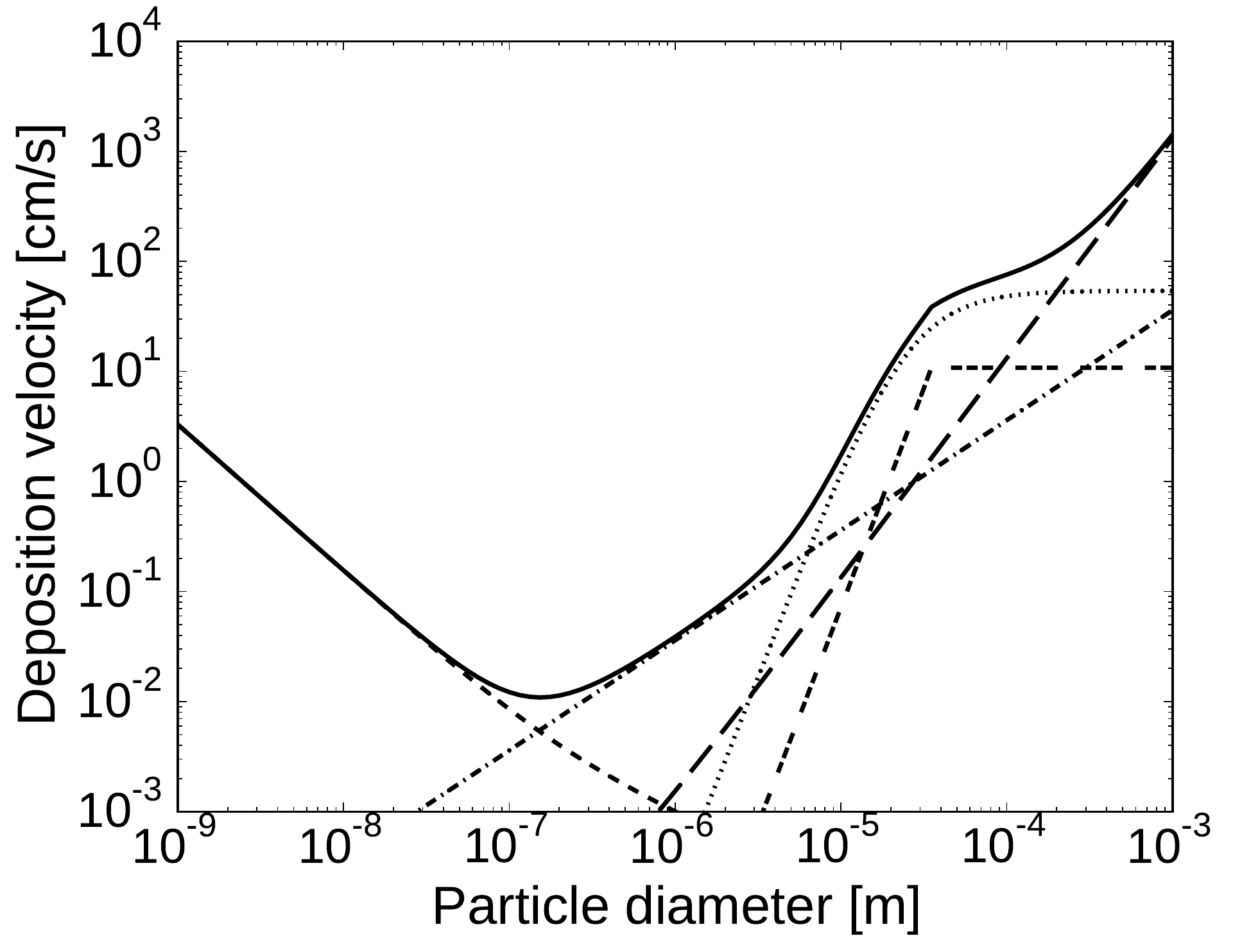} &
    \subfigimg[width=0.95\linewidth]{B)}{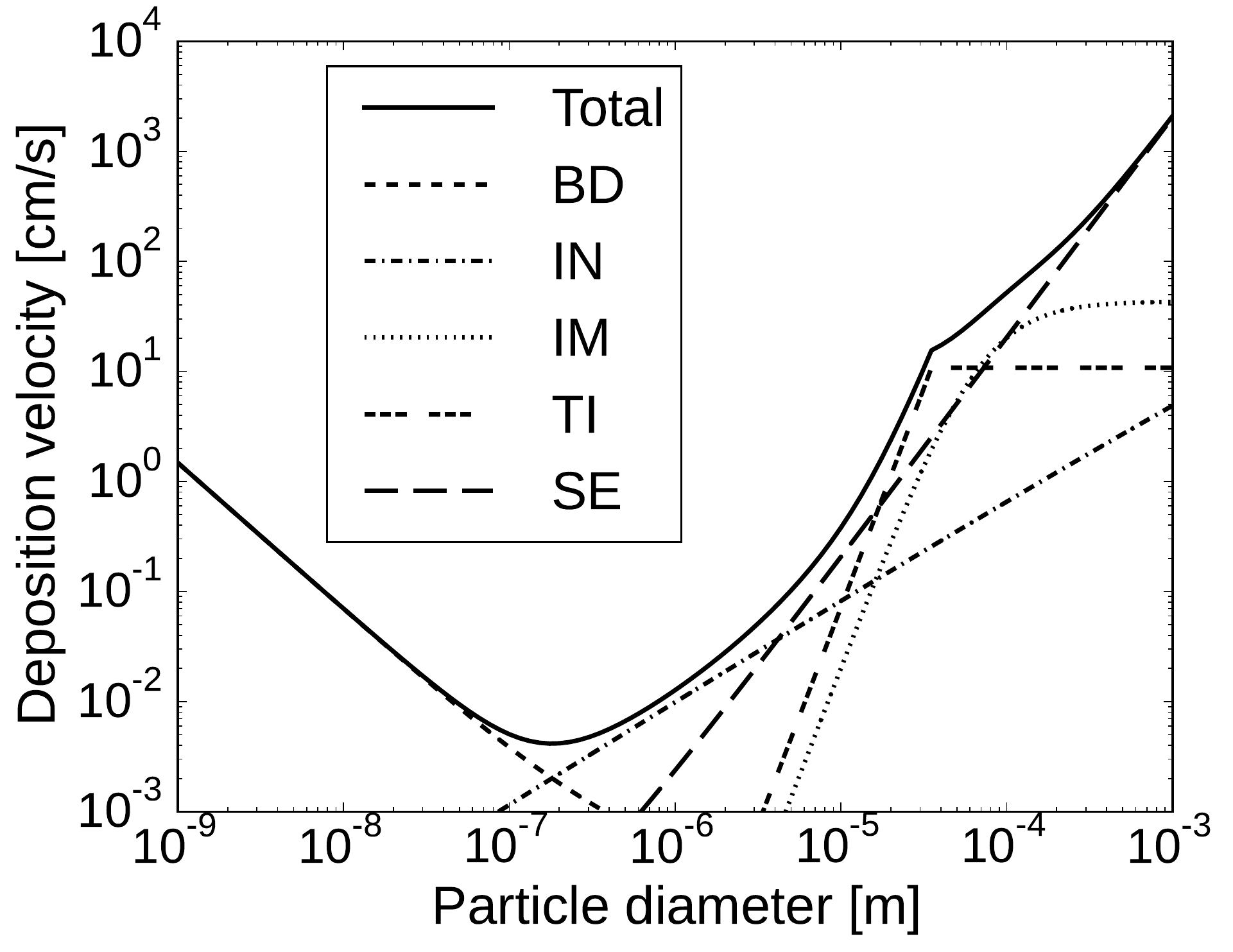} \\
  \end{tabular}
  \caption{Examples of the deposition velocity dependence on particle size, as given by the model of \citet{PetroffEtAl08b} and \citet{PetroffEtAl09}. Contributions of the underlying physical processes are shown as well. (A) Needle-like elements with diameter \SI{0.003}{\m}. (B) Broadleaf elements with diameter \SI{0.03}{\m}.
    Parameters of the model: Particle density $\rho_p = \SI{1000}{\kg\per\m\cubed}$, wind velocity $\umag = \SI{1}{\m\per\s}$, local friction velocity $u_f = \SI{0.1}{\m\per\s}$, plagiophile leaf distribution. Abbreviations: BD = Brownian diffusion, IN = interception, IM = inertial impaction, TI = turbulent impaction, SE = sedimentation.
  }
  \label{fig:depvel}
\end{figure}

We model the dry deposition through the term
\begin{equation}
  S_c = - \LAD u_d c
  \label{eq:dry-dep}
\end{equation}
in Eq. (\ref{eq:passive-scalar-eq}). Here $u_d$ is the \textit{deposition velocity}, usually given in \si{\cm\per\s}. Its values differ by orders of magnitudes depending on the properties of the particles, vegetation, and the environment \citep{LitschkeKuttler08}, and a detailed model capturing this behaviour is therefore desirable.
We have adopted the model described in \citep{PetroffEtAl08b} and \citep{PetroffEtAl09} for vegetation with needle-like leaves and broadleaves respectively, which includes all of the processes given above.
The authors presented formulas for the deposition velocities (or \textit{collection velocities} in their terminology) associated with each of the mechanical processes, and calculated the total deposition velocity as their sum. For brevity, we omit the detailed description of the model, and refer to the original publications instead. The dependence of the deposition velocity on the particle diameter when using this model is shown on Fig. \ref{fig:depvel}.

\subsection{Numerical methods}
\label{sec:numerical-methods}

\subsubsection{Artificial compressibility}

The method of artificial compressibility is employed for the numerical solution of system (\ref{eq:eqs-1}-\ref{eq:eqs-3}). Using this method, the system may be rewritten in the vector form,
\begin{equation}
  \bs{\Gamma} \dd{\bs{W}}{t}
  + \sum_{j=1}^{3}\dd{\bs{F}_j}{x_j}
  = \sum_{j=1}^{3} \dd{\bs{R}_j}{x_j}
  + \bs{Q},
\end{equation}
where $\bs{W} = (p^*, u_1, u_2, u_3, \theta)^T$ is the state vector, $\bs{F}_j$ are the inviscid fluxes (including the pressure term), $\bs{R}_j$ are the viscous fluxes, and $\bs{Q}$ are the sources and sinks. For the artificial compressibility matrix $\bs{\Gamma}$ we use the generalized formulation given by \citet{Turkel85},
\begin{equation}
  \bs{\Gamma} = \left(
    \begin{array}{ccccc}
      1/\beta      & 0 & 0 & 0 & 0 \\
      u_1/\beta    & 1 & 0 & 0 & 0 \\
      u_2/\beta    & 0 & 1 & 0 & 0 \\
      u_3/\beta    & 0 & 0 & 1 & 0 \\
      \theta/\beta & 0 & 0 & 0 & 1 \\
    \end{array}
  \right),
  \label{eq:ac-generalized}
\end{equation}
where $\beta$ is the artificial compressibility parameter, set to the value 100 everywhere in this work. Usage of the artificial compressibility method changes the temporal behaviour of the solution, and so the form given above is suitable only for steady-state problems, for which the temporal derivative vanishes when the solution is found.
For unsteady problems, a preconditioned time derivative in a pseudo time $\tau$ is added to the original equations,
\begin{equation}
  \bs{\Gamma} \dd{\bs{W}}{\tau}
  + \bs{A} \dd{\bs{W}}{t}
  + \sum_{j=1}^{3}\dd{\bs{F}_j}{x_j}
  = \sum_{j=1}^{3} \dd{\bs{R}_j }{x_j}
  + \bs{Q},
  \label{eq:dual-ts}
\end{equation}
where
\begin{equation}
  \bs{A} = \left(
    \begin{array}{ccccc}
      0 & 0 & 0 & 0 & 0 \\
      0 & 1 & 0 & 0 & 0 \\
      0 & 0 & 1 & 0 & 0 \\
      0 & 0 & 0 & 1 & 0 \\
      0 & 0 & 0 & 0 & 1 \\
    \end{array}
  \right).
  \label{eq:abl-mat-a}
\end{equation}
In every physical time step, we advance the solution in the pseudo time until the pseudo time derivative $\ddi{\bs{W}}{\tau}$ vanishes.

\subsubsection{Spatial discretization}

The governing equations are solved using a finite volume method on unstructured grids.
The AUSM$^+$-up numerical flux \citep{Liou06} is employed to calculate the inviscid fluxes. Second order scheme is obtained by utilizing the linear reconstruction process, where the gradients in the computational cells are calculated using the least square method \citep[Chap. 5]{Blazek01}. Artificial extrema are prevented by using the Venkatakrishnan limiter \citep{Venkatakrishnan95}.
Gradients on the faces of the computational cells necessary for evaluation of the viscous fluxes are calculated by the diamond cell scheme using the formulation given by \citet{Karel14}.

\subsubsection{Temporal discretization}

For steady-state problems where the accuracy in time is of no concern we use the implicit backward Euler method. For a system of ordinary differential equations $\dd{\bs{Y}}{t} = \bs{G} (\bs{Y}, t)$ the $n$-th step of the method is written as
\begin{equation}
 \bs{Y}_{n+1} - \bs{Y}_n  = \Delta t_n \bs{G} (\bs{Y}_{n+1}, t_{n+1}),
 \label{eq:impl-euler}
\end{equation}
where $\Delta t_n$ is the time step length. We adapt the time step based on the number of iterations needed for the solution of the linear systems. The time step is thus increased during the run, and the steady-state is reached faster.

Unsteady problems are solved by the second order BDF2 method,
\begin{equation}
  3 \bs{Y}_{n+1} - 4 \bs{Y}_n  + \bs{Y}_{n - 1} = 2 \Delta t_n \bs{G} (\bs{Y}_{n+1}, t_{n+1}).
\label{eq:bdf2}
\end{equation}
The history data needed by the method are missing in the first step, therefore the Euler method given above is used for the initialization of the method.

The use of an implicit method results in a need to solve a system of nonlinear equations in every time step. This is done by a Jacobian-free Newton-Krylov method \citep{KnollKeyes04}. The method allows us to avoid the computationally intensive evaluation of the Jacobian in every time step by using a Krylov method to solve the inner linear systems, for which only the ability to calculate the matrix-vector product (and not the knowledge of the Jacobian) is required. GMRES method \citep{SaadSchultz86} is employed as the inner system solver.
To accelerate its convergence, ILU(k) preconditioner \citep{ChanVanderVorst01} is utilized. To reduce the computational load, the preconditioner is calculated only in every 20-th time step, and the matrix coloring is used so that fewer function evaluation are needed for the Jacobian calculation.

\subsection{Implementation}
The solver is written in C++ programming language, using the in-house framework used previously for the problems of electric discharge propagation \citep{Karel14}.
For the solution of the system of nonlinear equation in every time step the PETSc library \citep{petsc-web-page} is employed.

\section{Test cases}

\subsection{Flow around a hill}
\label{sec:flow-around-hill}

The flow over an isolated hill is among the most used test cases for the CFD solvers aimed at atmospheric boundary layer flows. In various configurations, it was often investigated through the wind tunnel experiments as well as numerical simulations. Here we present a comparison of the results obtained by our solver with the experimental data from the RUSHIL wind tunnel study \citep{KhurshudyanEtAl81}, obtained from the ERCOFTAC QNET-CFD test case database \citep{ErcoftacQnetAC5-05}. In addition to the flow data over a 2D hill, the pollution dispersion over a hill ridge of the same shape is compared with the measured data.

\begin{figure}[ht]
  \centering
  \includegraphics[width=0.5\textwidth]{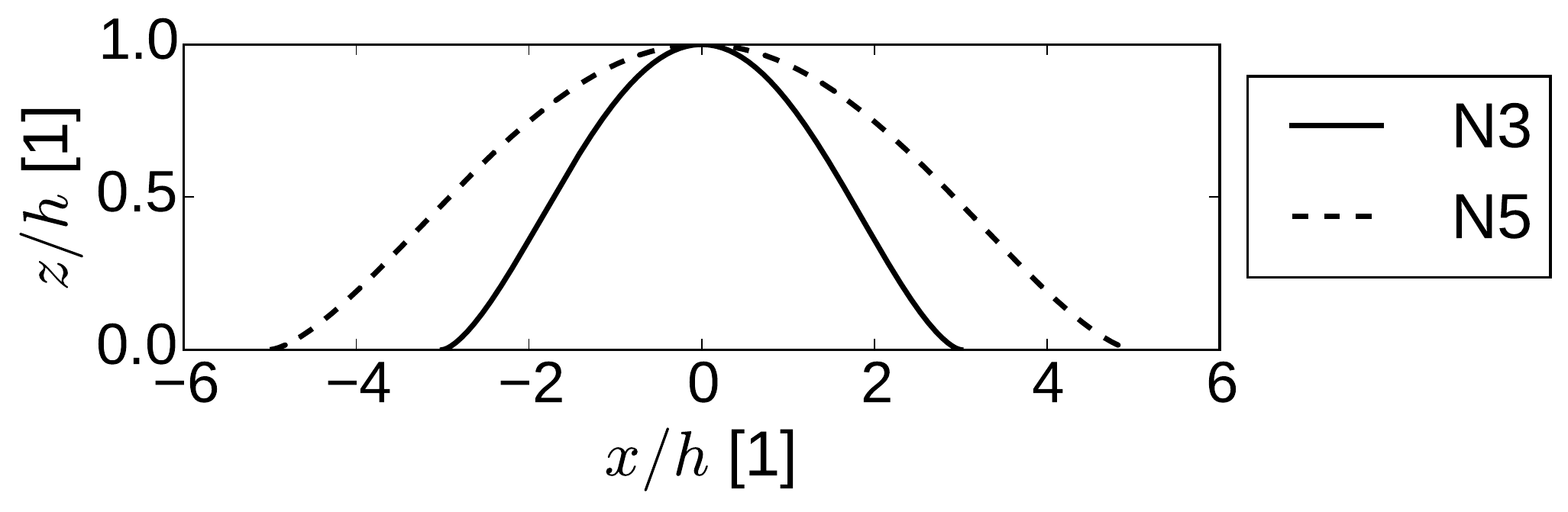}
  \caption{Shape of the N3 and N5 hills.}
  \label{fig:val-hill-shape}
\end{figure}

The performed numerical simulations reproduced the main aspects of the wind tunnel experiment. The 2D hill of a height $h$ and a half-width $a$ is described by the parametric equations,
\begin{align}
  x & = \frac{1}{2} \xi \left( 1 + \frac{a^2}{\xi^2 + m^2 (a^2 - \xi^2)} \right),  \nonumber \\
  z & =  \frac{1}{2} m \sqrt{a^2 - \xi^2} \left( 1 - \frac{a^2}{\xi^2 + m^2 (a^2 - \xi^2)} \right)
      \quad \textrm{for}\ \xi \in [-a; a],
\end{align}
where $m = \frac{h}{a} + \sqrt{\left(\frac{h}{a}\right)^2 + 1}$. Two geometrical variants with different aspect ratios $n = a/h$ of 3 and 5 (marked in the following text as N3 and N5 respectively) were investigated. In both cases, the height of the hill was $h = \SI{0.117}{\m}$. Maximal slope of the N3 and N5 hills was 26\degree\ and 16\degree\ respectively. Shape of the hills is depicted on Fig. \ref{fig:val-hill-shape}.
The computational domain spanned from -20$h$ to 40$h$ in the horizontal direction and from 0 to 13$h$ in the vertical direction, the centre of the hill was placed at $x = 0$.

The boundary conditions were specified similarly as in the numerical simulation of the same problem by \citet{CastroApsley97}. In the following text, $D$ marks the depth of the boundary layer, defined such that the velocity given by the log profile wind profile at the top of the boundary layer is equal to the free stream velocity, $u(D) = u_\infty$.
At the inlet, the log wind profile was prescribed for the velocity inside the boundary layer, i.e. for $z < D$, and $u(z) = u_\infty$ for $z \geq D$. Turbulence kinetic energy was given by $k(z) = C_\mu^{-0.5}u_*^2 (1 - \frac{z}{D})$ for $z < 0.9D$ and extended as a constant above, and its dissipation was set to $\epsilon = (C_\mu^{0.75} k^{1.5})/(\kappa z)$. Inlet potential temperature was set to constant $\theta = \SI{289}{K}$, and the homogeneous Neumann boundary condition (BC) was used for the pressure perturbation, i.e. $\ddi{p^*}{\bs{n}} = 0$.
At the outlet and at the top of the domain, zero pressure perturbation was prescribed, and all other variables were extrapolated from inside using the homogeneous Neumann BC.
At the ground, wall functions, as described in Sec. \ref{sec:bound-conds}, were employed, together the homogeneous Neumann BC for pressure perturbation and potential temperature.

Parameters of the boundary layer were as follows: friction velocity $u_* = \SI{0.178}{\m\per\s}$, von Kármán constant $\kappa = 0.4$, roughness length $z_0 = \SI{0.157}{\mm}$, free stream velocity $u_\infty = \SI{4}{\m\per\s}$. Depth of the boundary layer was thus $D = \SI{1.258}{\m}$.

Since in the wind tunnel measurements the pollutant was released from a point source, the 2D numerical model was not sufficient for the dispersion study.
It was therefore studied using the flow field calculated in 2D, which was then extended to 3D, so that the flow field represented the flow above the ridge of the same shape as the 2D hill. The domain was extended in the lateral direction to $[-8h; 8h]$.
A point source of the pollutant was placed on the midplane of the domain at the upwind base (horizontal position of the source $x_s = -a$), at the summit ($x_s = 0$), or at the downwind base ($x_s = a$) of the hill. The height of the source was $h_s = h/4$ in every case.
Zero mass concentration was prescribed at the inlet, and the homogeneous Neumann boundary conditions were used on all other boundaries.

The flow field was calculated on a 2D structured computational grid with 340 $\times$ 100 cells, graded so the grid was finer around the hill. Smallest cell had size 0.06$h$ $\times$ 0.043$h$, and the cells were expanded away from the hill with the expansion ratio 1.017 in horizontal direction and 1.020 in vertical direction.
For the pollutant dispersion, the grid was extruded to 3D. The number of cells in the lateral direction was 75. The grid was refined in the middle of the domain, so that the lateral size of the smallest cells was 0.051$h$, and the cells were expanded to the sides with the expansion factor 1.066.

\subsection{Warm bubble}
\label{sec:warm-bubble}

This test case serves to show that the solver can properly capture the unsteady thermally driven flow in the ABL. The settings replicate the rising thermal bubble test case from \citep{GiraldoRestelli08}, which was based on the previous formulation of a similar test by \citet{Robert93}.

A bubble of hot air is placed in the atmosphere with a constant potential temperature. The air in the two dimensional domain is initially at rest, and the thermal effects force the bubble to rise through the environment. At the beginning, the unperturbed atmosphere has the potential temperature $\theta = \SI{300}{\K}$, and the bubble is created by increasing the potential temperature by
\begin{equation}
  \theta^* = \frac{\theta_c}{2} \left( 1 + \cos \left(\frac{\pi r}{r_c} \right) \right) \quad \textrm{when } r \leq r_c,
\end{equation}
where $\theta_c = \SI{0.5}{\K}$, $r = \sqrt{(x - x_c)^2  + (z - z_c)^2}$, the centre of the bubble is placed to $(x_c, z_c) = (500, 350)\ \si{ \m}$, and its diameter to $r_c = \SI{250}{\m}$. The initial velocity is set to zero, and the initial pressure is set using the barometric formula, so that the air is at hydrostatic balance.

The computational domain had size $[0, 1000]\ \si{\m} \times [0, 1000]\ \si{\m}$. Its boundary conditions were all set as slip walls, i.e. the velocity normal to the wall was set to zero, and for the velocity parallel to the wall as well as for all other variables the homogeneous Neumann BC was applied. The flow was modelled as inviscid.

The dependence on the mesh resolution was assessed using four Cartesian meshes with uniform spatial resolution of 20, 10, 5, and 2.5 m in both vertical and horizontal direction.

The evolution of the system was simulated for $t \in [0, 700]\ \si{\s}$.
The length of one time step was set as $\Delta t = \SI{1}{\s}$.

\subsection{Forest canopy flow}
\label{sec:forest-canopy-flow}

The described \ke model of the vegetation flow was tested on a problem of the flow in and above a forest canopy.
\citet{DupontEtAl11} presented field measurements and large eddy simulations of the flow over a maritime pine forest. The forest of an average height $h = \SI{22}{\m}$ had a dense crown layer roughly \SI{8}{\m} thick and an open trunk space. A 41.5 m high measurement tower was located $9h$ from the edge of the forest in the north-west direction, while a homogeneous forest with a fetch greater than 1 km stood in the opposite direction from the tower.
In addition to the tower, a smaller mast of height 13 m was located $4h$ from the edge of the forest.
This configuration allowed to investigate both the flow over a homogenous forest as well as the edge effects based on the wind direction.

In their validation of the vegetation model we adopted, \citet{KatulEtAl04} changed the constant of the \ke model $C_\mu$ to 0.03 to provide a better match of the turbulent viscosity in the unperturbed atmosphere to the measured values of typical neutral ABL flows.
Here we have tested both this value (with constant $\sigma_\epsilon$ changed accordingly to satisfy Eq. (\ref{eq:keps-constants-abl})), as well as the constants given before with $C_\mu = 0.09$. Both sets of constants are listed in Tab. \ref{tab:veg-fc-ke-consts}.

\begin{table}[h]
  \centering
  \begin{tabular}{rrrrr}\hline
           $C_\mu$ & $C_{\epsilon_1}$ & $C_{\epsilon_2}$ & $\sigma_k$ & $\sigma_\epsilon$ \\
     \textbf{0.03} &             1.44 &             1.92 &        1.0 &             1.92  \\
     \textbf{0.09} &             1.44 &             1.92 &        1.0 &             1.167 \\ \hline
  \end{tabular}
  \caption{Tested sets of constants of the \ke model for the forest canopy flow.}
  \label{tab:veg-fc-ke-consts}
\end{table}

\begin{figure}[ht]
  \centering
  \begin{tabular}{@{}p{0.25\linewidth}@{\qquad}p{0.6\linewidth}@{}}
    \subfigimg[height=0.2\textwidth]{A)}{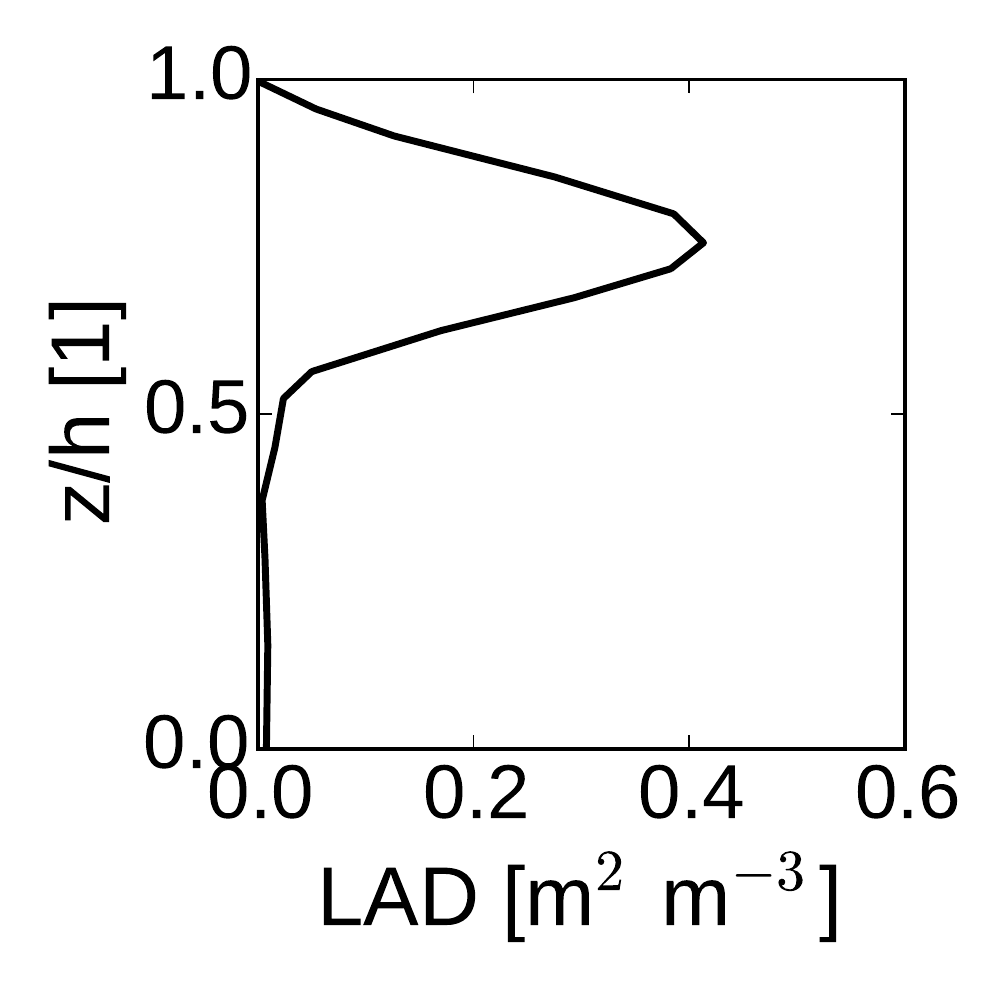} &
    \subfigimg[height=0.2\textwidth]{B)}{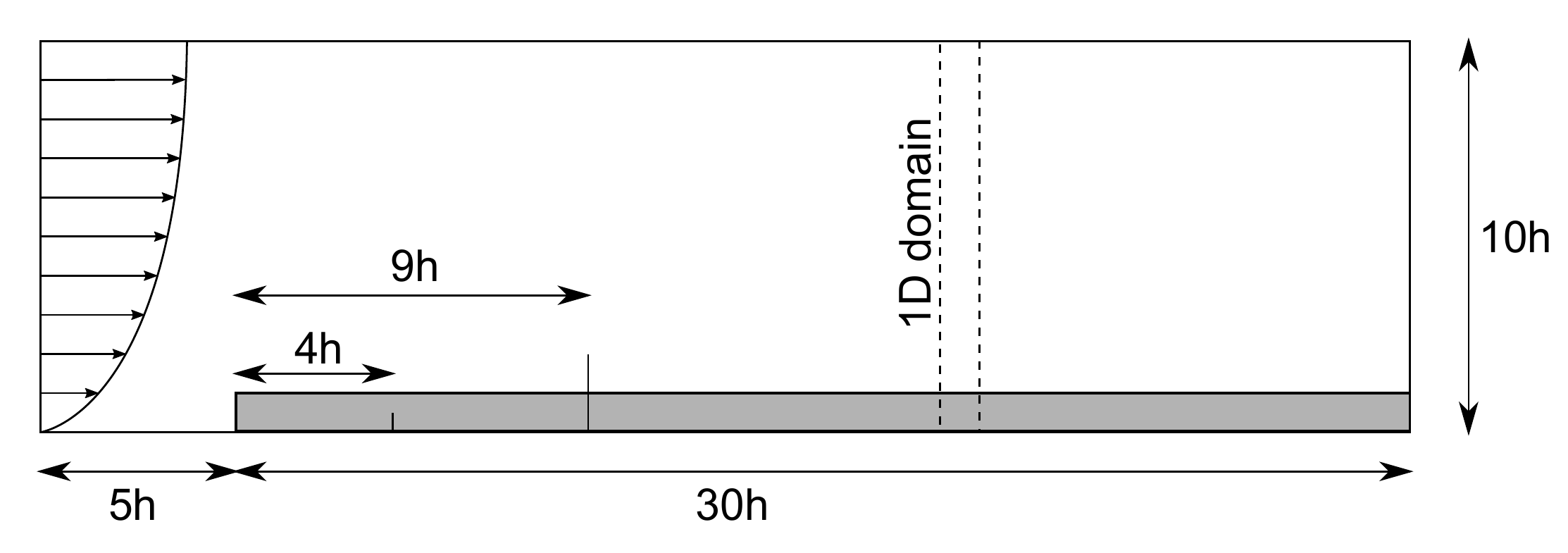} \\
  \end{tabular}
  \caption{Flow over a forest canopy. (A) LAD profile of the pine forest. Vertical coordinate is normalized by the height of the forest $h = \SI{22}{\m}$. Taken from \citep{DupontEtAl11}. (B) Sketch of the 2D domain. Positions of the measurement masts at $4h$ and $9h$ from the edge of the forest are marked by solid lines. Computational domain of the 1D model of the flow over a homogeneous forest is represented by the dashed lines.}
  \label{fig:veg-fc}
\end{figure}

We investigated the flow over a homogenous forest using a 1D model, and the edge flow using a 2D model (Fig. \ref{fig:veg-fc}B).

\paragraph{Homogeneous forest}
The 1D vertical model was constructed as follows.
The flow was modelled between the ground and a height $10h$.
At the ground the wall functions were prescribed with roughness length $z_0 = \SI{0.03}{\m}$. Pressure perturbation and potential temperature were extrapolated from inside using homogeneous Neumann BC.
The upper boundary was modelled as a slip wall.
The flow was driven by prescribed horizontal pressure gradient acceleration
$\frac{1}{\rho_\mathrm{ref}}\dd{p^*}{x} = \SI{0.001}{\m\per\s\squared}$.
The leaf area density of the canopy is pictured in Fig. \ref{fig:veg-fc}A. The vegetation drag coefficient was set to $C_d = 0.26$. Atmosphere was considered to be neutrally stratified with potential temperature $\theta = \SI{300}{K}$.

The vertical interval was discretized by 100 cells. The cells inside the canopy (i.e. for $z < h$) had height $0.023 h$, and the cells above were continuously expanded with an expansion factor 1.06.

\paragraph{Edge flow}
To capture the behaviour of the flow over the edge of the forest a 2D model was employed (Fig. \ref{fig:veg-fc}B). The size of the computational domain was chosen to allow the flow to stabilize before reaching the outlet. \citet{DupontEtAl11} evaluated that the adjustment region extends to around $22h$ from the forest edge in this case. Based on this, the computational domain was set to extend to $30h$ downstream from the edge of the forest, and $5h$ upstream. The height $10h$ is same as in the 1D model.

Boundary conditions at the ground were the same as in the 1D model.  At the outlet zero pressure fluctuation was prescribed, and the homogeneous Neumann BC for all other variables was used. Log wind profile was prescribed at the inlet with the friction velocity $u_* = \SI{0.23}{\m\per\s}$, and it was complemented by the turbulence inlet profiles described in Sec. \ref{sec:bound-conds}. Potential temperature was set to $\theta = \SI{300}{K}$. Finally, the homogeneous Neumann BC was prescribed at the top of the domain for all variables except for the pressure, which was calculated so that the total pressure $p_0 = p + \frac{1}{2}\rho\umag^2$ was constant at the top boundary.
The same leaf area density profile as in 1D case (Fig. \ref{fig:veg-fc}A) and the same drag coefficient $C_d = 0.26$ were used.

The domain was discretized by 100 cells in vertical direction, using the same grading as in the 1D model. In horizontal direction 300 cells were used with width $0.023 h$ at the edge of the forest, and expanding upstream with factor 1.05 and downstream with factor 1.011.

\subsection{Particle collection by a hedgerow}
\label{sec:part-coll-windbr}

The dry deposition model was tested on the problem of a hedgerow filtering the particle-laden flow that was originally investigated by \citet{TiwaryEtAl05}.
In their field experiments, the authors measured concentrations of polystyrene particles of diameters between \SI{0.8}{\um} and \SI{15}{\um} upwind and downwind of the hawthorn hedge. From these measurement, the collection efficiency of the barrier was determined.
The authors investigated the problem also numerically, using a detailed vegetation model.

In our study, we have constructed a 2D numerical model reproducing the experiment, and evaluated the influence of several parameters of the model on the results, namely of the drag coefficient $C_d$ and of the properties of the leaves.

\begin{figure}[ht]
  \centering
  \begin{tabular}{@{}p{0.25\linewidth}@{\qquad}p{0.6\linewidth}@{}}
    \subfigimg[height=0.2\textwidth]{A)}{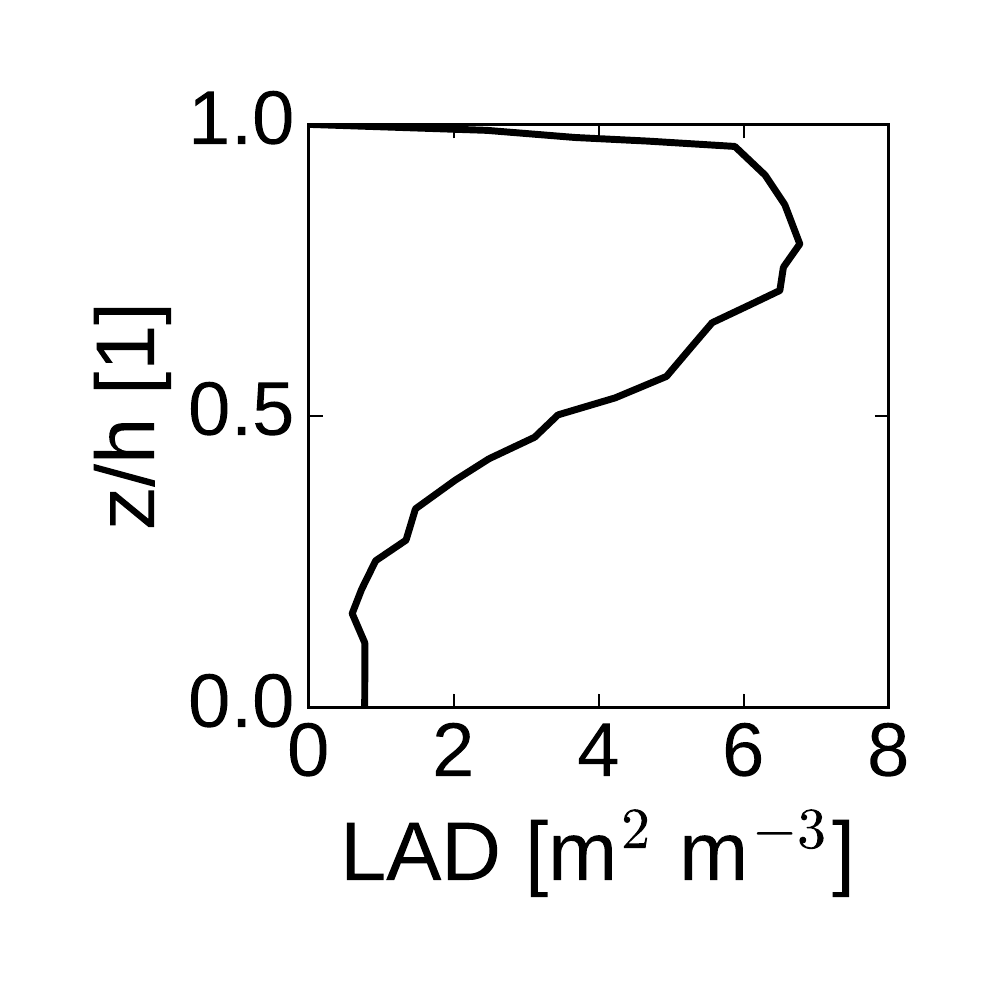} &
    \subfigimg[height=0.2\textwidth]{B)}{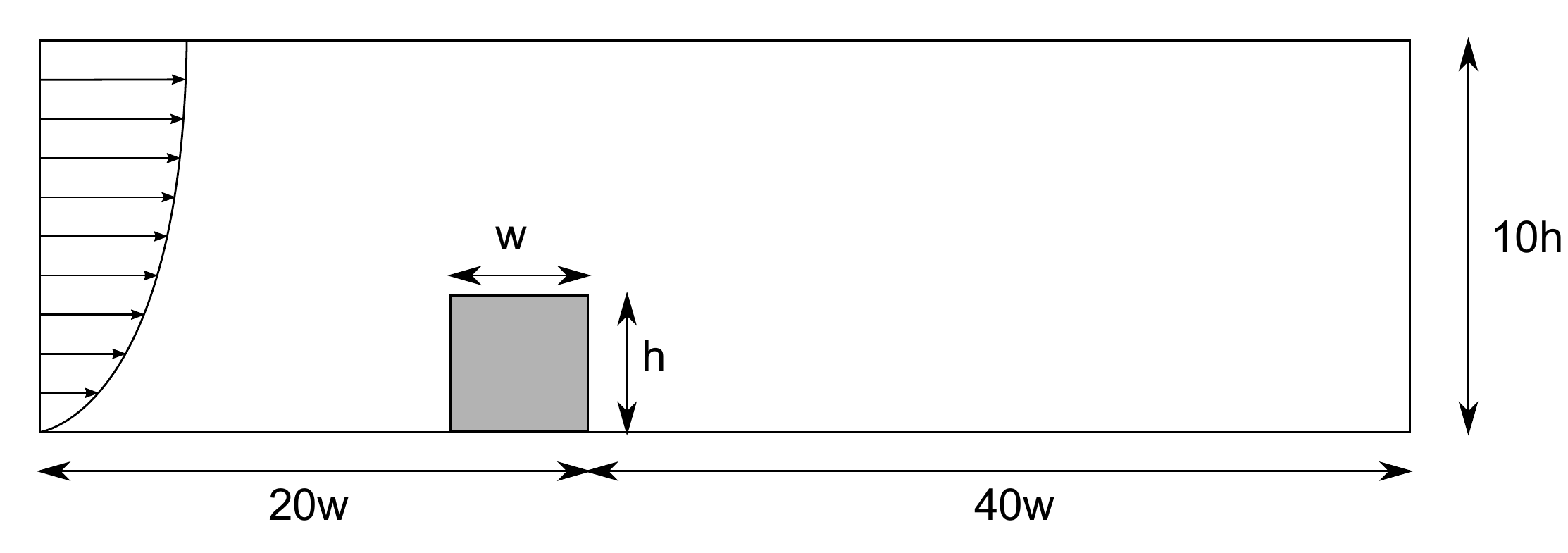} \\
  \end{tabular}
  \caption{Flow through a vegatation barrier. (A) LAD profile of the vegetation. Vertical coordinate is normalized by the height of the barrier $h = \SI{2.2}{\m}$. Taken from \citep{TiwaryEtAl05}. (B) Sketch of the computational domain (not to scale).}
  \label{fig:veg-pc}
\end{figure}

The vegetation barrier of width $w = \SI{1.6}{\m}$ and height $h = \SI{2.2}{\m}$ was placed inside the computational domain spanning $20w$ upwind and $40w$ downwind from the end of the barrier and with height $10h$ (Fig. \ref{fig:veg-pc}B). The barrier was porous, described by its leaf area density profile (Fig. \ref{fig:veg-pc}A), obtained from the original paper.

The boundary conditions for the flow equations were set as follows: At the inlet and at the top of the domain, log wind profile with $u_* = \SI{0.198}{\m\per\s}$ and $z_0 = \SI{0.0189}{\m}$ was prescribed. The reference velocity at $z = h$ was thus ${u_{\mathrm{ref}} = \SI{2.3}{\m\per\s}}$. Potential temperature $\theta = \SI{293}{\K}$ was set to a constant value to model the neutrally stratified atmosphere. Neumann BC was prescribed for the pressure. Profiles of the turbulence variables were given by equations (\ref{eq:inlet-k-eps}).
At the outlet, zero pressure fluctuation $p^*$ was prescribed, and the homogeneous Neumann BC was used for all other variables.
At the ground, the wall functions were used, together with the homogeneous Neumann BC for pressure fluctuation and potential temperature.

The transport and the collection of the particles of the diameters 0.875, 1.5, 2.75, 4.25, 6.25, 8.75, 12.5 and 15 \si{\um} and of the density $\rho_p = \SI{1050}{\kg\per\m\cubed}$ was investigated. Turbulent Schmidt number was set to $Sc_T = 0.7$.
The ambient background concentration was obtained by prescribing the concentration \SI{1}{\mg\per\m\cubed} at the inlet and at the top of the domain. At all other boundaries, the homogeneous Neumann BC was used for the particle concentration. No resuspension of the particles was allowed.

\begin{figure}[ht]
  \centering
  \includegraphics[width=0.4\linewidth]{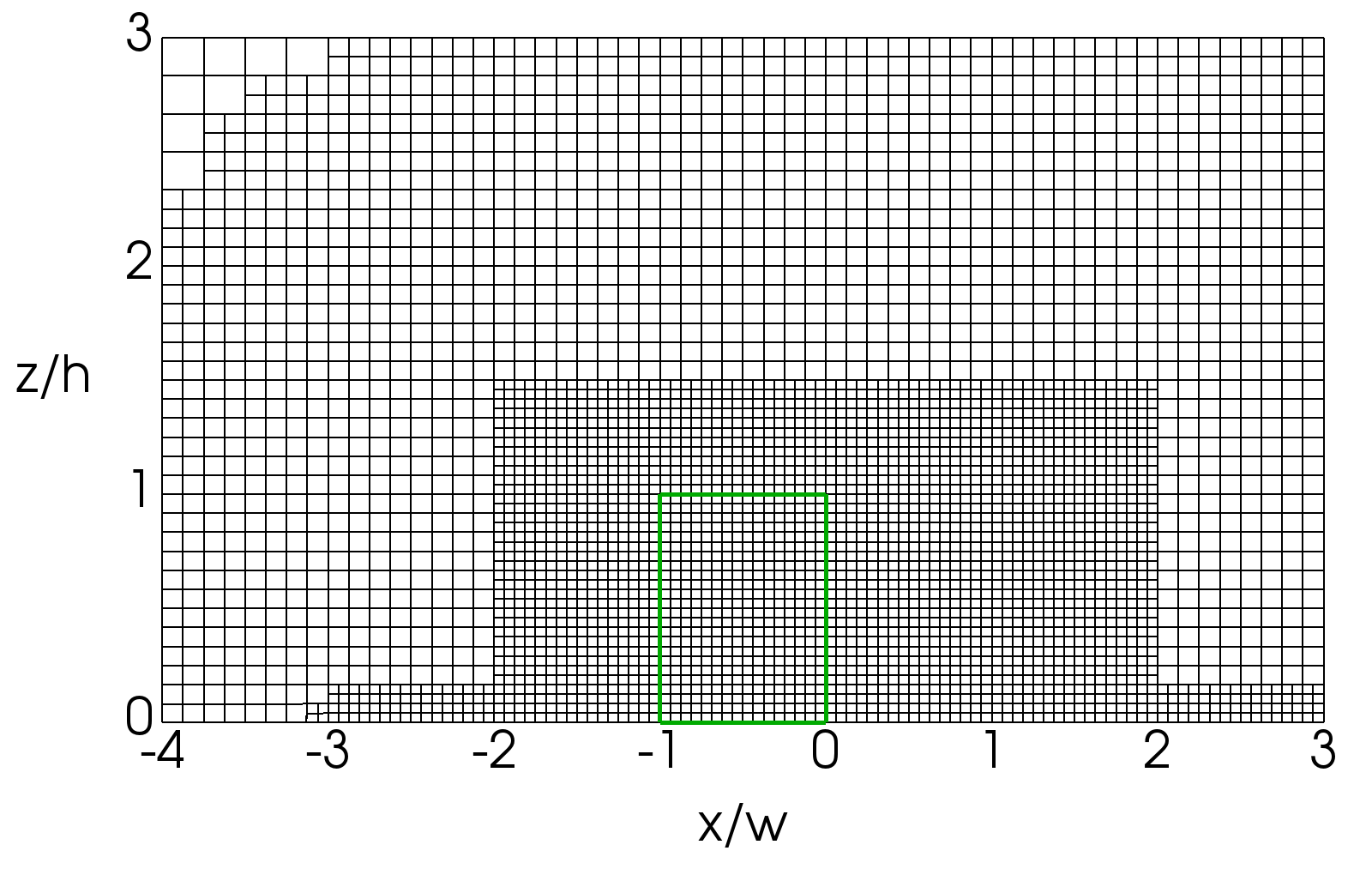}
  \caption{Computational mesh around the hedgerow.}
  \label{fig:veg-pc-mesh}
\end{figure}

The unstructured computational mesh was generated using the \textit{snappyHexMesh} generator from the OpenFOAM software package \citep{OpenFoamUserGuide}. The mesh, consisting of approximately nineteen thousand cells, was refined around the vegetation barrier (Fig. \ref{fig:veg-pc-mesh}). The largest cells in the domain were \SI{0.8}{\m} $\times$ \SI{0.73}{\m} large and the smallest were \SI{0.1}{\m} $\times$ \SI{0.092}{\m} large, so that the vegetation block itself was discretized into 16 $\times$ 24 cells.

\section{Results}

\subsection{Flow around a hill}
\label{sec:res-flow-around-hill}

\paragraph{Flow field}

\begin{figure}[ht!]
  \centering
  \footnotesize
  \begin{tabular}{@{}P{0.24\linewidth}@{}P{0.24\linewidth}@{}P{0.24\linewidth}@{}}
    $x/a = 0$ & $x/a = 1$ & $x/a = 2$ \\
    \includegraphics[width=\linewidth]{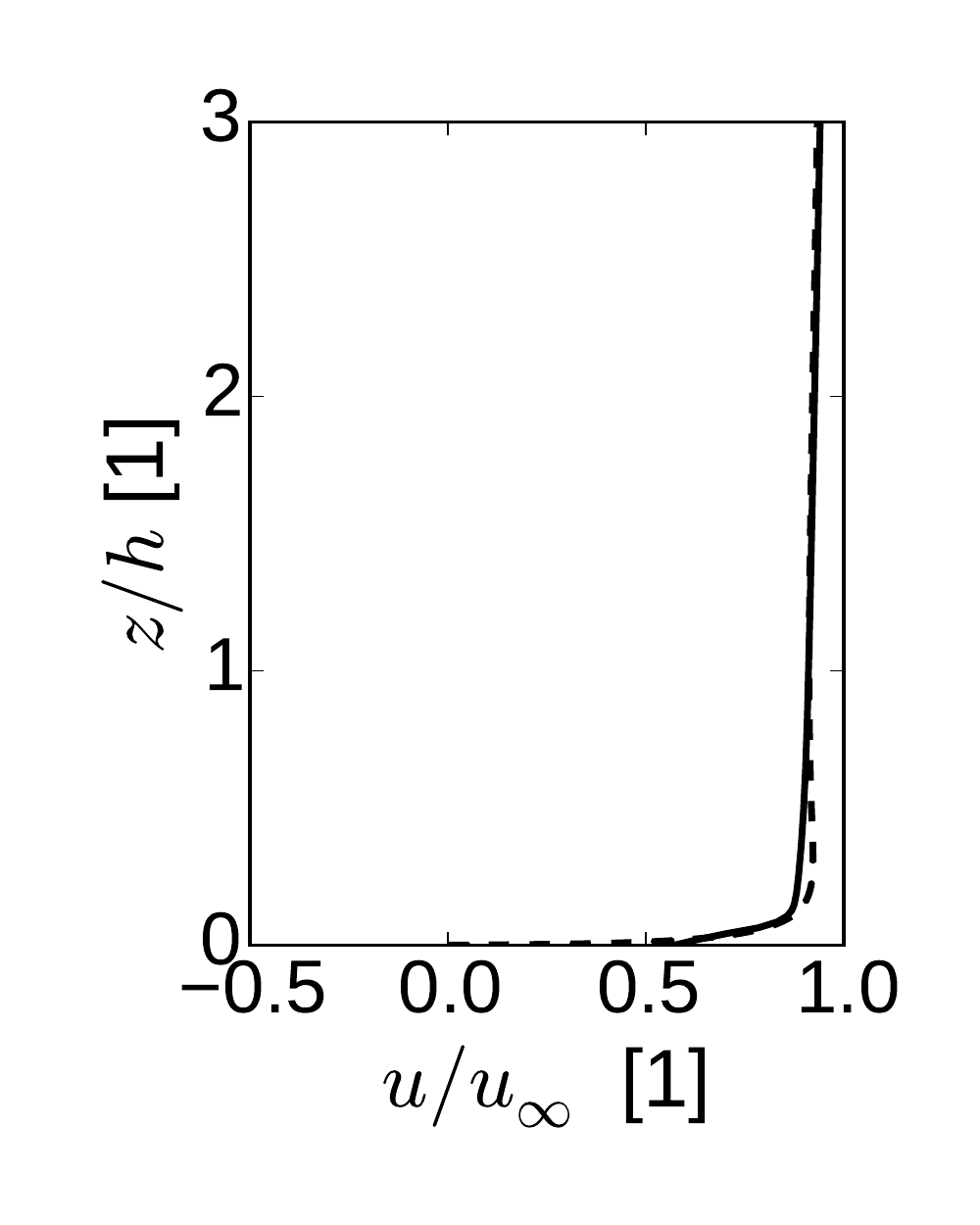}  &
    \includegraphics[width=\linewidth]{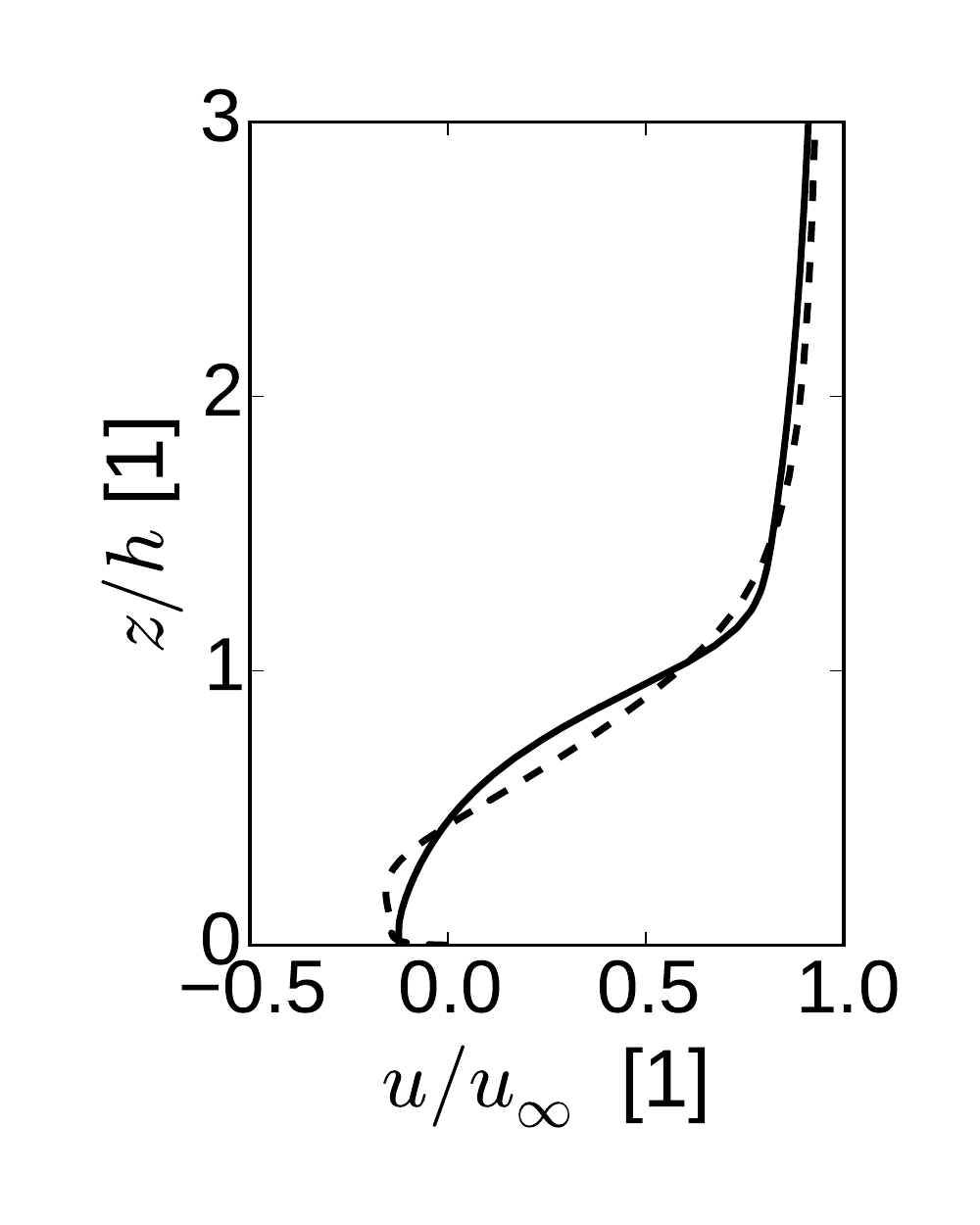} &
    \includegraphics[width=\linewidth]{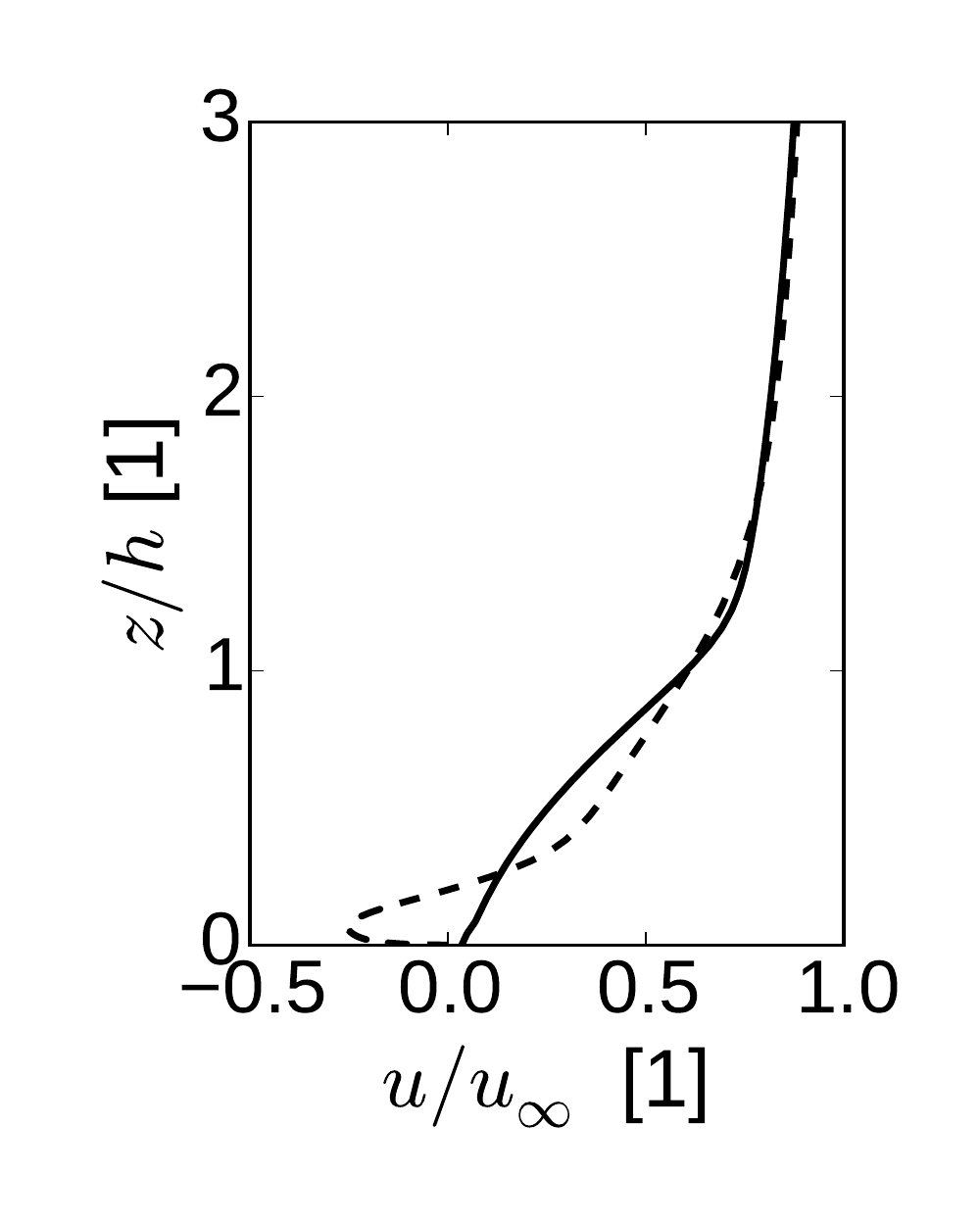} \\
    \includegraphics[width=\linewidth]{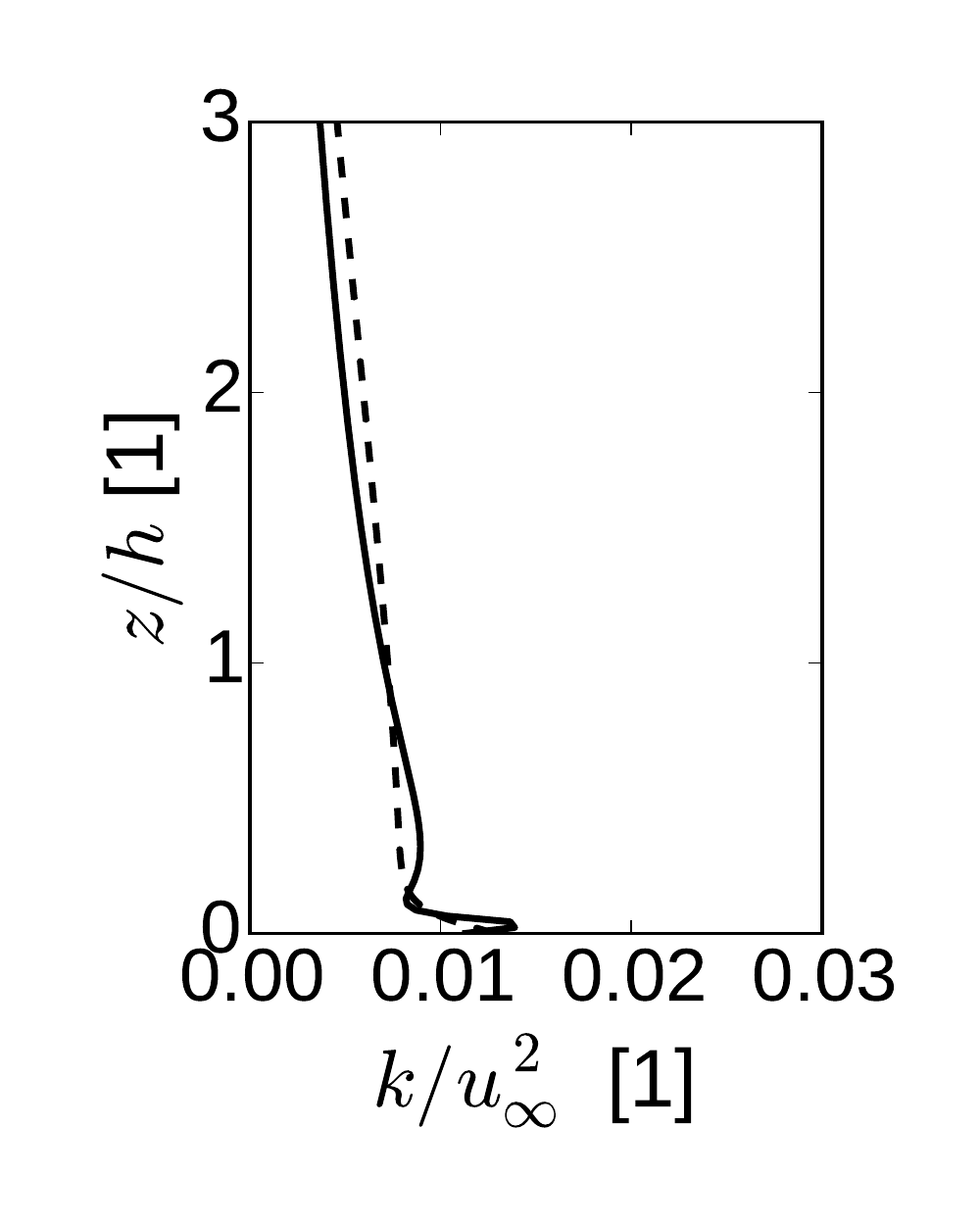}  &
    \includegraphics[width=\linewidth]{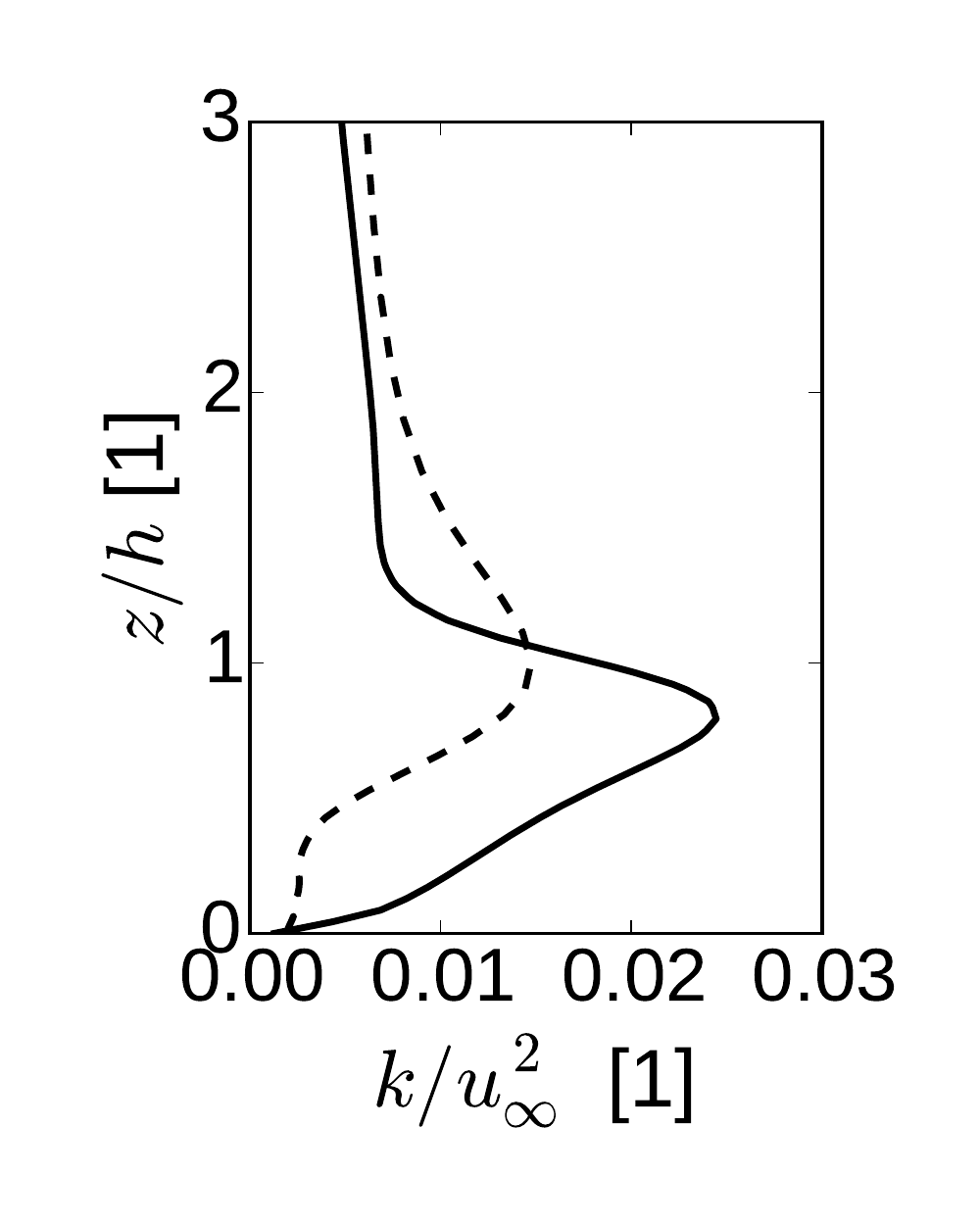} &
    \includegraphics[width=\linewidth]{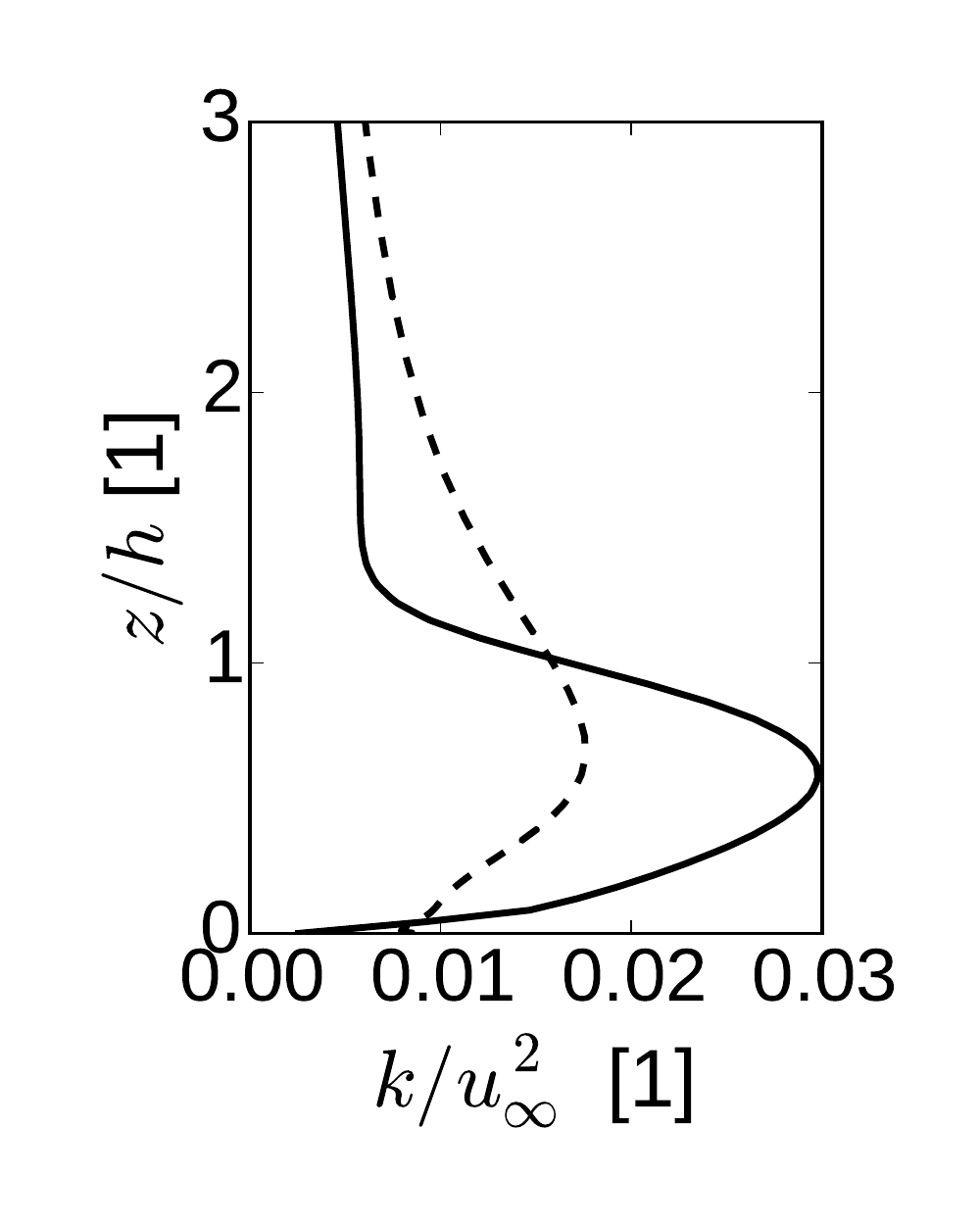} \\
  \end{tabular}
  \caption{Flow around the N3 hill. Vertical profiles of the normalized horizontal velocity (top row) and the normalized turbulence kinetic energy (bottom row) at the hill summit ($x/a = 0$), at the downstream base ($x/a = 1$), and downstream from the hill ($x/a = 2$). Computation (solid lines) and measurement (dashed line).}
  \label{fig:val-hill-N3}
\end{figure}

\begin{figure}[ht!]
  \centering
  \footnotesize
  \begin{tabular}{@{}P{0.24\linewidth}@{}P{0.24\linewidth}@{}P{0.24\linewidth}@{}}
    $x/a = 0$ & $x/a = 1$ & $x/a = 2$ \\
    \includegraphics[width=\linewidth]{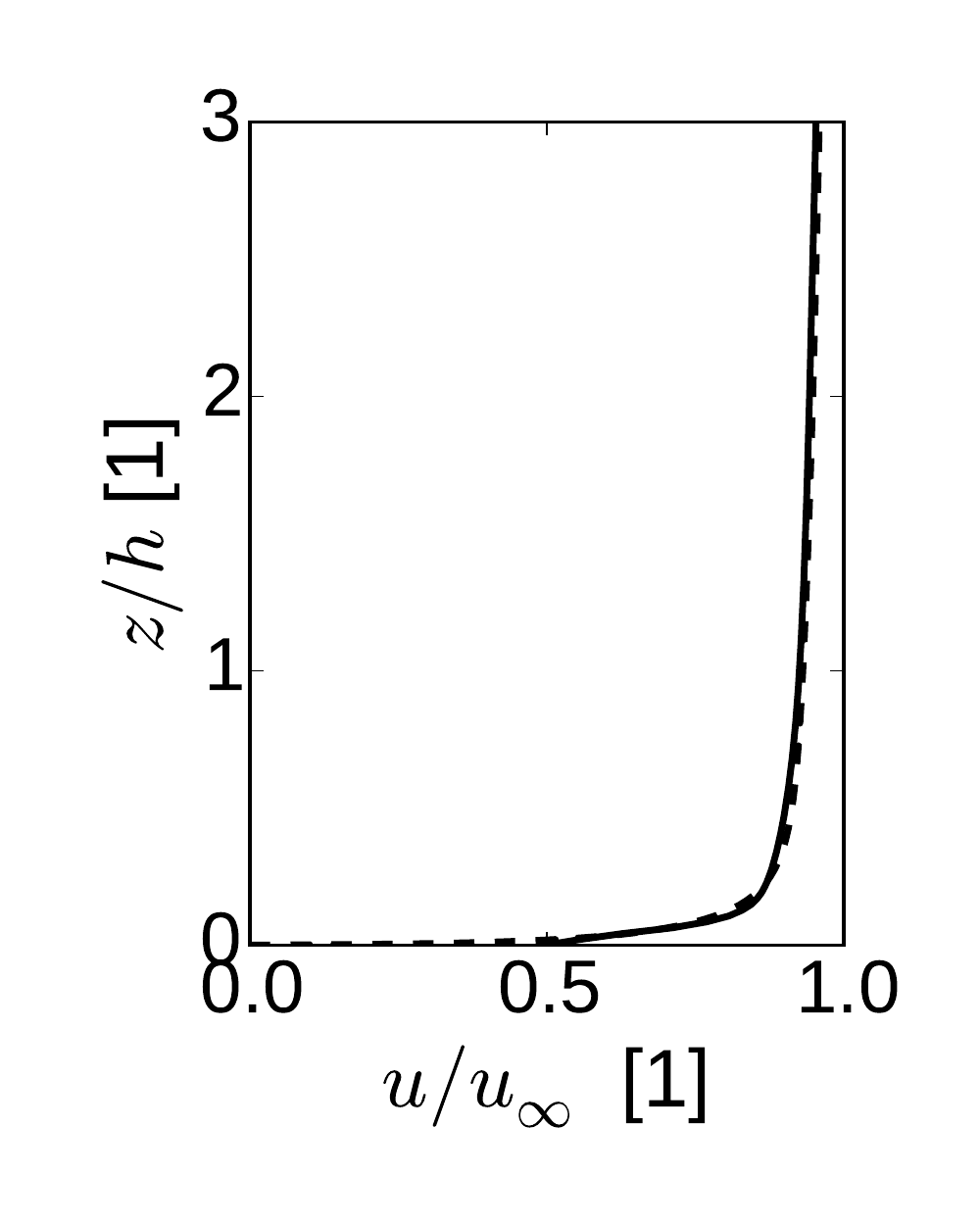}  &
    \includegraphics[width=\linewidth]{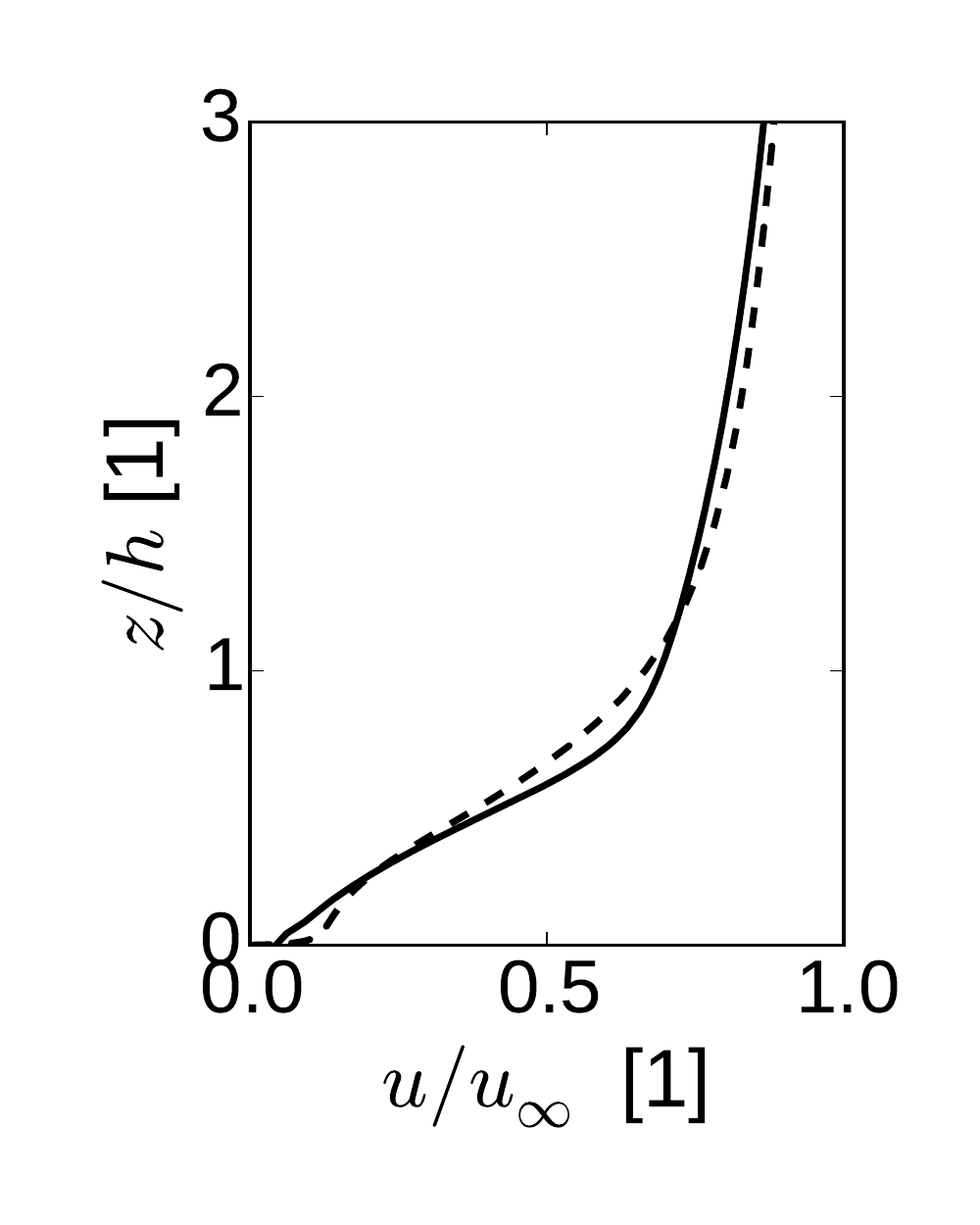} &
    \includegraphics[width=\linewidth]{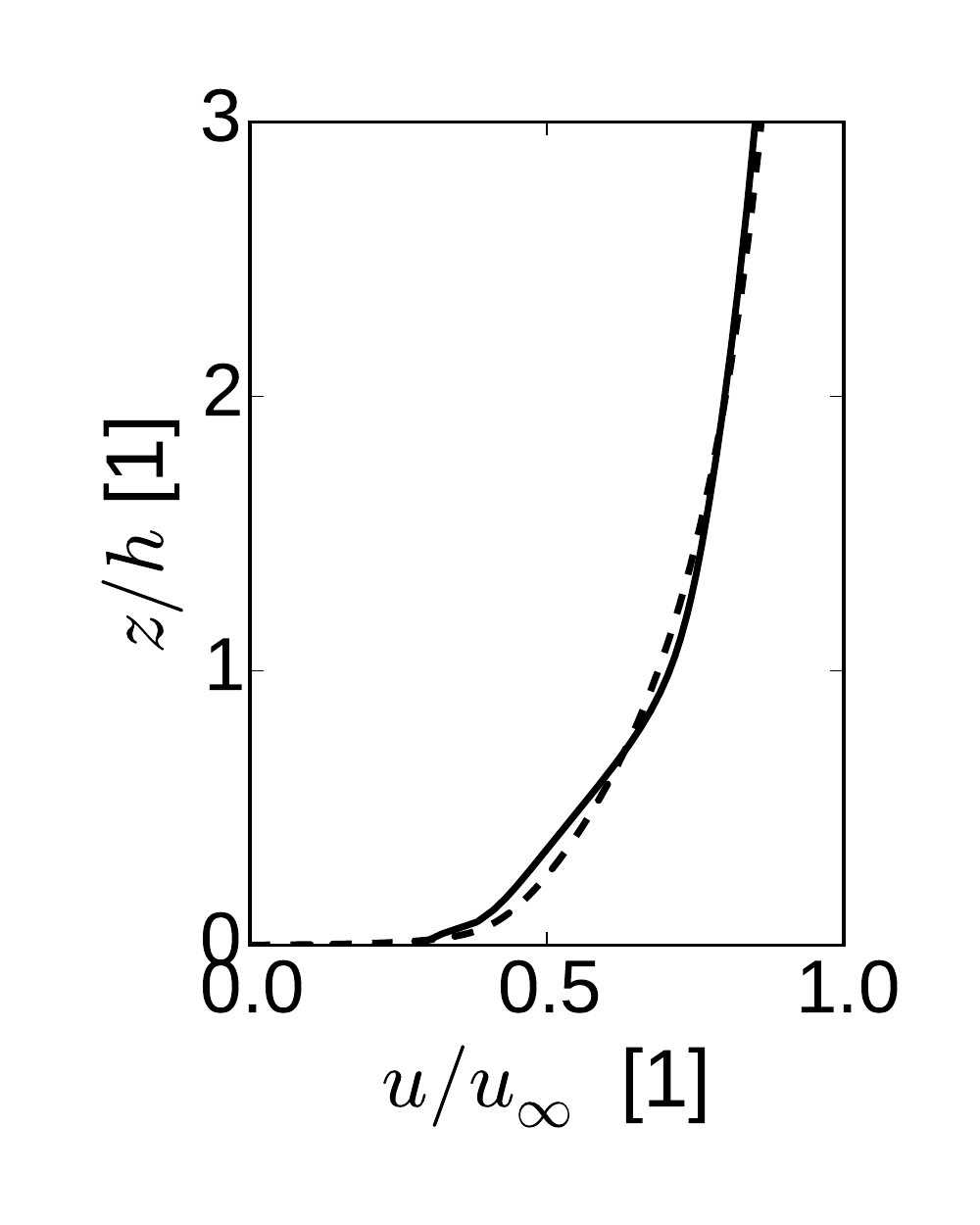} \\
    \includegraphics[width=\linewidth]{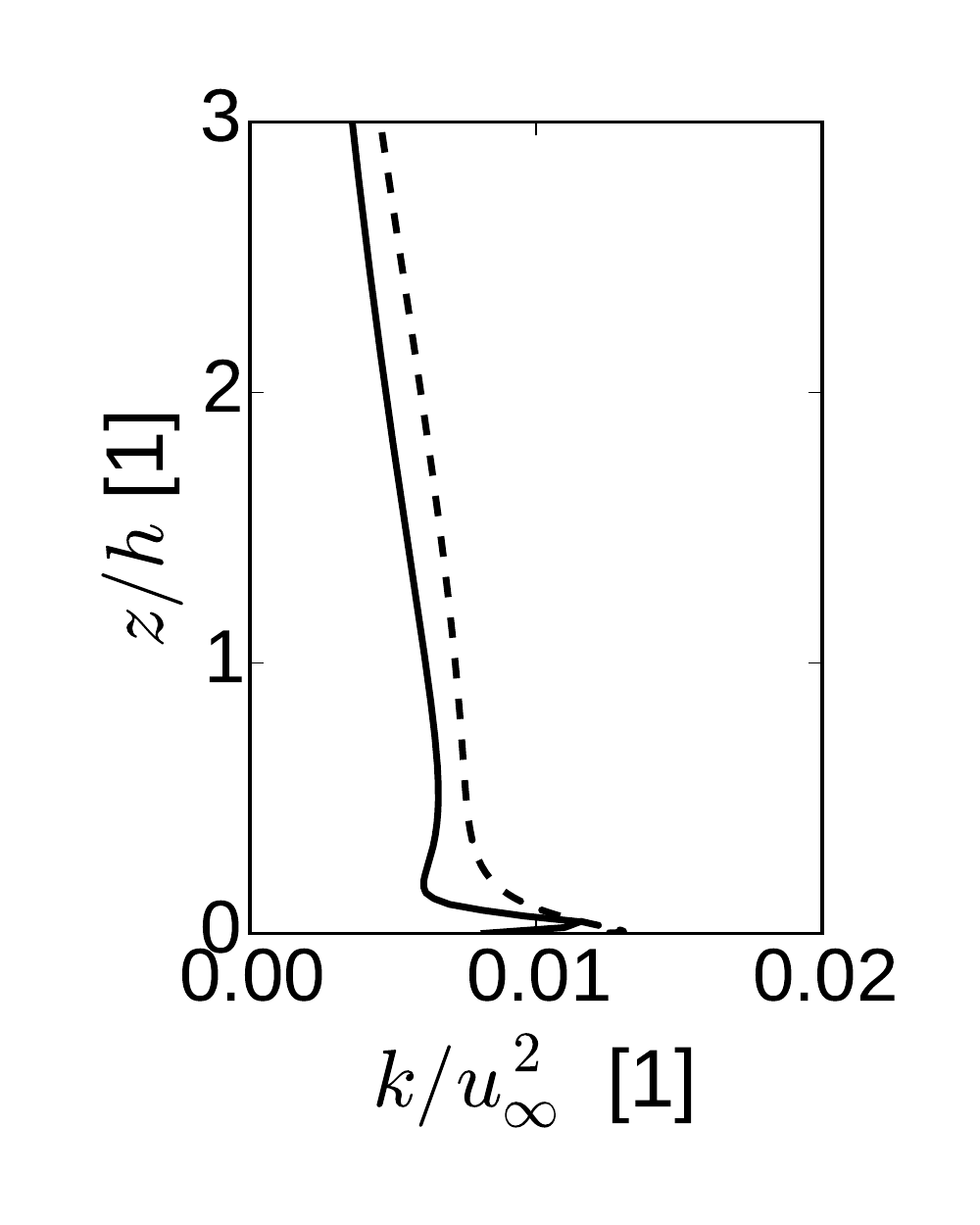}  &
    \includegraphics[width=\linewidth]{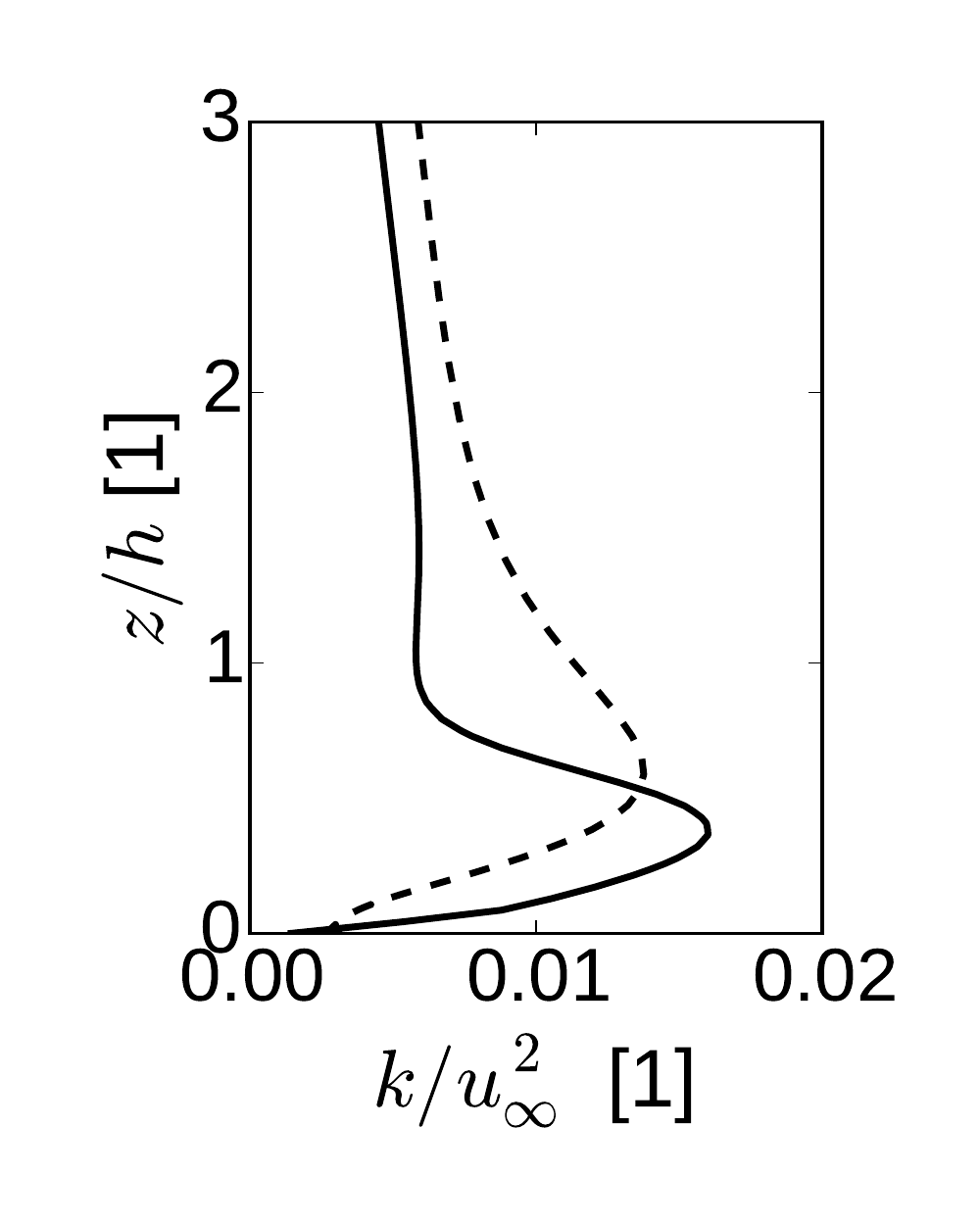} &
    \includegraphics[width=\linewidth]{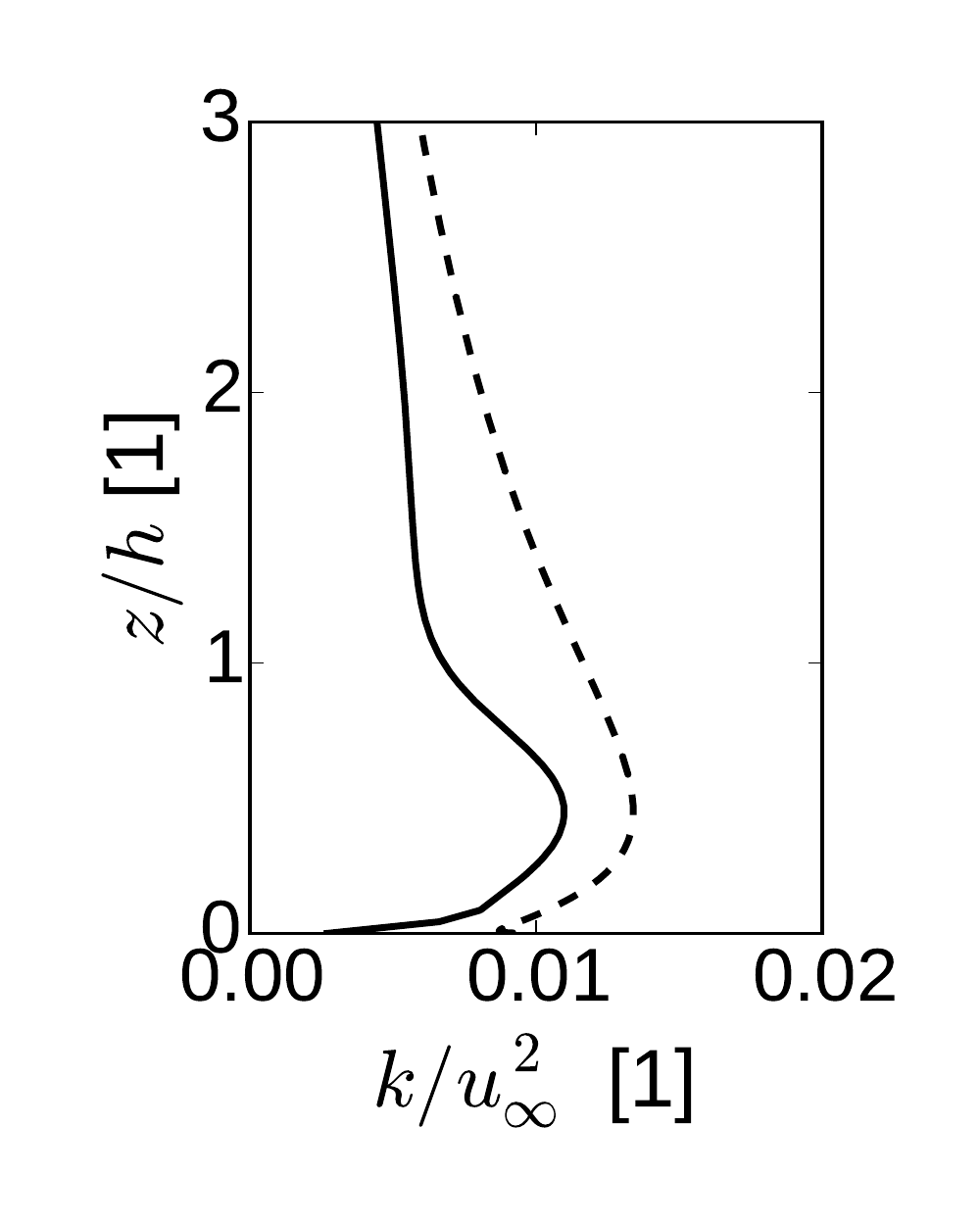} \\
  \end{tabular}
  \caption{Flow around the N5 hill. Quantities shown are as in Fig. \ref{fig:val-hill-N3}.}
  \label{fig:val-hill-N5}
\end{figure}

Figures \ref{fig:val-hill-N3} and \ref{fig:val-hill-N5} show the vertical profiles of the normalized horizontal velocity and the turbulent kinetic energy for both N3 and N5 hills.
The solution agrees reasonably with the measurements. The calculated solution shows a flow separation for the N3 hill and no separation for the N5 hill, as was observed in the experiment.
The reattachment point for the N3 hill is however closer to the hill in our computations ($x/h = 5.6$) than what was indicated by the measurement ($x/h = 6.4$).
A near ground increase in the turbulent kinetic energy downstream of the hill is reproduced for both geometrical variants (Fig. \ref{fig:val-hill-N3} and Fig. \ref{fig:val-hill-N5}, bottom rows), however, the maximum of TKE is overpredicted for the separated flow. At the same time, TKE is generally underpredicted further from the N5 hill.

Overall, the solution shows a good level of agreement, especially looking at the calculated flow field.
The choice of the turbulence model is expected to have a very significant influence on the results, and more complex turbulence models (such as the Reynolds Stress Model) might provide better agreement even in the calculated TKE.

\paragraph{Pollutant dispersion}

\begin{figure}[ht!]
  \centering
  \footnotesize
  \begin{tabular}{@{}m{0.07\linewidth}@{}P{0.28\linewidth}@{}P{0.28\linewidth}@{}P{0.28\linewidth}@{}}
    & Upwind base & Summit & Downwind base \\
    N3 &
    \raisebox{-.5\height}{\includegraphics[width=\linewidth]{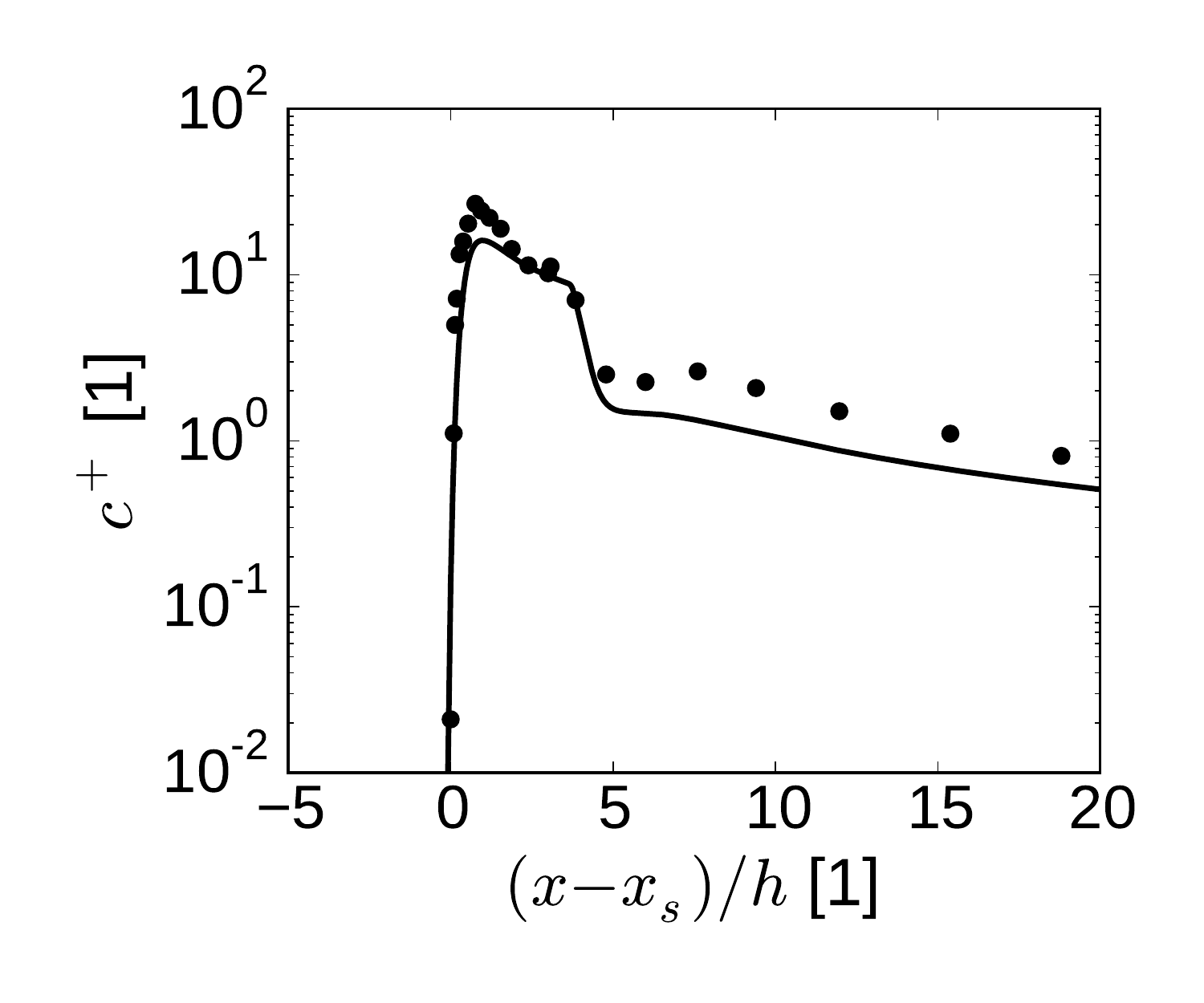}}  &
    \raisebox{-.5\height}{\includegraphics[width=\linewidth]{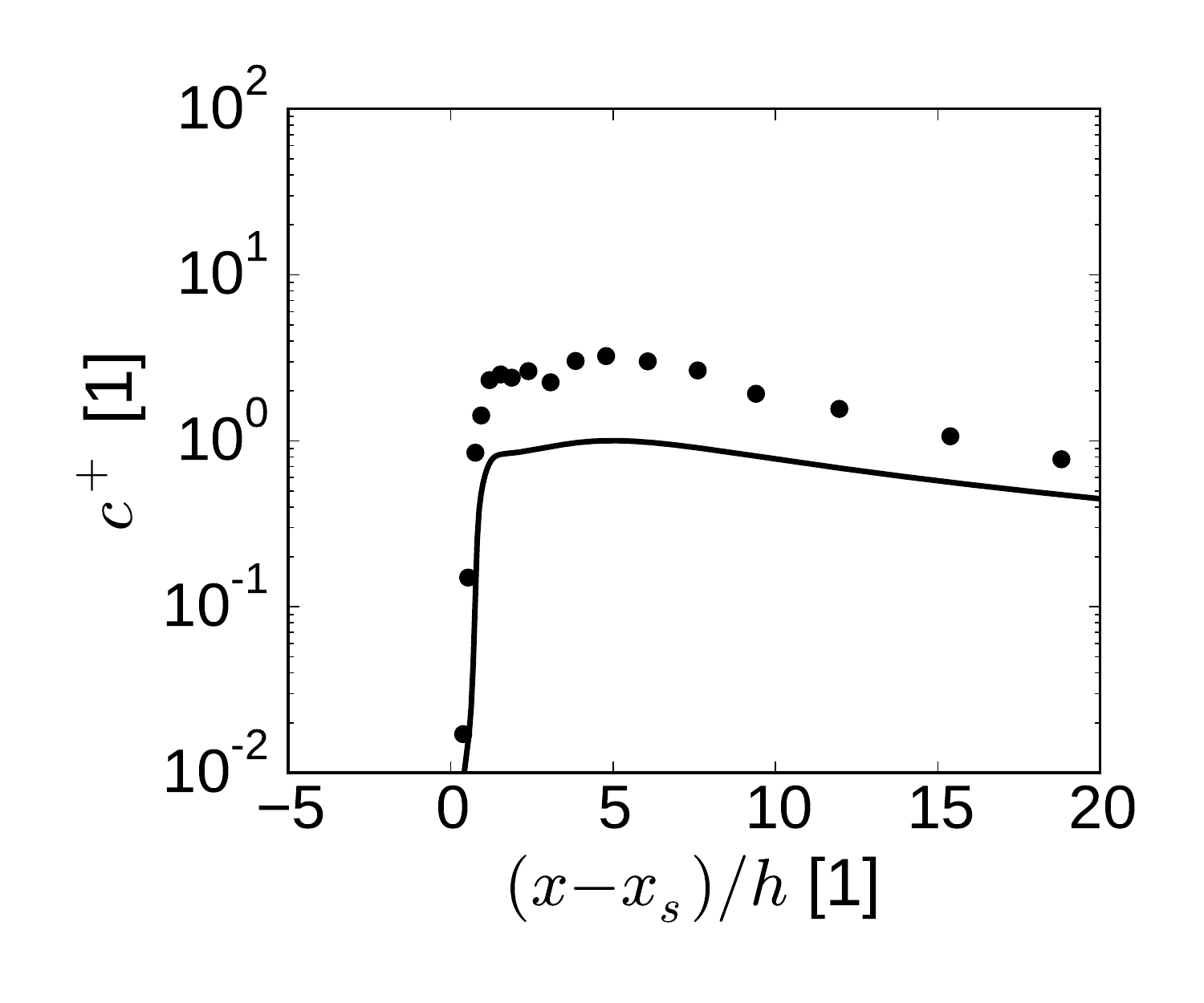}} &
    \raisebox{-.5\height}{\includegraphics[width=\linewidth]{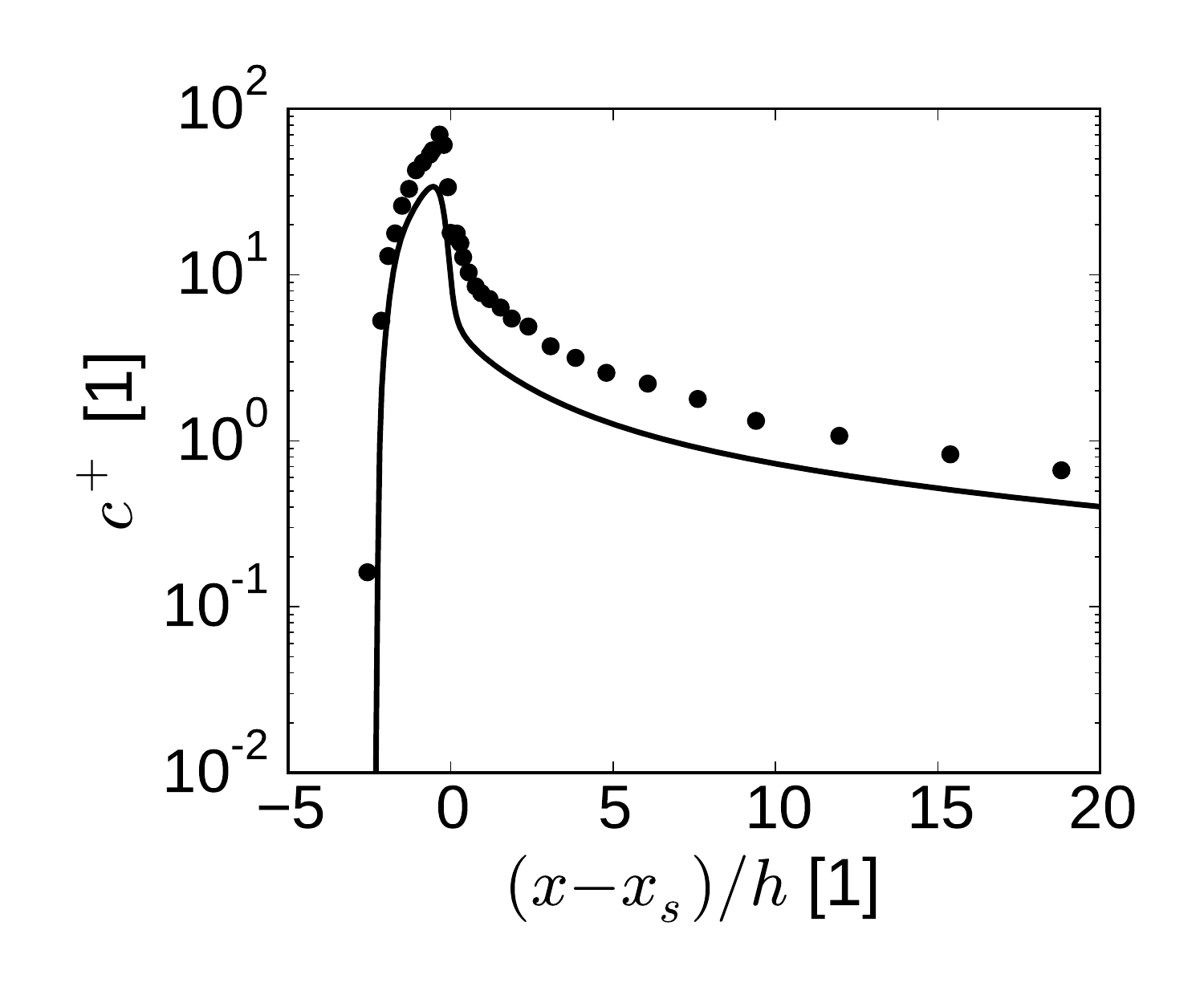}} \\
    N5 &
    \raisebox{-.5\height}{\includegraphics[width=\linewidth]{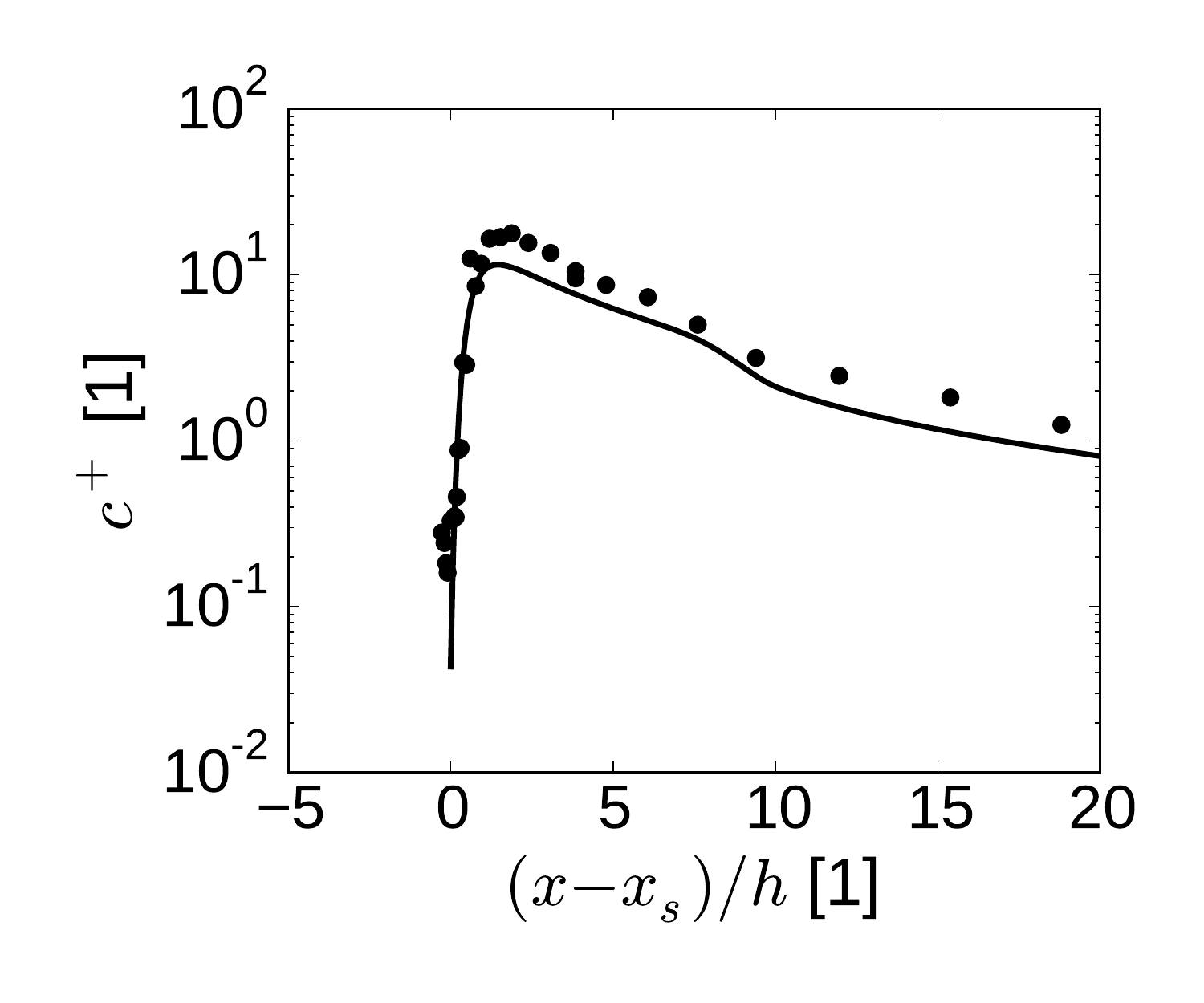}}  &
    \raisebox{-.5\height}{\includegraphics[width=\linewidth]{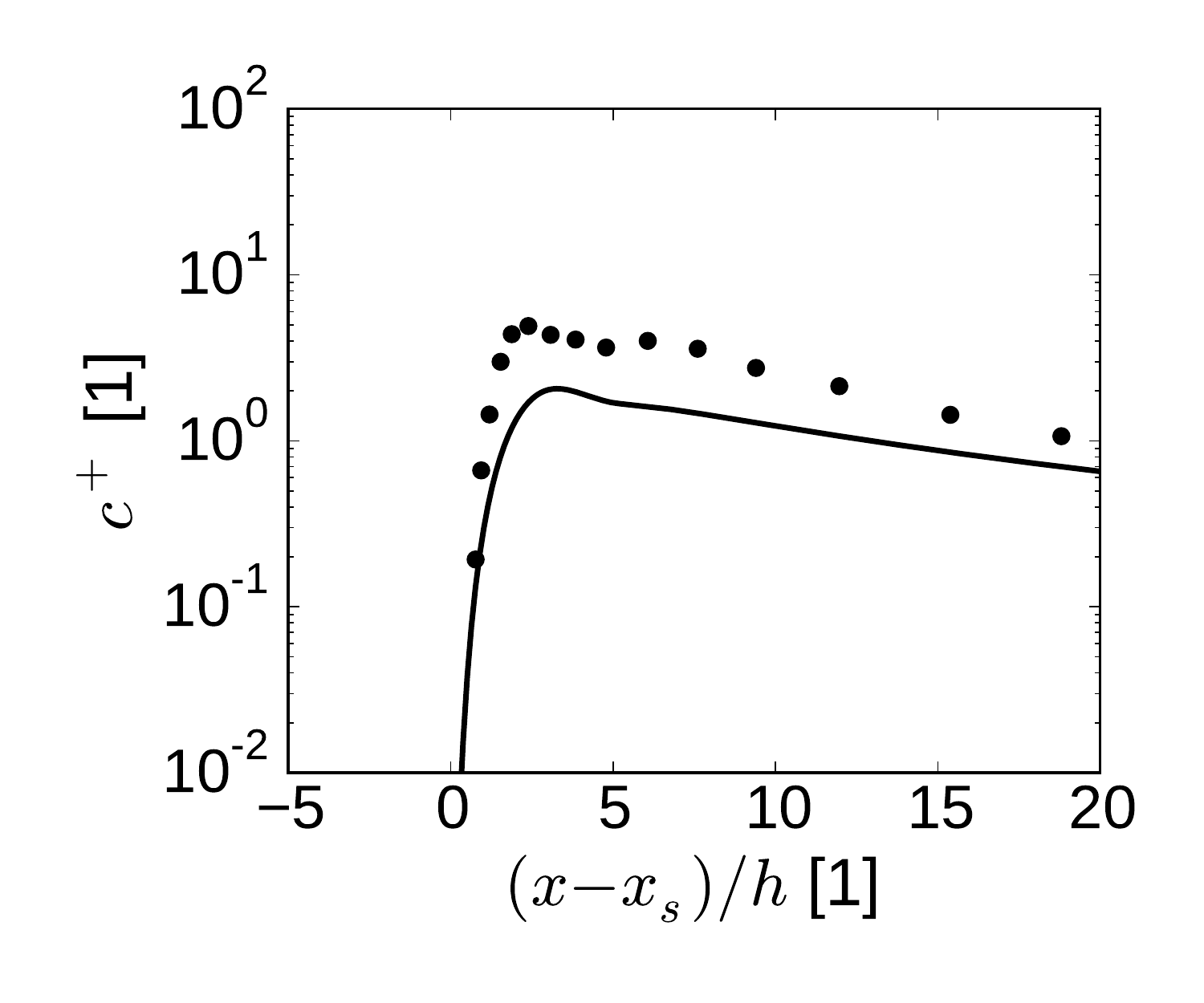}} &
    \raisebox{-.5\height}{\includegraphics[width=\linewidth]{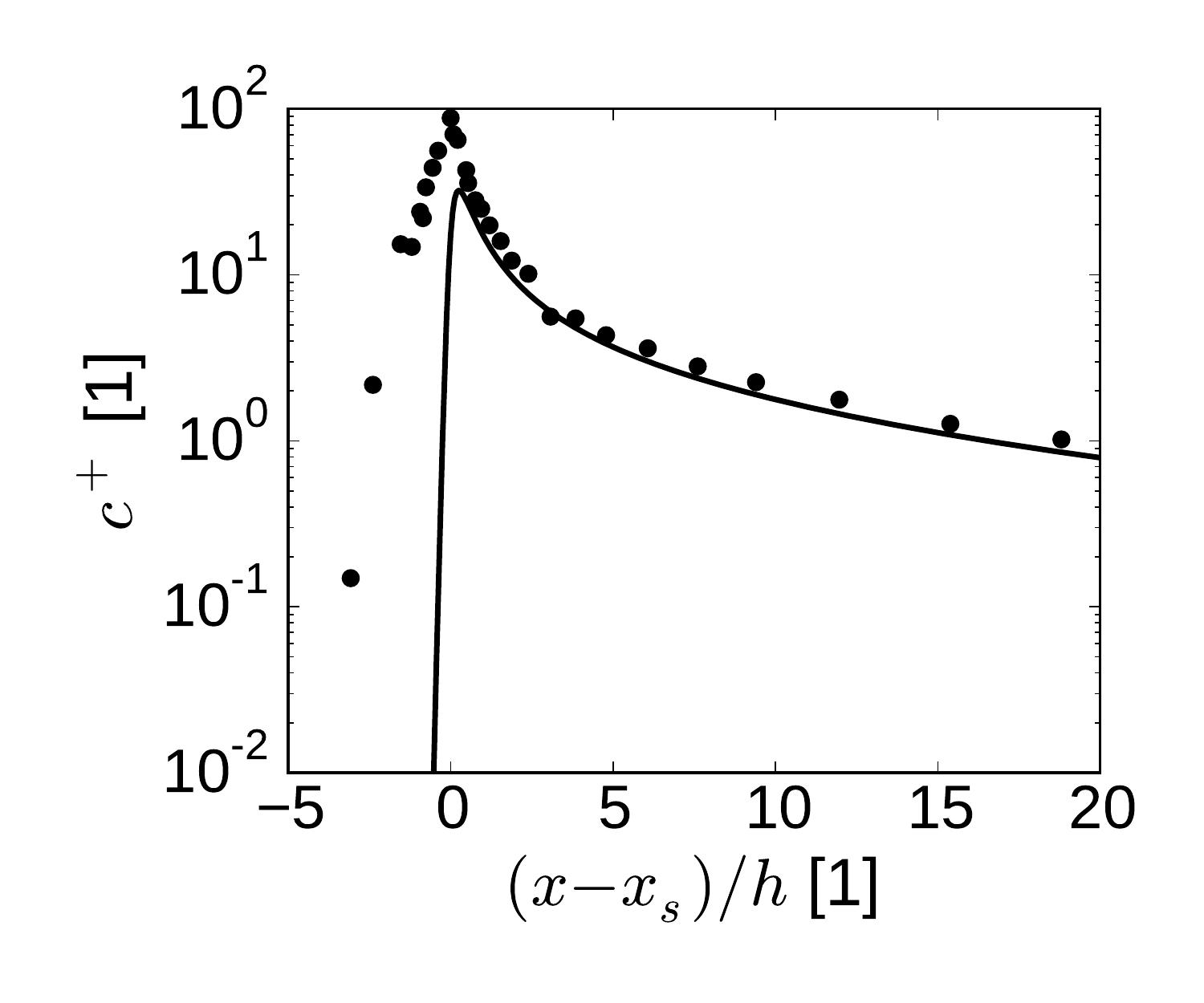}} \\
  \end{tabular}
  \caption{Flow around a hill. Ground level normalized concentration with the pollutant source at the upwind base (left column), at the summit (middle column), or at the downwind base (right column) of the N3 and N5 hills. Height of the source is $h_s = h/4$ in all cases. Computed (solid lines) and measured (symbols) values.}
  \label{fig:val-hill-surfaceConc}
\end{figure}

The calculated and measured concentrations are presented here in normalized form,
\begin{equation}
  c^+ = \frac{c u_\infty h^2}{Q},
\end{equation}
where $Q$ is the source intensity in \si{\kg\per\s}.
Fig. \ref{fig:val-hill-surfaceConc} shows the ground level concentrations for all calculated source positions and for both hill shapes. Several discrepancies between the measured and calculated values are present, and deserve some commentary.

First, the calculated concentration of the pollutant released at the summit of the hill is well below the measured values for both hill shapes (Fig. \ref{fig:val-hill-surfaceConc}, middle column). Cause of this error is unclear. On possible explanation may lie in the fact that the wind speed at the release point is higher at the summit than at the bases due to the flow speedup. Lower levels of the calculated turbulent diffusion at the summit would thus lead to the pollutant being advected faster, producing the observed underprediction.

Secondly, the concentration is underpredicted further away from the N3 hill for all source positions (Fig. \ref{fig:val-hill-surfaceConc}, upper row). This may be caused by the overpredicted TKE close to the hill (see Fig. \ref{fig:val-hill-N3}), and thus increased turbulent mixing in that area, leading to a faster dispersion of the pollutant.

And lastly, we note that the measurements show high concentration values upstream from the source placed at the downwind base of the N5 hill, which is not reproduced by the computation. \citet{CastroApsley97} speculated that this is caused by the flow separation occurring intermittently, which was not captured by the measurements (nor our RANS model). In that case, the source would be occasionally placed in the separation bubble, and the pollutant would be advected upstream.
Such increase of the pollutant upstream of the release point may be observed on the concentration values for the release point at the downwind base of the N3 hill. Our calculation places the release point in the separation bubble, and the concentration measurements agree well with the calculations.

\subsection{Warm bubble}
\label{sec:res-warm-bubble}

Our calculated results are compared with what we will call a reference solution by \citet{GiraldoRestelli08}. This reference solution was calculated by a Discontinuous Galerkin solver that used 10th order polynomials and was based on the equations for density perturbations, momentum, and total energy perturbation. The solver was denoted by DG3 in the original paper. The reference results were obtained on the mesh with the spatial resolution of 5 m in both directions.

\begin{figure}[ht]
  \centering
  \begin{tabular}{@{}p{0.2\linewidth}@{\quad}p{0.2\linewidth}@{\quad}p{0.2\linewidth}@{\quad}p{0.2\linewidth}@{}}
    \subfigimg[width=0.95\linewidth]{A)}{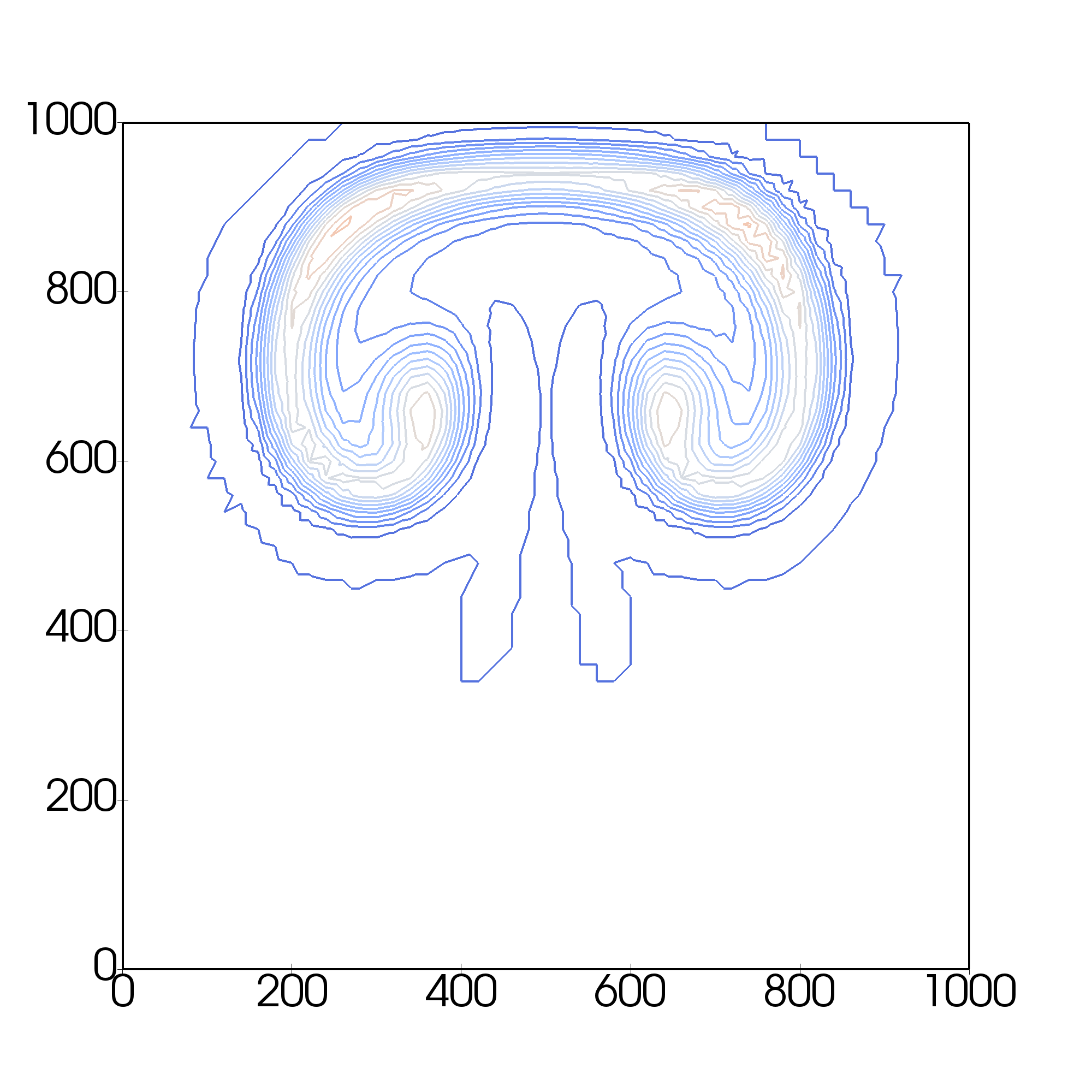} &
    \subfigimg[width=0.95\linewidth]{B)}{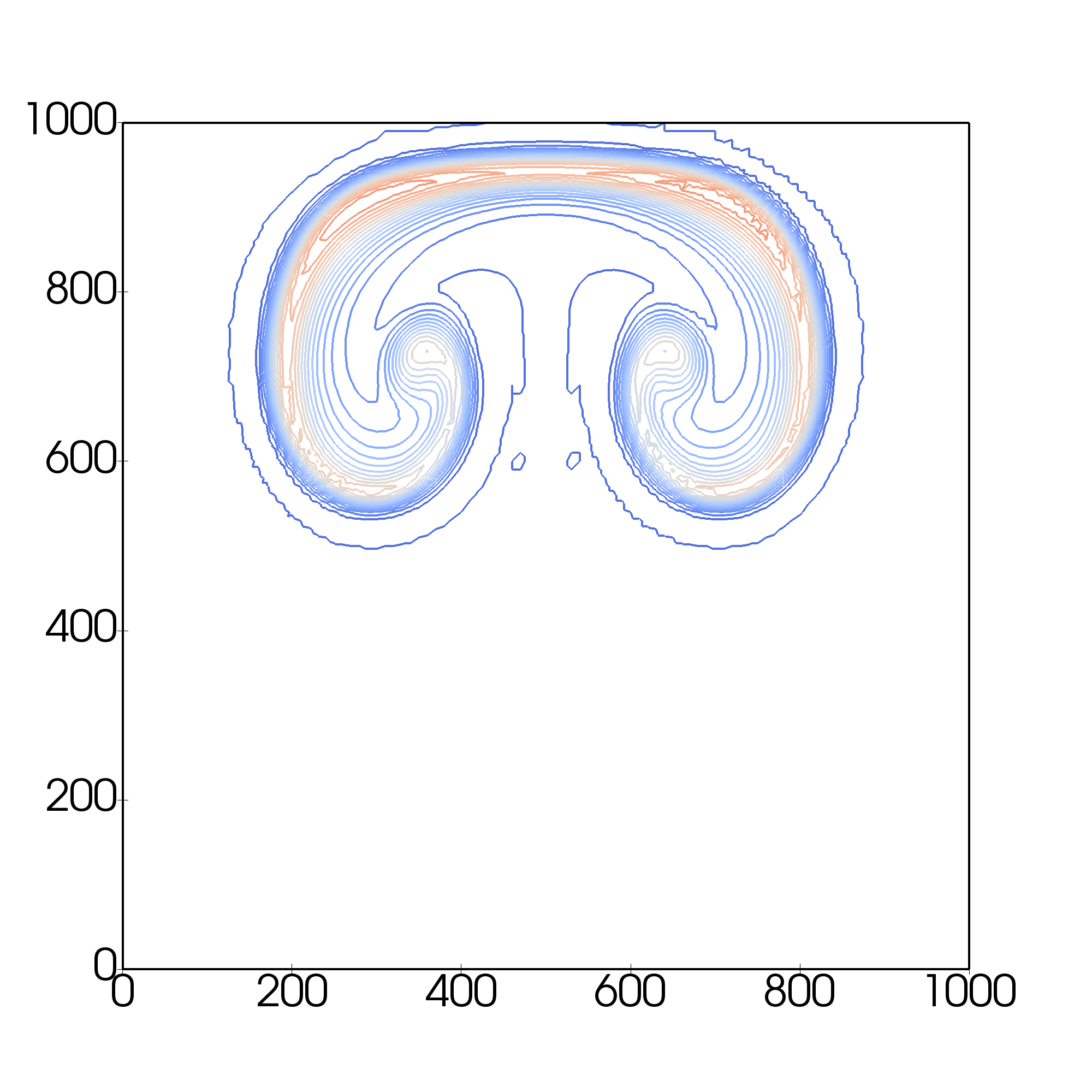} &
    \subfigimg[width=0.95\linewidth]{C)}{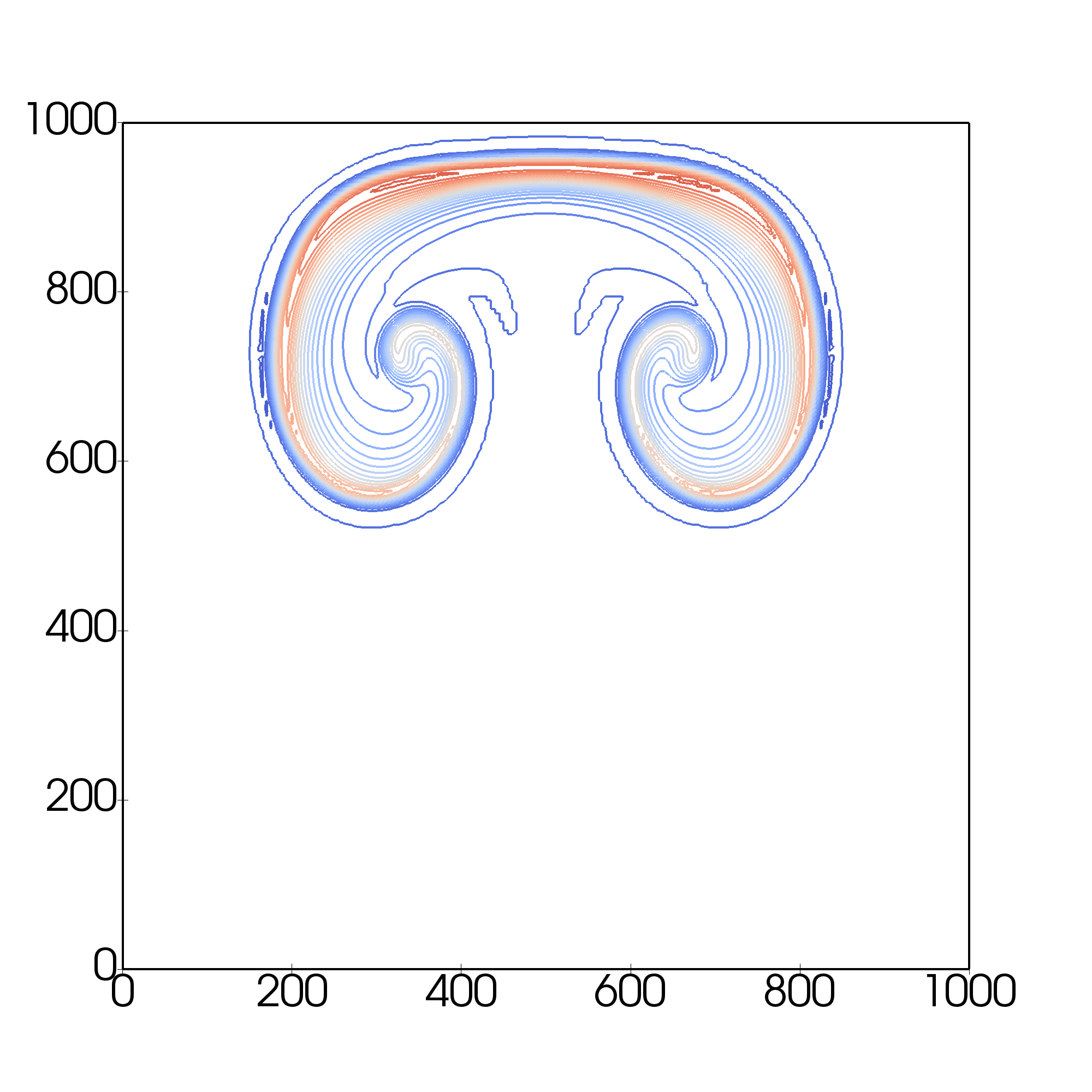} &
    \subfigimg[width=0.95\linewidth]{D)}{img/warmBubble/{ABL2.5}.png}
  \end{tabular}
  \caption{Warm bubble test case. Potential temperature perturbation at $t = \SI{700}{\s}$. Mesh resolution: (A) 20 m (B) 10 m (C) 5 m (D) 2.5 m. Interval between contours is 0.025 K.}
  \label{fig:val-wb-mesh-theta2d}
\end{figure}

\begin{figure}[ht]
  \centering
  \includegraphics[width=0.45\textwidth]{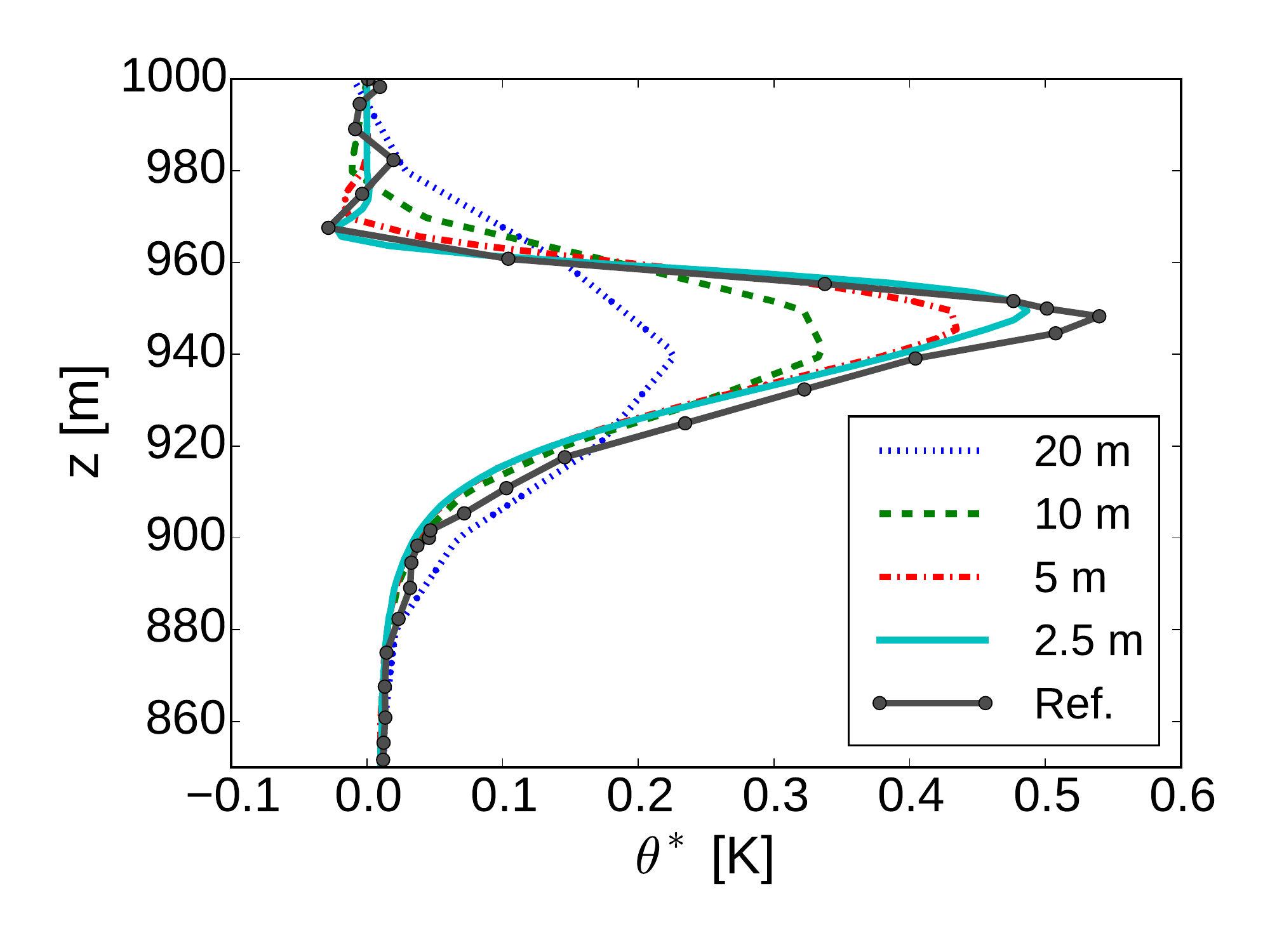}
  \caption{Warm bubble test case. Vertical profile of the potential temperature perturbation at the centreline at $t = \SI{700}{\s}$, compared with the reference solution from \citep{GiraldoRestelli08}.}
  \label{fig:val-wb-mesh-thetaLine}
\end{figure}

Contours of the potential temperature perturbation at the final time are shown on Fig. \ref{fig:val-wb-mesh-theta2d}, and its vertical profile at the centreline is shown on Fig. \ref{fig:val-wb-mesh-thetaLine}.
At the final time, the bubble has risen to the top of the domain, and its mushroom shape is partially affected by its interaction with the upper boundary.
The vertical profile of the potential temperature perturbation reveals that even at the highest employed resolution of 2.5 m, the peak of the profile is slightly underpredicted compared to the reference solution calculated on mesh with resolution of 5 m. This should be however expected, as the reference solution was calculated using a high order Discontinous Galerkin method, compared to our second-order finite volume solver. Further refinement of the computational mesh might increase the precision of our solution, although at excessive computational cost: our finest mesh consisted of 160 thousand computational cells, halving the spatial resolution would quadruple this number.

Furthermore, our solver places the lower end of the bubble above the position given by the reference solution. This might be attributed to the simplifications made in our physical model, of which the most notable is the use of the reference density $\rho_\mathrm{ref}$ instead of the actual density $\rho$ in the pressure term in the velocity equation (\ref{eq:eqs-2}). This shows that there are limits to its accuracy in the domains spanning more than few hundred meters in the vertical direction.

\begin{table}[ht]
\setlength{\tabcolsep}{15pt}
\centering
\footnotesize
\begin{tabular}{@{}lrr@{}crr@{}crr@{}}\toprule
 & \multicolumn{2}{c}{$\theta^*$ [K]} && \multicolumn{2}{c}{$u_x$ [$\si{m\per\s}$]} && \multicolumn{2}{c}{$u_z$ [$\si{m\per\s}$]}\\
\cmidrule{2-3} \cmidrule{5-6} \cmidrule{8-9}
                 & min     & max     &&  min     & max     && min      & max    \\
\midrule
Computation      & -0.029  &  0.491  &&  -1.980  & 1.980   && -1.855   & 2.565  \\
Reference        & -0.093  &  0.538  &&  -2.081  & 2.081   && -1.915   & 2.543  \\
\bottomrule
\end{tabular}
\caption{Minima and maxima of potential temperature perturbation, horizontal velocity, and vertical velocity at time $t = \SI{700}{\s}$.}
\label{tab:val-wb-minmax}
\end{table}

As a further comparison, Tab. \ref{tab:val-wb-minmax} lists the minima and maxima of the selected variables at the final time. The underprediction of the maximal potential temperature perturbation, visible at Fig. \ref{fig:val-wb-mesh-thetaLine}, is again exhibited here. However, the overall qualitative as well as quantitative agreement of our solution with the reference one is demonstrated.

\subsection{Forest canopy flow}
\label{sec:res-forest-canopy-flow}

\begin{figure}[p]
  \centering
  \begin{tabular}{@{}P{0.24\linewidth}@{\qquad}P{0.24\linewidth}@{\qquad}P{0.24\linewidth}@{}}
    Homogeneous forest & Edge flow - $4h$ & Edge flow - $9h$ \\
    \includegraphics[width=\linewidth]{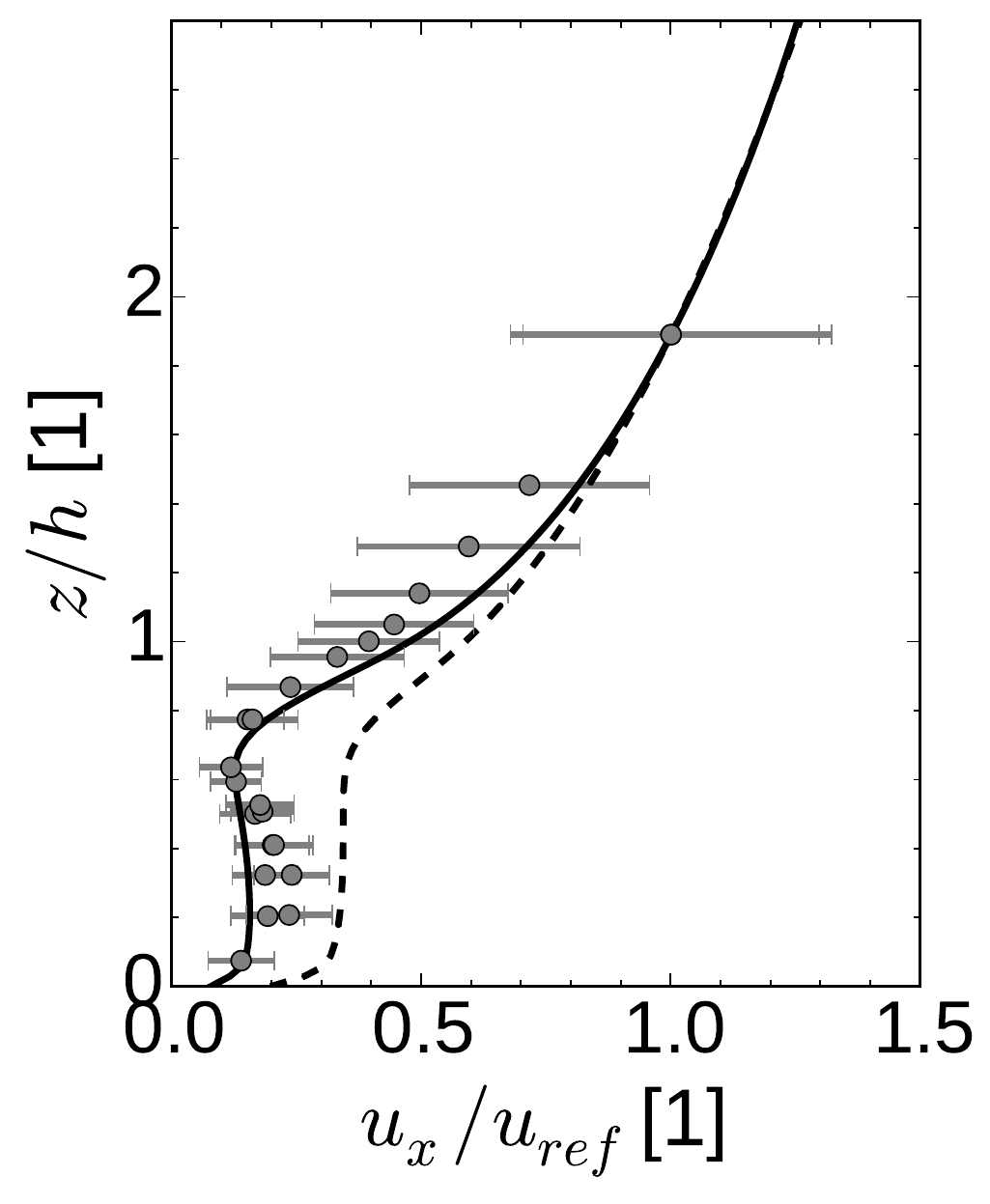}  &
    \includegraphics[width=\linewidth]{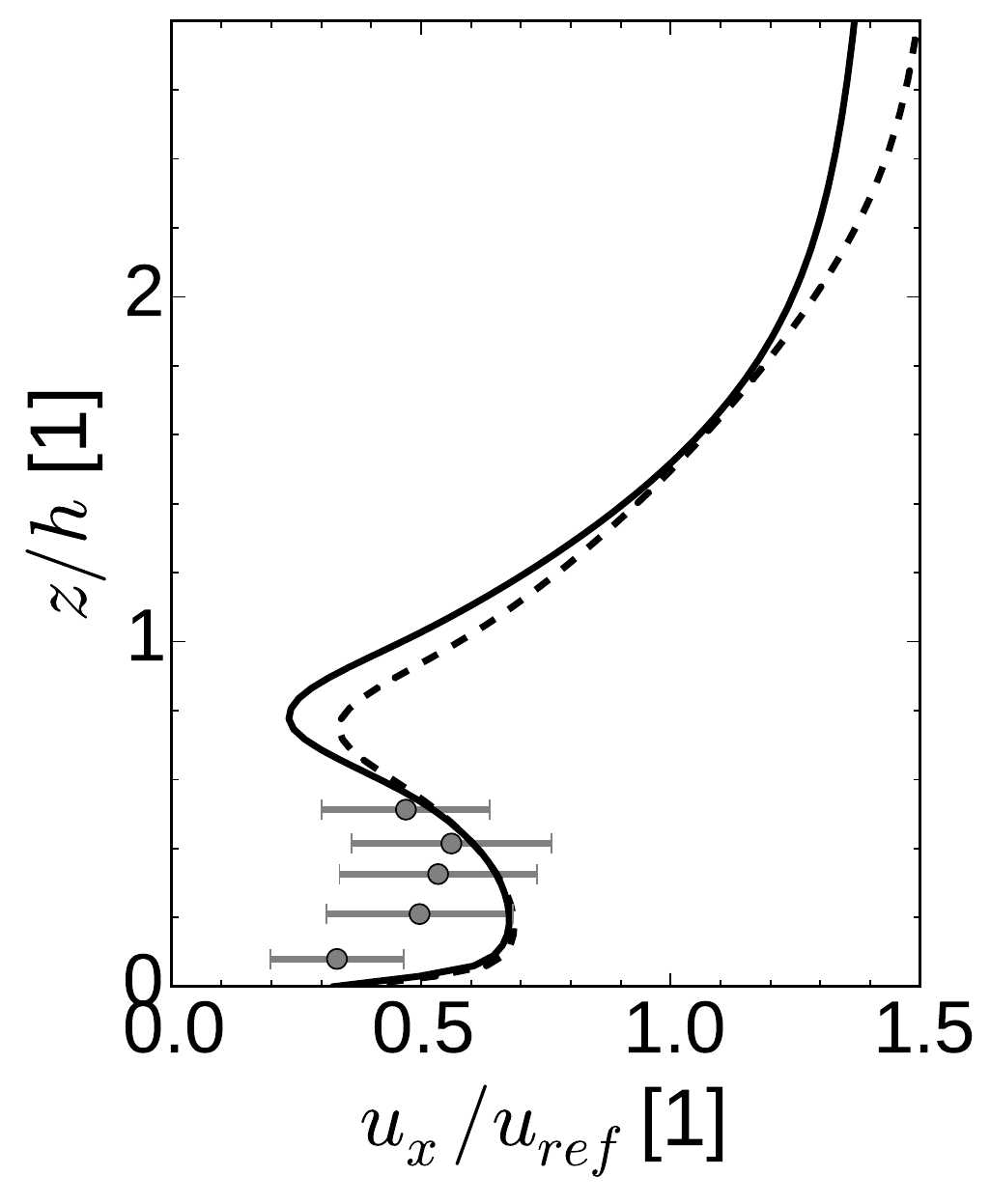} &
    \includegraphics[width=\linewidth]{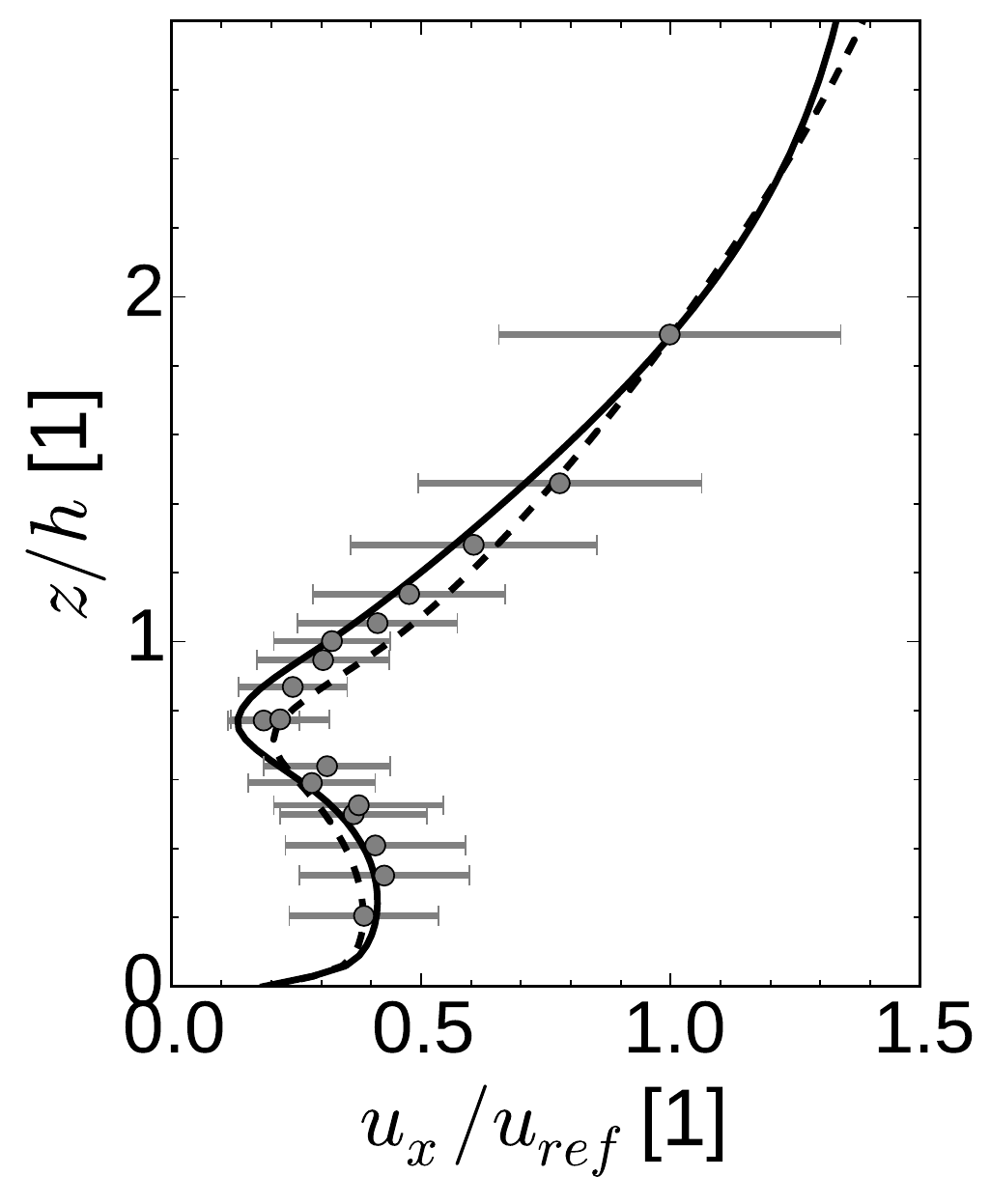} \\
    \includegraphics[width=\linewidth]{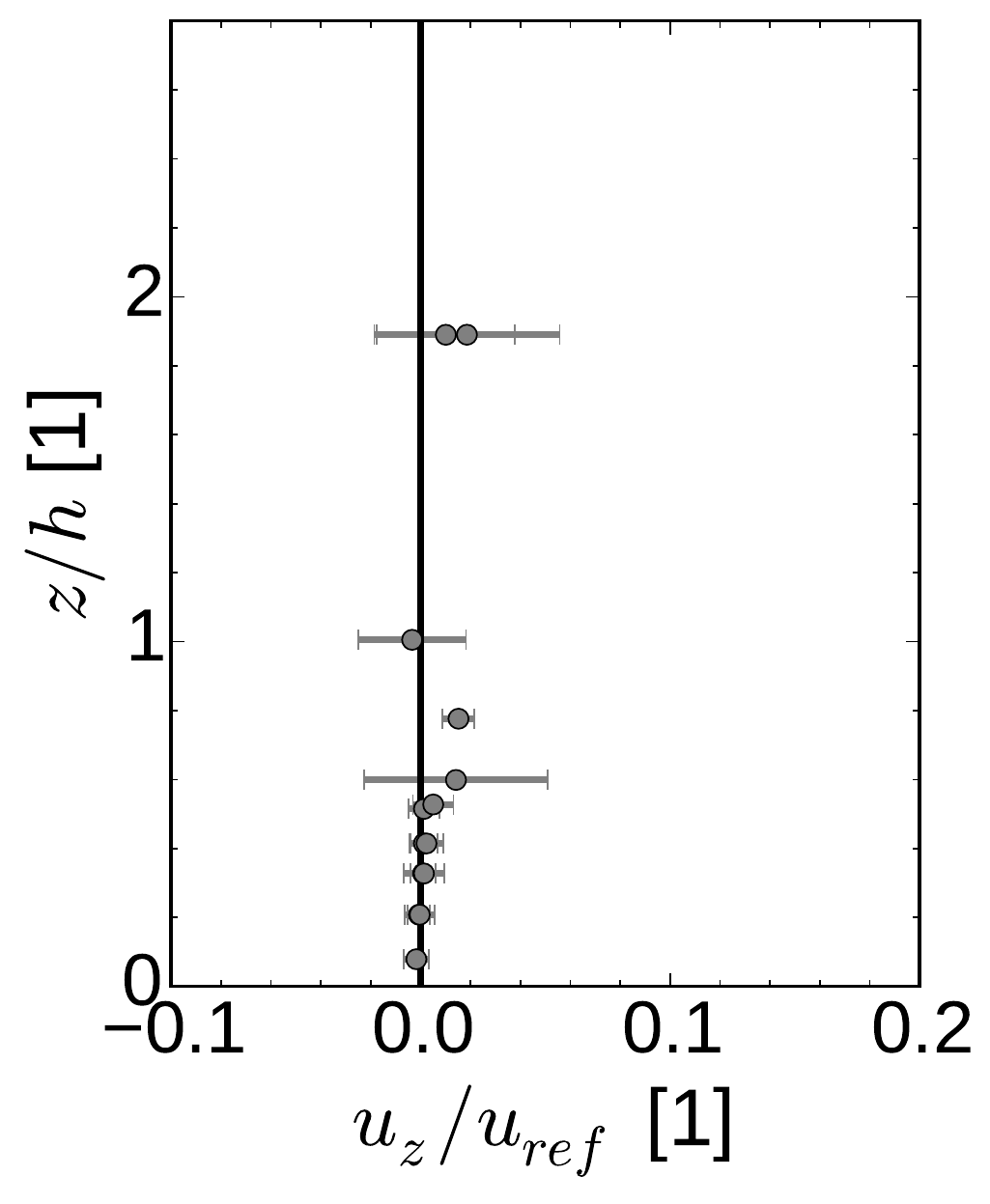}  &
    \includegraphics[width=\linewidth]{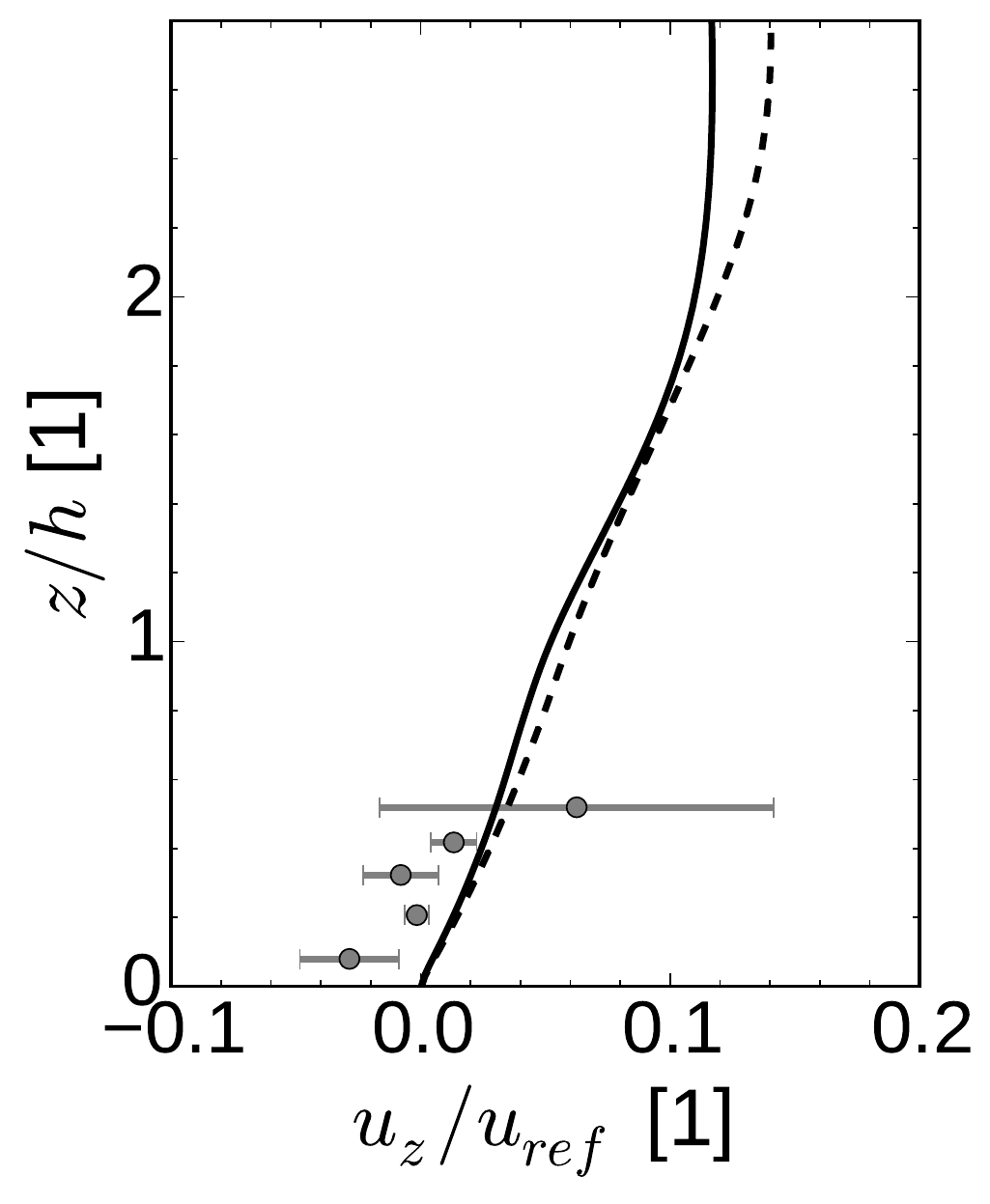} &
    \includegraphics[width=\linewidth]{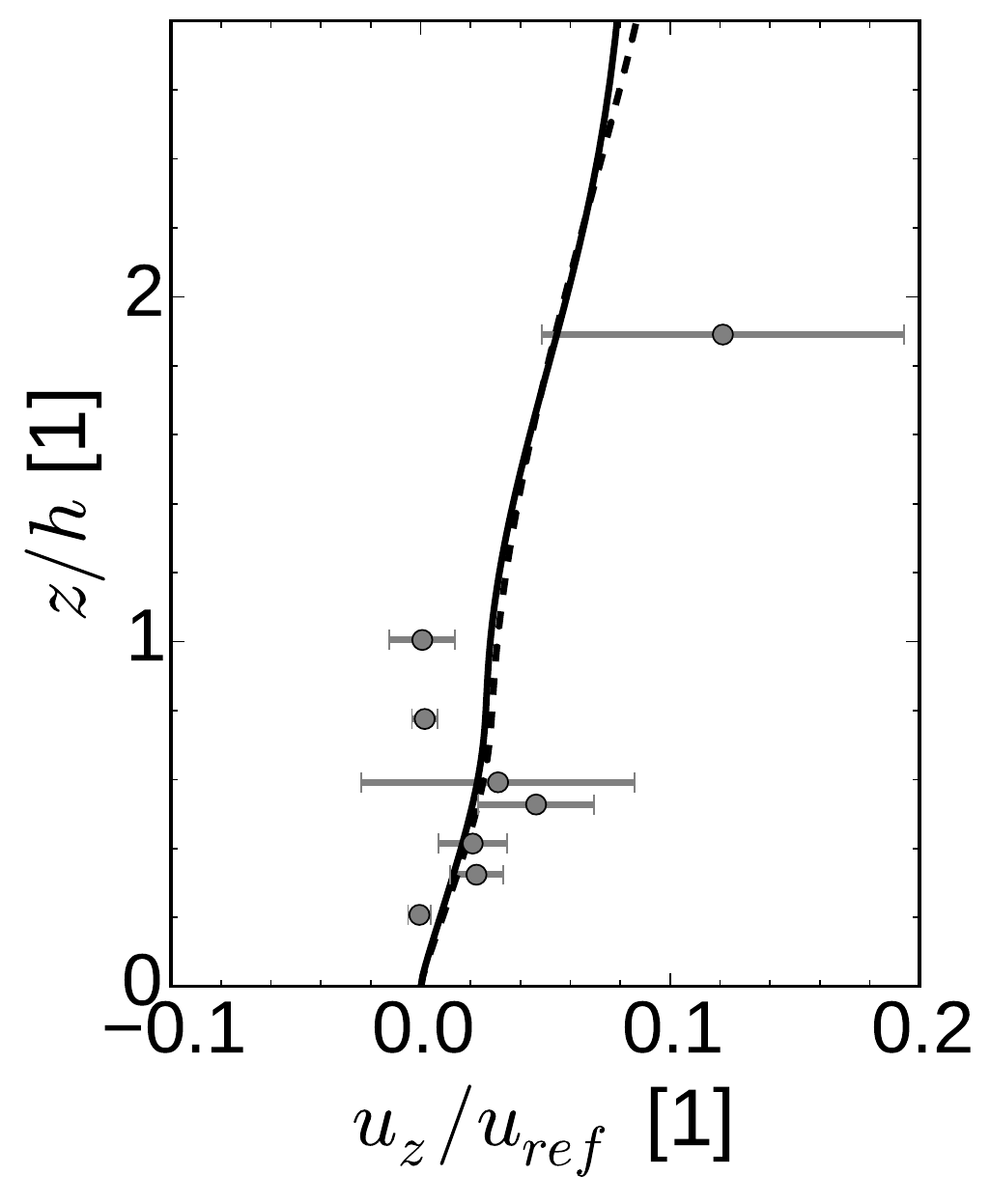} \\
    \includegraphics[width=\linewidth]{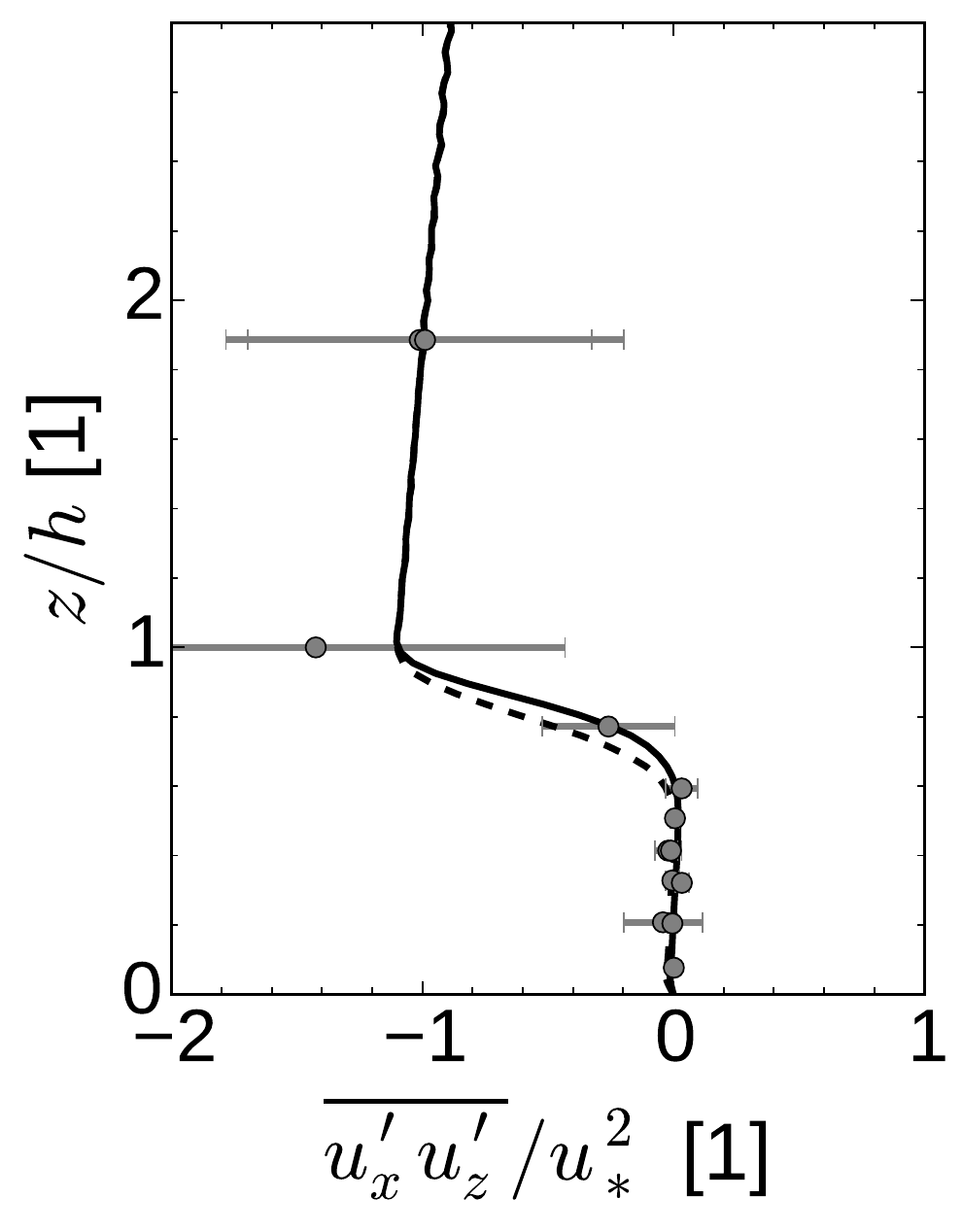}  &
    \includegraphics[width=\linewidth]{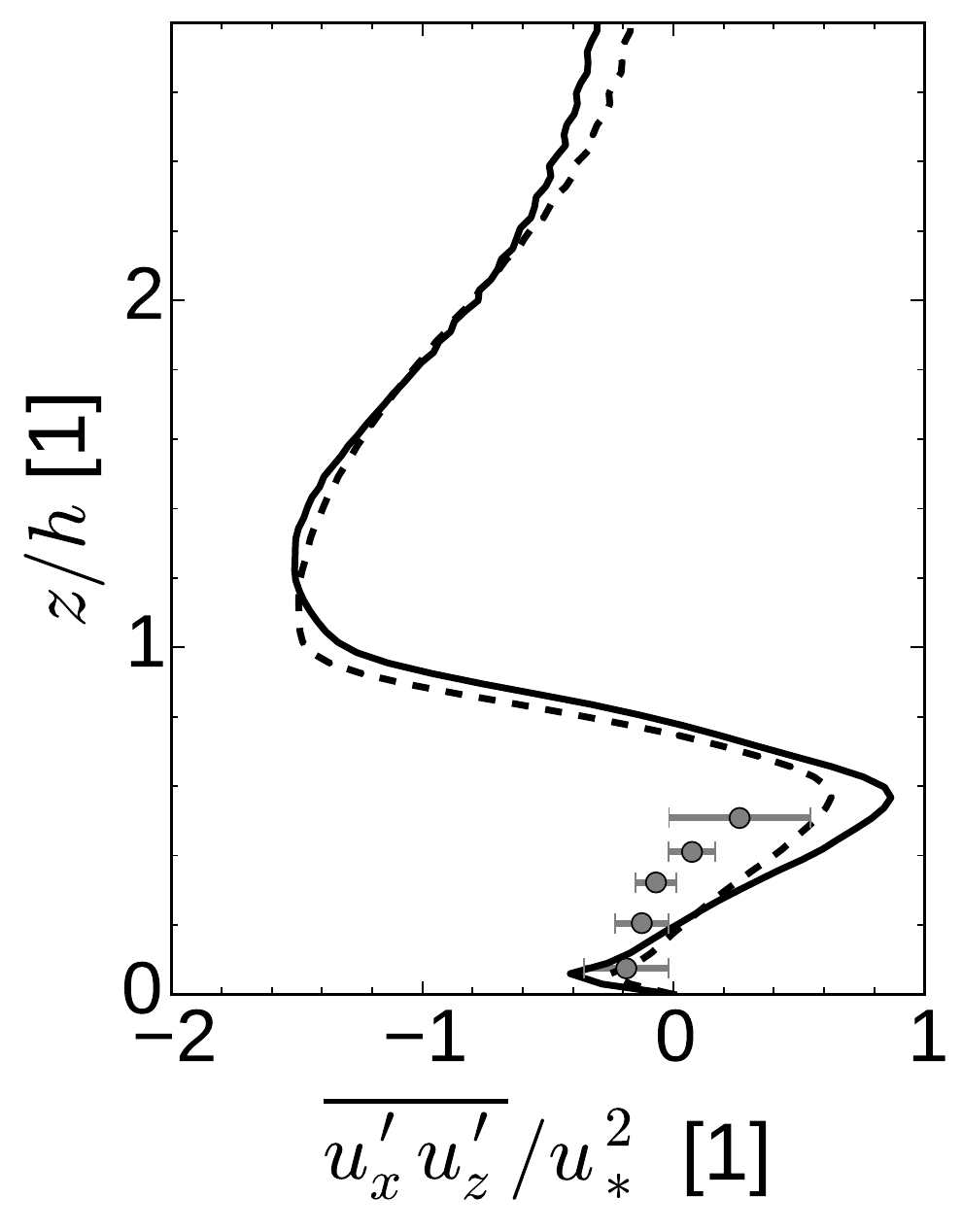} &
    \includegraphics[width=\linewidth]{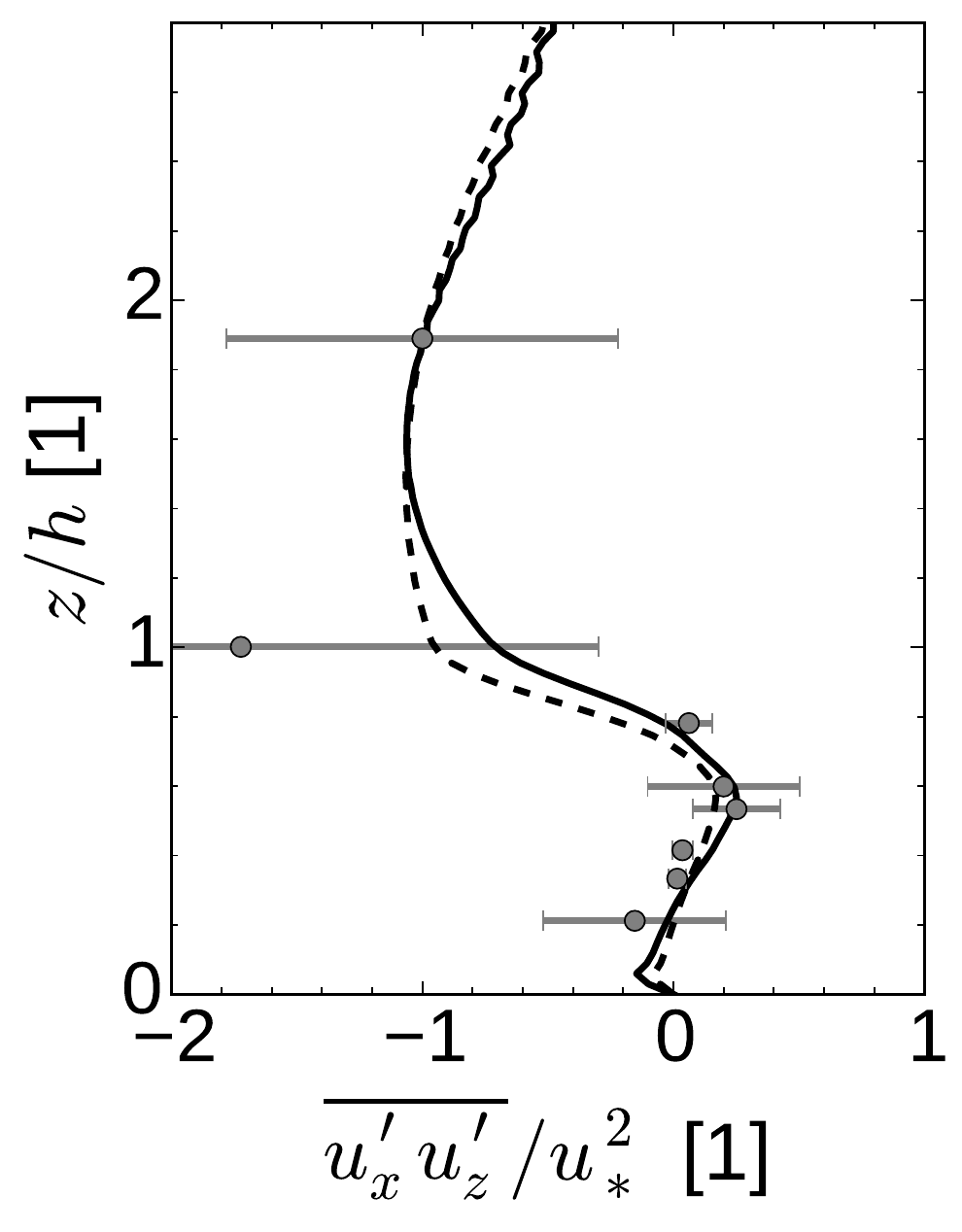} \\
    \includegraphics[width=\linewidth]{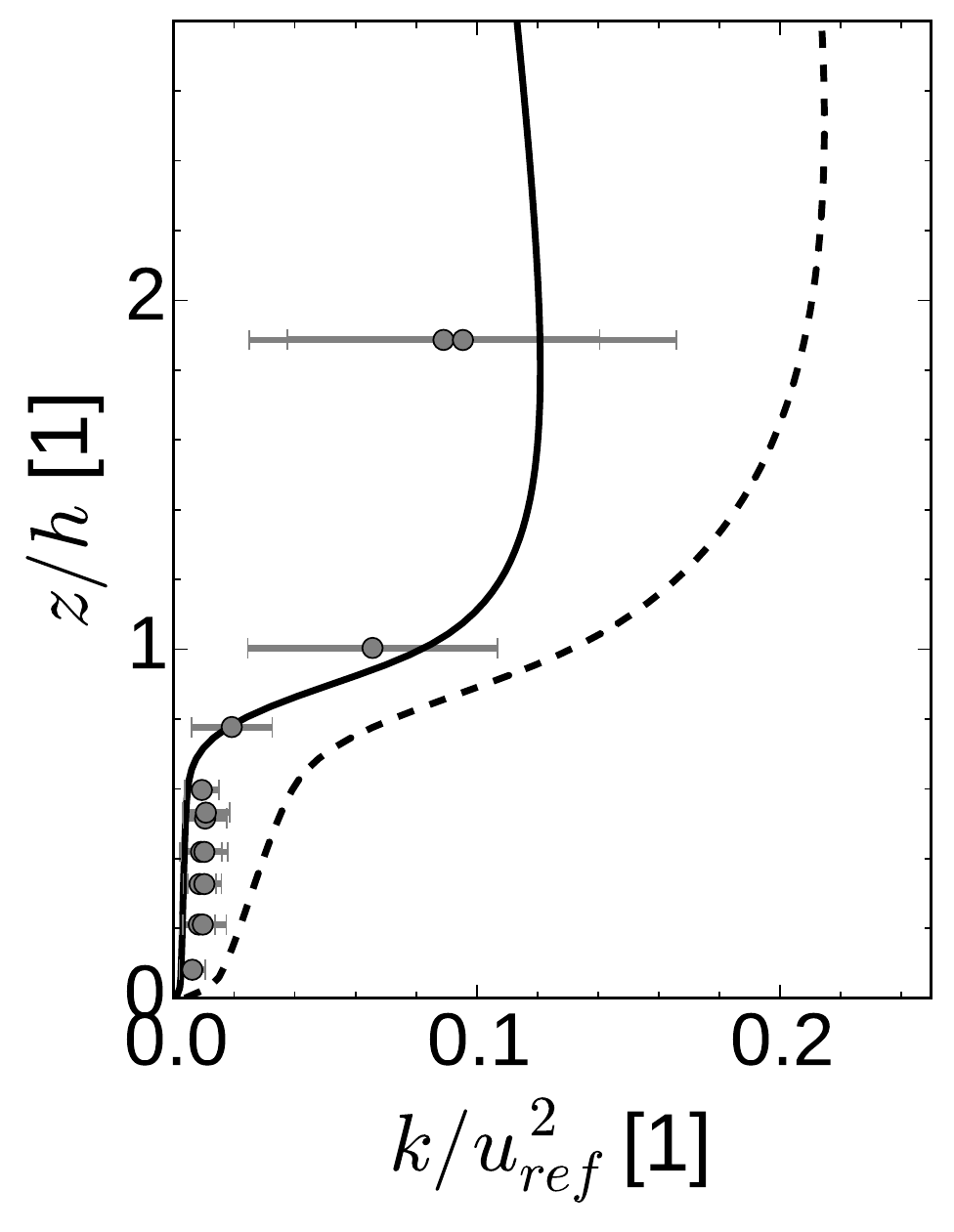}  &
    \includegraphics[width=\linewidth]{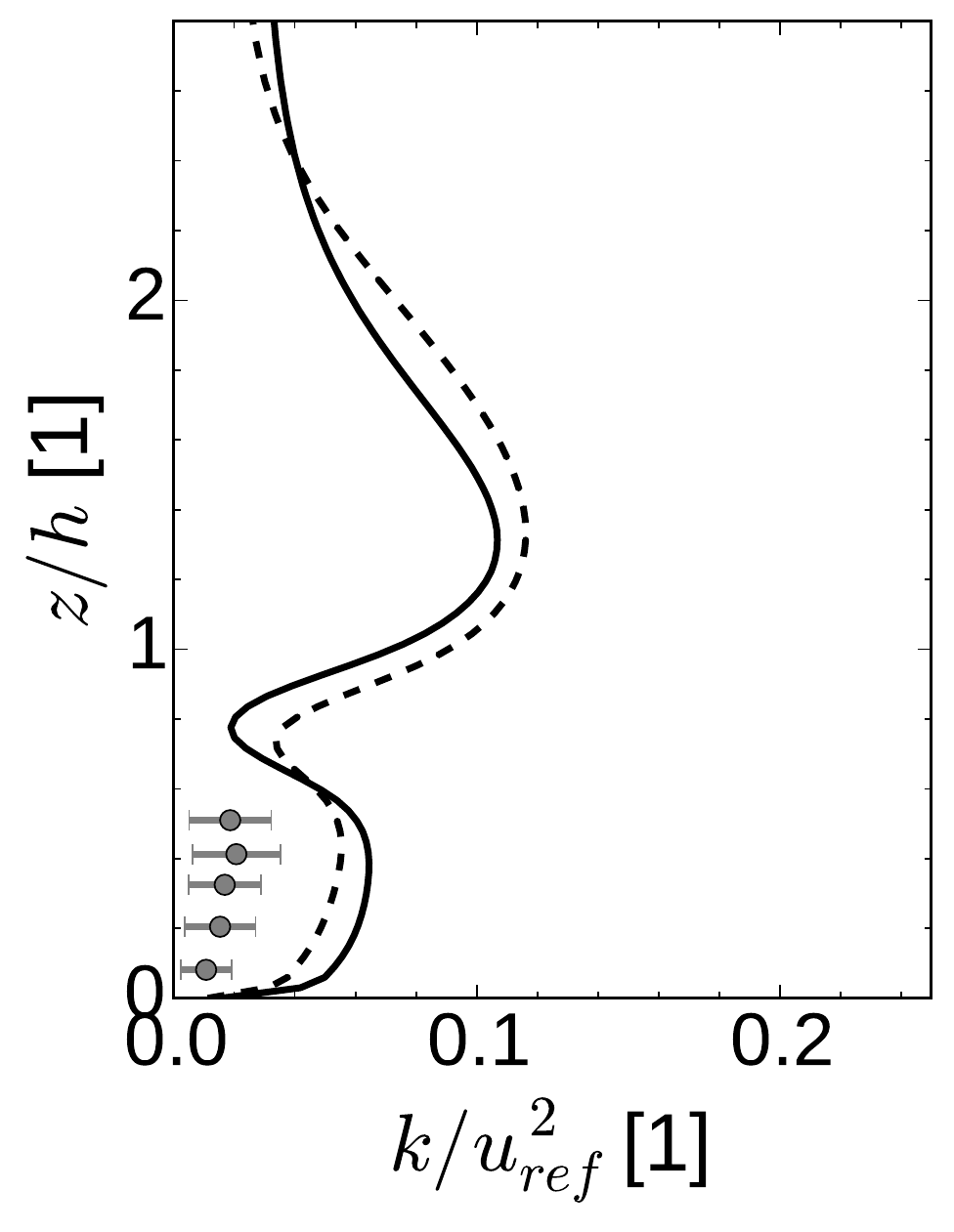} &
    \includegraphics[width=\linewidth]{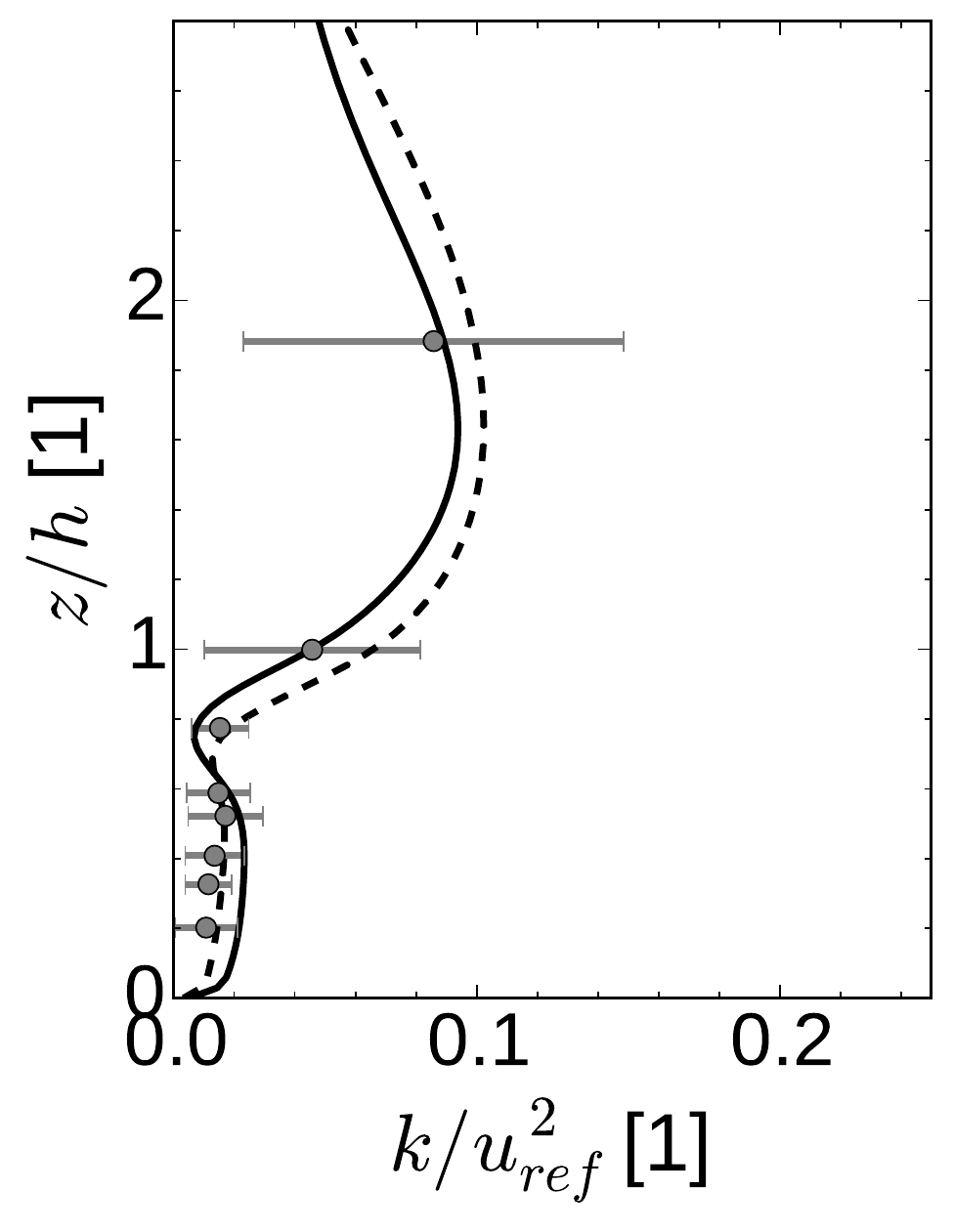} \\
  \end{tabular}
  \caption{Vertical profiles of horizontal velocity $u_x$, vertical velocity $u_z$, Reynolds stresses and turbulence kinetic energy $k$ inside and above the canopy. Values are normalized by the reference velocity $u_{\mathrm{ref}}$ or friction velocity $u_*$, both measured at height $z = \SI{41.5}{\m}$.
    Solid lines: $C_\mu = 0.03$, dashed lines: $C_\mu = 0.09$, symbols: measurements by \citet{DupontEtAl11}.
  }
  \label{fig:veg-fc-res}
\end{figure}

\paragraph{Homogeneous forest}
Fig. \ref{fig:veg-fc-res}, left column, shows the vertical profiles of normalized horizontal and vertical velocities, Reynolds stresses and turbulence kinetic energy, together with the measured values for the homogeneous forest case.
The values are normalized by a reference flow velocity $u_{\mathrm{ref}}$ and friction velocity $u_*$, both measured at the top of the tower, i.e. at height $z = \SI{41.5}{\m}$.

Above the canopy the velocity profile has a typical logarithmic profile. The velocity is quickly reduced inside the canopy, and reaches a secondary maximum in the open trunk space.
The secondary maximum is however reproduced only for $C_\mu = 0.03$, and not for $C_\mu = 0.09$. In that case, the reduction of the velocity is not as extensive, and the horizontal velocity is overpredicted inside the canopy and the trunk space.
The momentum fluxes are reduced to negligible values below the crown layer, signifying minimal momentum transfer between the flow above and below the crown layer. Turbulence kinetic energy is overpredicted for $C_\mu =0.09$, while the model with $C_\mu = 0.03$ shows good agreement with the measurements.

\paragraph{Edge flow}
The middle and the right columns of Fig. \ref{fig:veg-fc-res} show the vertical profiles of the same quantities as in the case of homogeneous forest at a distance $4h$ and $9h$ from the edge of the forest. The measured values at $4h$ are available only in the lower half of the canopy due to the smaller mast.

Compared to the homogeneous forest case, the secondary maximum of the horizontal velocity inside the trunk space is much more prominent at both locations. That is reproduced well, especially at $9h$ from the edge.
Upward motion caused by the deceleration of the flow, observed both at $4h$ and $9h$, is typical for the adjustment region close the edge of the forest \citep{DupontBrunet08a}.
The positive momentum flux inside the canopy, noted as ``striking'' by \citet{DupontEtAl11}, is reproduced at $4h$ and especially well at $9h$. Turbulence kinetic energy inside the canopy is overpredicted at $4h$, but reaches a good agreement at $9h$.

\paragraph{Discussion}

In the 2D edge flow case, the flow is well reproduced by the model with both sets of constants.
The similarity of the solutions in this case can be explained by the fact that the flow near the edge of the forest is heavily influenced by the inlet profile of the turbulent viscosity, which is independent of $C_\mu$ with our employed boundary conditions for $k$ and $\epsilon$. This is not the case for the 1D problem, and the difference of the calculated profiles for the two choices of $C_\mu$ is thus much more pronounced.
The model with $C_\mu = 0.09$ performs considerably worse, nevertheless, main features of the flow are still captured.

Arguably, the performance in the edge flow case is more relevant to the intended application of our model, which is mainly aimed at the problems of urban flows. In these settings, small, separated patches of vegetation are more typical than the continuous vegetation cover represented by the 1D case.
Therefore, considering comparable performance of the models with both set of constants in the 2D case, better performance with $C_\mu = 0.03$ in 1D does not justify the change of the universally accepted constant $C_\mu = 0.09$, so often used in the atmospheric modelling community for the flows without the vegetation \citep{CastroApsley97,HargreavesWright07,BaloghEtAl12,VranckxEtAl15} as well as with the vegetation present \citep{SvenssonHaeggkvist90,Green92,KenjeresKuile13,GromkeBlocken15a}.

\subsection{Particle collection by a hedgerow}
\label{sec:res-part-coll-windbr}

\paragraph{Influence of the drag coefficient}

\begin{figure}[ht]
  \centering
  \includegraphics[width=0.4\linewidth]{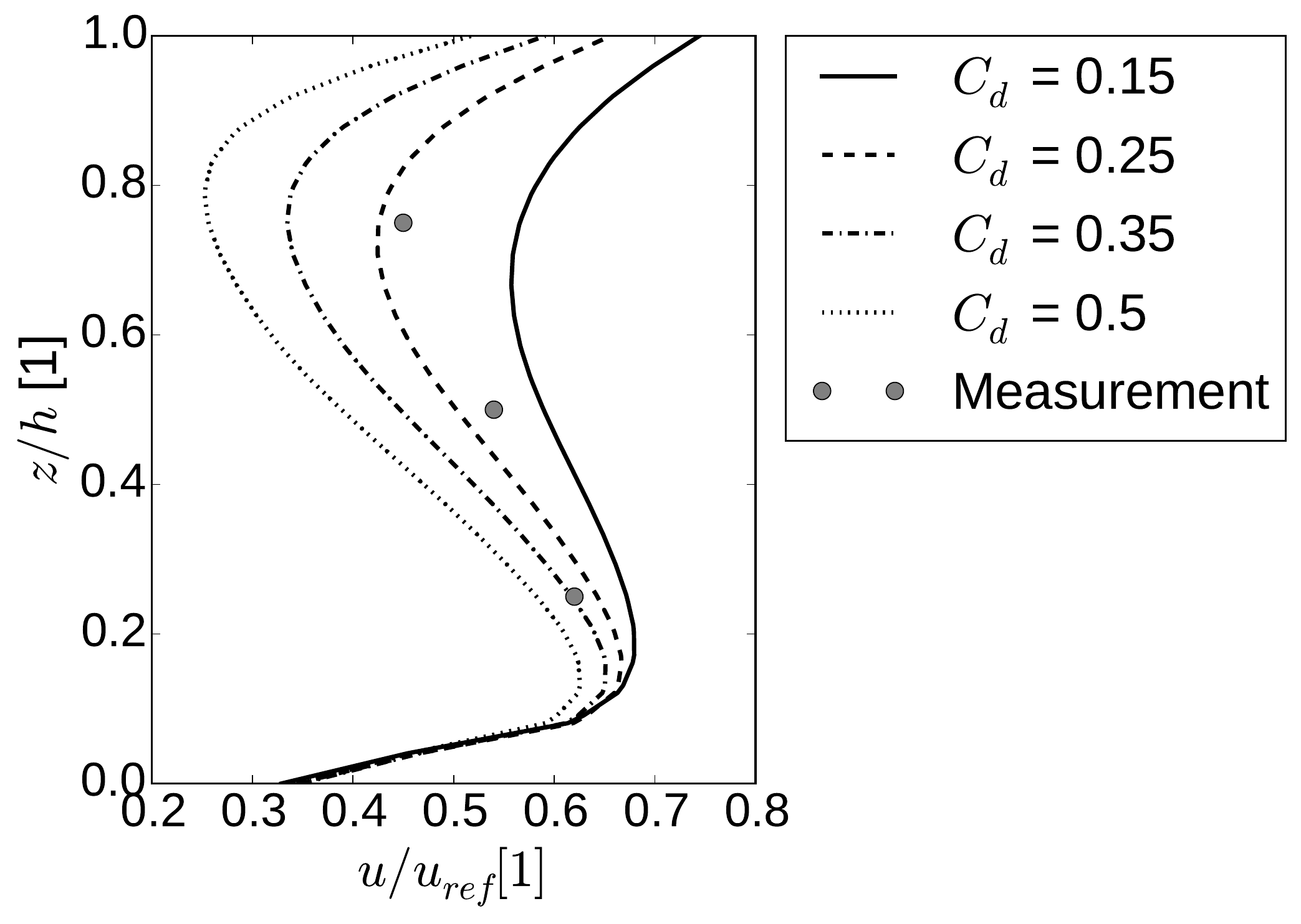}
  \caption{Vertical profiles of normalized velocity magnitude behind the vegetation barrier. Measured values taken from \citep{TiwaryEtAl05}.}
  \label{fig:veg-pc-uvert}
\end{figure}

The flow through and around the barrier was calculated for four values of the drag coefficient $C_d$, spanning the interval from 0.15 to 0.5 of realistic drag coefficient values \citep{EndalewEtAl09,KatulEtAl04}.
Fig. \ref{fig:veg-pc-uvert} shows the vertical profiles of the velocity magnitude behind the barrier normalized by the reference inlet velocity at height $h$, compared with the measured values.

As expected, the largest slowdown is in all cases observed around $z/h = 0.8$, where the LAD profile attains its maximal value. Local maximum of the velocity profile is visible around $z/h = 0.15$, to where is the blocked flow deflected.
Near-ground behaviour is affected mostly by the ground shear stress and is independent on the choice of the drag coefficient.

Choice of the drag coefficient $C_d = 0.25$ provides a reasonable agreement with the measured values at $z/h = 0.25$, $0.5$ and $0.75$. This choice is within the range usually given as realistic for vegetation barriers.
It is worth noting that the authors of the original paper \citep{TiwaryEtAl05} used the value $C_d = 0.5$ in their simulations and obtained a good agreement as well. This may be caused by the different vegetation model: while the source term in the momentum equation is the same in our and in their formulation, the authors of the original paper did not modify the turbulence model to include the vegetation effects.

\paragraph{Parameterization of the leaves}

Let us now turn to the filtering properties of the hedgerow. From the experiment, the filtering capacity was described via the \textit{collection efficiency} (CE), defined as
\begin{equation}
  CE = \frac{c_{in} - c_{out}}{c_{in}},
\end{equation}
where $c_{in}$ and $c_{out}$ are the values of the mass concentration measured $0.1h$ upwind and $0.1h$ downwind from the barrier at height $0.75h$. The collection velocity generally falls into the range between 0 and 100\%, but may reach negative values if the pollutant accumulates behind the barrier so that $c_{out} > c_{in}$.

\begin{figure}[ht]
  \centering
  \begin{tabular}{@{}p{0.35\linewidth}@{\qquad}p{0.35\linewidth}@{}}
    \subfigimg[width=0.95\linewidth]{A)}{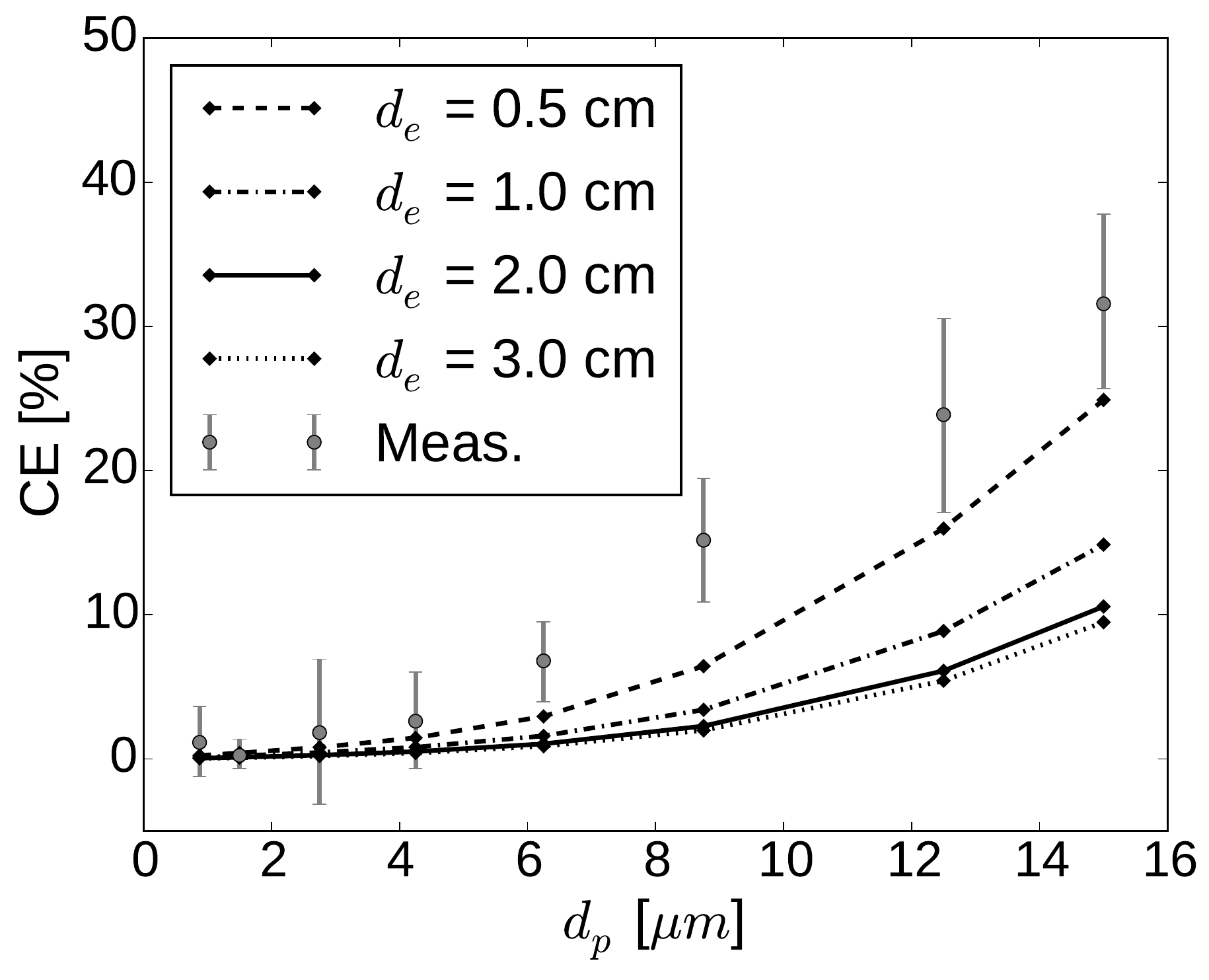} &
    \subfigimg[width=0.95\linewidth]{B)}{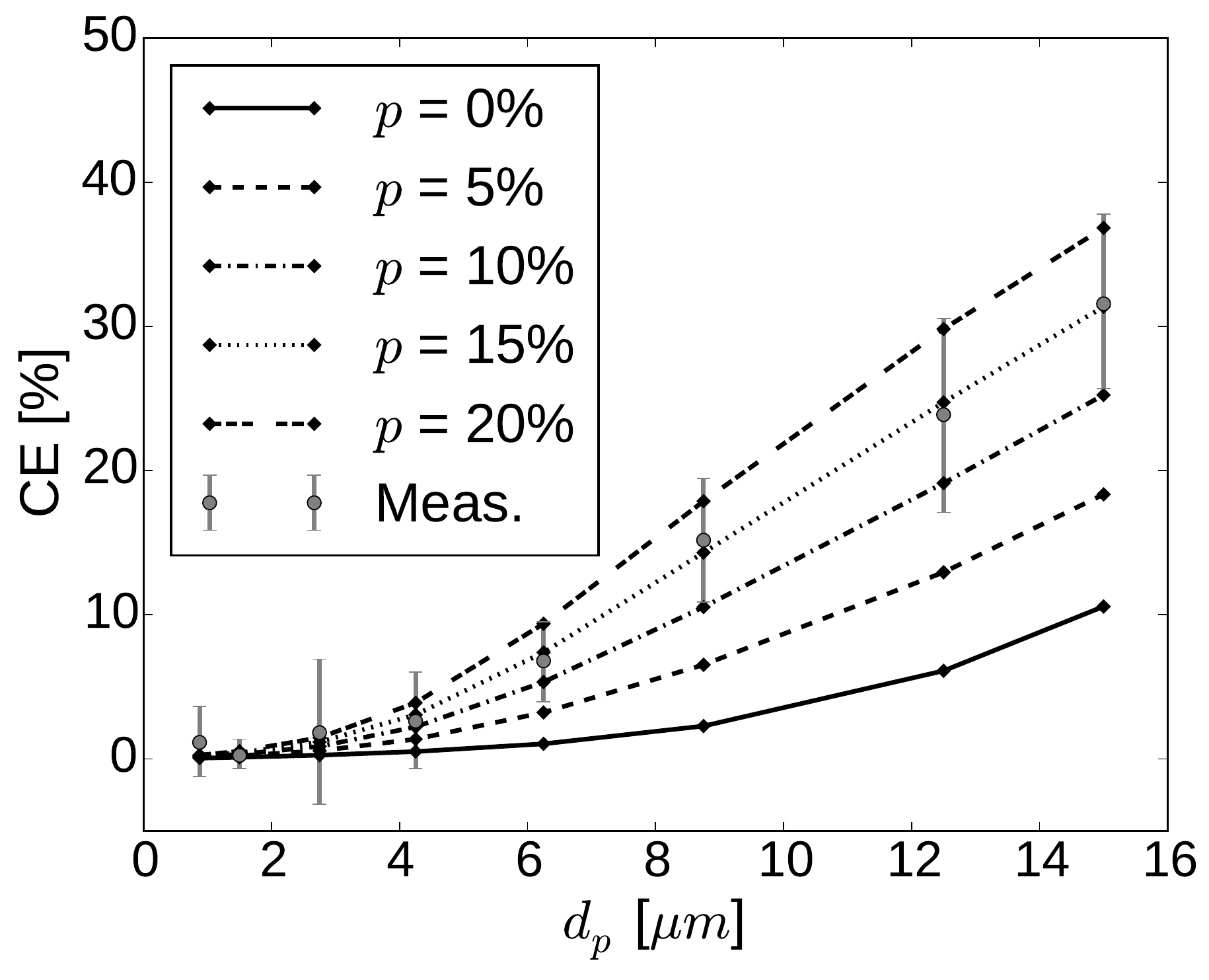} \\
  \end{tabular}
  \caption{Particle collection by a hedgerow test case. Collection efficiency and its dependency on the particle diameter and vegetation properties.
  (A) Leaves of different diameters $d_e$ with smooth surface. (B) Thorny leaves, modelled as a mixture of smooth leaves of diameter $d_e = \SI{2}{\cm}$ and needles of diameter $d_e = \SI{0.5}{\mm}$ with the parameterized proportion $p$ of the needle surface area to the total surface area. On both panels, the solid line references the same case of the vegetation with smooth leaves of diameter $d_e = \SI{2}{\cm}$. Measured data from \citep{TiwaryEtAl05}.}
  \label{fig:veg-pc-ce}
\end{figure}

In the adopted deposition velocity model, the vegetation is described by its type and typical size of the vegetation elements. Fig. \ref{fig:veg-pc-ce}A shows the calculated collection efficiencies when the vegetation elements are modelled as leaves with different diameters $d_e$. The increasing collection efficiency for particles of larger size, observed in the experiment, is clearly reproduced by our model.
Furthermore, the CE increases for smaller leaf sizes. \citet{TiwaryEtAl05} state that the range of the size of the hawthorn is between 1.1 and 3.2 cm. However, even when the leaf sizes are set to half of the value at the lower end of the range, the calculated collection efficiencies are still below the measured values.
This may be attributed to the neglected fine needle-like collectors, such as the leaf hairs and thorns, which increase the deposition velocity \citep{BeckettEtAl00,TiwaryEtAl05,Janhall15}. To reflect this, we further modelled the vegetation as a mix of planar leaf elements of diameter $d_e = \SI{2}{\cm}$ and fine needle-like collectors of diameter $d_e = \SI{0.5}{\mm}$. The proportion of the surface area of the fine collectors to the total surface area is denoted as $p$, and the deposition velocity is calculated as
\begin{equation}
  u_d = p u_d^{needle} + (1 - p) u_d^{leaf},
\end{equation}
where $u_d^{needle}$ and $u_d^{leaf}$ are the deposition velocities calculated for the needle and leaf elements respectively.

Comparison of the collection efficiencies calculated with this model and with the parameter $p$ ranging from 0\% to 20\% is shown on Fig. \ref{fig:veg-pc-ce}B. The collection efficiency rises with higher proportion of needles, reflecting the higher deposition velocity on the fine collectors. Best agreement with the measurement is obtained for $p = 15\%$.
It is therefore demonstrated that this approach, grounded in the realistic assumption of mixed leaves and needles, is capable of reproducing the observed behaviour. However, the exact proportion of the needle surface area is difficult to directly compare with the biological data, as we are not aware of any study on this topic.

\section{Conclusions}

The solver focused on small scale atmospheric flows in areas with vegetation present was described and tested.
The 2D hill test case showed that the solver is capable of correctly predicting the separation of the flow in the examined configurations, although there were deficiencies in the shape of the turbulence kinetic energy profiles. This also negatively affected the computed pollutant concentrations.
Isotropic turbulence model is however unlikely to predict the dispersion perfectly in any case, as it cannot predict the different rates of turbulent mixing in horizontal and vertical directions, which were observed in the atmosphere.

The warm bubble test case proved the applicability of the solver for the thermally driven flows. The inclusion of the potential temperature equation also opens the way for modelling the thermal effects of the vegetation, which are not yet included in the model formulation.

Modelling the flow through the forest canopy revealed the difficulties associated with the choice of constants of the \ke model. The problem was however less pronounced in the edge flow case, for which the usual choice $C_\mu = 0.09$ produced good agreement with the measurements.
If we wanted to reduce the effects of the choice of model constants, we may rather use large eddy simulations. The subgrid-scale model constants in LES have smaller influence on the results, and LES may thus produce more reliable results.

Finally, the hedgerow test case showed that the dry deposition model can reproduce the measured collection efficiencies well, although it required some fine tuning of the parameters describing the vegetation. In the choice of the vegetation parameters lies a significant obstacle for any modelling studies. Unless we have field measurements available, these parameters must be estimated, and as demonstrated here, their choice may profoundly influence the results.
Usage of suitable methods for managing this parameter uncertainty is therefore desirable.

\section*{Acknowledgements}

This work was supported by the grant SGS16/206/OHK2/3T/12 of the Czech Technical University in~Prague.


\begin{thebibliography}{45}
\expandafter\ifx\csname natexlab\endcsname\relax\def\natexlab#1{#1}\fi
\providecommand{\url}[1]{\texttt{#1}}
\providecommand{\href}[2]{#2}
\providecommand{\path}[1]{#1}
\providecommand{\DOIprefix}{doi:}
\providecommand{\ArXivprefix}{arXiv:}
\providecommand{\URLprefix}{URL: }
\providecommand{\Pubmedprefix}{pmid:}
\providecommand{\doi}[1]{\href{http://dx.doi.org/#1}{\path{#1}}}
\providecommand{\Pubmed}[1]{\href{pmid:#1}{\path{#1}}}
\providecommand{\bibinfo}[2]{#2}
\ifx\xfnm\relax \def\xfnm[#1]{\unskip,\space#1}\fi
\bibitem[{Balay et~al.(2015)Balay, Abhyankar, Adams, Brown, Brune, Buschelman,
  Dalcin, Eijkhout, Gropp, Kaushik, Knepley, McInnes, Rupp, Smith, Zampini and
  Zhang}]{petsc-web-page}
\bibinfo{author}{Balay, S.}, \bibinfo{author}{Abhyankar, S.},
  \bibinfo{author}{Adams, M.}, \bibinfo{author}{Brown, J.},
  \bibinfo{author}{Brune, P.}, \bibinfo{author}{Buschelman, K.},
  \bibinfo{author}{Dalcin, L.}, \bibinfo{author}{Eijkhout, V.},
  \bibinfo{author}{Gropp, W.}, \bibinfo{author}{Kaushik, D.},
  \bibinfo{author}{Knepley, M.}, \bibinfo{author}{McInnes, L.},
  \bibinfo{author}{Rupp, K.}, \bibinfo{author}{Smith, B.},
  \bibinfo{author}{Zampini, S.}, \bibinfo{author}{Zhang, H.},
  \bibinfo{year}{2015}.
\newblock \bibinfo{title}{{PETS}c {W}eb page}.
\newblock \bibinfo{howpublished}{\url{http://www.mcs.anl.gov/petsc}}.
\bibitem[{Balogh et~al.(2012)Balogh, Parente and Benocci}]{BaloghEtAl12}
\bibinfo{author}{Balogh, M.}, \bibinfo{author}{Parente, A.},
  \bibinfo{author}{Benocci, C.}, \bibinfo{year}{2012}.
\newblock \bibinfo{title}{{RANS} simulation of {ABL} flow over complex terrains
  applying an enhanced k-{$\epsilon$} model and wall function formulation:
  Implementation and comparison for fluent and {OpenFOAM}}.
\newblock \bibinfo{journal}{J. Wind Eng. Ind. Aerodyn.}
  \bibinfo{volume}{104-106}, \bibinfo{pages}{360--368}.
\newblock \DOIprefix\doi{10.1016/j.jweia.2012.02.023}.
\bibitem[{Beckett et~al.(2000)Beckett, Freer-Smith and Taylor}]{BeckettEtAl00}
\bibinfo{author}{Beckett, K.P.}, \bibinfo{author}{Freer-Smith, P.},
  \bibinfo{author}{Taylor, G.}, \bibinfo{year}{2000}.
\newblock \bibinfo{title}{Particulate pollution capture by urban trees: effect
  of species and windspeed}.
\newblock \bibinfo{journal}{Global Change Biol.} \bibinfo{volume}{6},
  \bibinfo{pages}{995--1003}.
\newblock \DOIprefix\doi{10.1046/j.1365-2486.2000.00376.x}.
\bibitem[{Blazek(2001)}]{Blazek01}
\bibinfo{author}{Blazek, J.}, \bibinfo{year}{2001}.
\newblock \bibinfo{title}{Computational fluid dynamics: principles and
  applications}.
\newblock \bibinfo{publisher}{Elsevier}.
\bibitem[{Blocken(2015)}]{Blocken15}
\bibinfo{author}{Blocken, B.}, \bibinfo{year}{2015}.
\newblock \bibinfo{title}{Computational fluid dynamics for urban physics:
  Importance, scales, possibilities, limitations and ten tips and tricks
  towards accurate and reliable simulations}.
\newblock \bibinfo{journal}{Build. Environ.} \bibinfo{volume}{91},
  \bibinfo{pages}{219--245}.
\newblock \DOIprefix\doi{10.1016/j.buildenv.2015.02.015}.
\bibitem[{Blocken et~al.(2012)Blocken, Janssen and van Hooff}]{BlockenEtAl12}
\bibinfo{author}{Blocken, B.}, \bibinfo{author}{Janssen, W.},
  \bibinfo{author}{van Hooff, T.}, \bibinfo{year}{2012}.
\newblock \bibinfo{title}{{CFD} simulation for pedestrian wind comfort and wind
  safety in urban areas: General decision framework and case study for the
  eindhoven university campus}.
\newblock \bibinfo{journal}{Environ. Modell. Software} \bibinfo{volume}{30},
  \bibinfo{pages}{15--34}.
\newblock \DOIprefix\doi{10.1016/j.envsoft.2011.11.009}.
\bibitem[{Bruse(2007)}]{Bruse07}
\bibinfo{author}{Bruse, M.}, \bibinfo{year}{2007}.
\newblock \bibinfo{title}{Particle filtering capacity of urban vegetation: a
  microscale numerical approach}.
\newblock \bibinfo{journal}{Berliner Geographische Arbeiten}
  \bibinfo{volume}{109}, \bibinfo{pages}{61--70}.
\bibitem[{Buccolieri et~al.(2011)Buccolieri, Salim, Leo, Sabatino, Chan, Ielpo,
  de~Gennaro and Gromke}]{BuccolieriEtAl11}
\bibinfo{author}{Buccolieri, R.}, \bibinfo{author}{Salim, S.M.},
  \bibinfo{author}{Leo, L.S.}, \bibinfo{author}{Sabatino, S.D.},
  \bibinfo{author}{Chan, A.}, \bibinfo{author}{Ielpo, P.},
  \bibinfo{author}{de~Gennaro, G.}, \bibinfo{author}{Gromke, C.},
  \bibinfo{year}{2011}.
\newblock \bibinfo{title}{Analysis of local scale tree{\textendash}atmosphere
  interaction on pollutant concentration in idealized street canyons and
  application to a real urban junction}.
\newblock \bibinfo{journal}{Atmos. Environ.} \bibinfo{volume}{45},
  \bibinfo{pages}{1702--1713}.
\newblock \DOIprefix\doi{10.1016/j.atmosenv.2010.12.058}.
\bibitem[{Castro and Apsley(1997)}]{CastroApsley97}
\bibinfo{author}{Castro, I.P.}, \bibinfo{author}{Apsley, D.D.},
  \bibinfo{year}{1997}.
\newblock \bibinfo{title}{Flow and dispersion over topography: A comparison
  between numerical and laboratory data for two-dimensional flows}.
\newblock \bibinfo{journal}{Atmos. Environ.} \bibinfo{volume}{31},
  \bibinfo{pages}{839--850}.
\newblock \DOIprefix\doi{10.1016/s1352-2310(96)00248-8}.
\bibitem[{Chan and van~der Vorst(2001)}]{ChanVanderVorst01}
\bibinfo{author}{Chan, T.}, \bibinfo{author}{van~der Vorst, H.},
  \bibinfo{year}{2001}.
\newblock \bibinfo{title}{Approximate and incomplete factorizations}.
\newblock \bibinfo{journal}{Parallel Numerical Algorithms, ICASE/LaRC
  Interdisciplinary Series in Science and Engeneering} ,
  \bibinfo{pages}{167--202}.
\bibitem[{Dupont et~al.(2011)Dupont, Bonnefond, Irvine, Lamaud and
  Brunet}]{DupontEtAl11}
\bibinfo{author}{Dupont, S.}, \bibinfo{author}{Bonnefond, J.M.},
  \bibinfo{author}{Irvine, M.R.}, \bibinfo{author}{Lamaud, E.},
  \bibinfo{author}{Brunet, Y.}, \bibinfo{year}{2011}.
\newblock \bibinfo{title}{Long-distance edge effects in a pine forest with a
  deep and sparse trunk space: In situ and numerical experiments}.
\newblock \bibinfo{journal}{Agric. For. Meteorol.} \bibinfo{volume}{151},
  \bibinfo{pages}{328--344}.
\newblock \DOIprefix\doi{10.1016/j.agrformet.2010.11.007}.
\bibitem[{Dupont and Brunet(2008)}]{DupontBrunet08a}
\bibinfo{author}{Dupont, S.}, \bibinfo{author}{Brunet, Y.},
  \bibinfo{year}{2008}.
\newblock \bibinfo{title}{Edge flow and canopy structure: A large-eddy
  simulation study}.
\newblock \bibinfo{journal}{Boundary-Layer Meteorol.} \bibinfo{volume}{126},
  \bibinfo{pages}{51--71}.
\newblock \DOIprefix\doi{10.1007/s10546-007-9216-3}.
\bibitem[{Endalew et~al.(2009)Endalew, Hertog, Delele, Baetens, Persoons,
  Baelmans, Ramon, Nicola{\"\i} and Verboven}]{EndalewEtAl09}
\bibinfo{author}{Endalew, A.M.}, \bibinfo{author}{Hertog, M.},
  \bibinfo{author}{Delele, M.}, \bibinfo{author}{Baetens, K.},
  \bibinfo{author}{Persoons, T.}, \bibinfo{author}{Baelmans, M.},
  \bibinfo{author}{Ramon, H.}, \bibinfo{author}{Nicola{\"\i}, B.},
  \bibinfo{author}{Verboven, P.}, \bibinfo{year}{2009}.
\newblock \bibinfo{title}{{CFD} modelling and wind tunnel validation of airflow
  through plant canopies using {3D} canopy architecture}.
\newblock \bibinfo{journal}{Int. J. Heat Fluid Flow} \bibinfo{volume}{30},
  \bibinfo{pages}{356--368}.
\newblock \DOIprefix\doi{10.1016/j.ijheatfluidflow.2008.12.007}.
\bibitem[{{ERCOFTAC}(2004)}]{ErcoftacQnetAC5-05}
\bibinfo{author}{{ERCOFTAC}}, \bibinfo{year}{2004}.
\newblock \bibinfo{title}{{ERCOFTAC} {QNET-CFD} {Wiki}, application challenge
  5-05}.
\newblock
  \bibinfo{howpublished}{\url{http://qnet-ercoftac.cfms.org.uk/w/index.php/AC_5-05}}.
\newblock \bibinfo{note}{[Online; accessed 22-August-2016]}.
\bibitem[{Giraldo and Restelli(2008)}]{GiraldoRestelli08}
\bibinfo{author}{Giraldo, F.X.}, \bibinfo{author}{Restelli, M.},
  \bibinfo{year}{2008}.
\newblock \bibinfo{title}{A study of spectral element and discontinuous
  {Galerkin} methods for the {Navier}--{Stokes} equations in nonhydrostatic
  mesoscale atmospheric modeling: {Equation} sets and test cases}.
\newblock \bibinfo{journal}{J. Comput. Phys.} \bibinfo{volume}{227},
  \bibinfo{pages}{3849--3877}.
\newblock \DOIprefix\doi{10.1016/j.jcp.2007.12.009}.
\bibitem[{Green(1992)}]{Green92}
\bibinfo{author}{Green, S.}, \bibinfo{year}{1992}.
\newblock \bibinfo{title}{Modelling turbulent air flow in a stand of
  widely-spaced trees}.
\newblock \bibinfo{journal}{Phoenics J.} \bibinfo{volume}{5},
  \bibinfo{pages}{294--312}.
\bibitem[{Greenshields(2015)}]{OpenFoamUserGuide}
\bibinfo{author}{Greenshields, C.J.}, \bibinfo{year}{2015}.
\newblock \bibinfo{title}{Open{FOAM} - {T}he {O}pen {S}ource {CFD} {T}oolbox -
  {U}ser's {G}uide. Version 3.0.0}.
\newblock \bibinfo{organization}{CFD Direct Ltd.}
\bibitem[{Gromke and Blocken(2015)}]{GromkeBlocken15a}
\bibinfo{author}{Gromke, C.}, \bibinfo{author}{Blocken, B.},
  \bibinfo{year}{2015}.
\newblock \bibinfo{title}{Influence of avenue-trees on air quality at the urban
  neighborhood scale. {Part I}: Quality assurance studies and turbulent schmidt
  number analysis for {RANS} {CFD} simulations}.
\newblock \bibinfo{journal}{Environ. Pollut.} \bibinfo{volume}{196},
  \bibinfo{pages}{214--223}.
\newblock \DOIprefix\doi{10.1016/j.envpol.2014.10.016}.
\bibitem[{Hargreaves and Wright(2007)}]{HargreavesWright07}
\bibinfo{author}{Hargreaves, D.}, \bibinfo{author}{Wright, N.},
  \bibinfo{year}{2007}.
\newblock \bibinfo{title}{On the use of the k{\textendash} model in commercial
  {CFD} software to model the neutral atmospheric boundary layer}.
\newblock \bibinfo{journal}{J. Wind Eng. Ind. Aerodyn.} \bibinfo{volume}{95},
  \bibinfo{pages}{355--369}.
\newblock \DOIprefix\doi{10.1016/j.jweia.2006.08.002}.
\bibitem[{Janhäll(2015)}]{Janhall15}
\bibinfo{author}{Janhäll, S.}, \bibinfo{year}{2015}.
\newblock \bibinfo{title}{Review on urban vegetation and particle air pollution
  - deposition and dispersion}.
\newblock \bibinfo{journal}{Atmos. Environ.} \bibinfo{volume}{105},
  \bibinfo{pages}{130--137}.
\newblock \DOIprefix\doi{10.1016/j.atmosenv.2015.01.052}.
\bibitem[{Karel(2014)}]{Karel14}
\bibinfo{author}{Karel, J.}, \bibinfo{year}{2014}.
\newblock \bibinfo{title}{Numerical simulation of streamer propagation on
  unstructured dynamically adapted grids}.
\newblock Ph.D. thesis. Czech Technical University in Prague and Universit{\'e}
  Paris 13.
\bibitem[{Katul et~al.(2004)Katul, Mahrt, Poggi and Sanz}]{KatulEtAl04}
\bibinfo{author}{Katul, G.}, \bibinfo{author}{Mahrt, L.},
  \bibinfo{author}{Poggi, D.}, \bibinfo{author}{Sanz, C.},
  \bibinfo{year}{2004}.
\newblock \bibinfo{title}{One- and two-equation models for canopy turbulence}.
\newblock \bibinfo{journal}{Bound. Layer Meteor.} \bibinfo{volume}{113},
  \bibinfo{pages}{81--109}.
\newblock \DOIprefix\doi{10.1016/0167-6105(93)90124-7}.
\bibitem[{Kenjereš and ter Kuile(2013)}]{KenjeresKuile13}
\bibinfo{author}{Kenjereš, S.}, \bibinfo{author}{ter Kuile, B.},
  \bibinfo{year}{2013}.
\newblock \bibinfo{title}{Modelling and simulations of turbulent flows in urban
  areas with vegetation}.
\newblock \bibinfo{journal}{J. Wind Eng. Ind. Aerodyn.} \bibinfo{volume}{123},
  \bibinfo{pages}{43--55}.
\newblock \DOIprefix\doi{10.1016/j.jweia.2013.09.007}.
\bibitem[{Khurshudyan et~al.(1981)Khurshudyan, Snyder and
  Nekrasov}]{KhurshudyanEtAl81}
\bibinfo{author}{Khurshudyan, L.H.}, \bibinfo{author}{Snyder, W.H.},
  \bibinfo{author}{Nekrasov, I.V.}, \bibinfo{year}{1981}.
\newblock \bibinfo{title}{Flow and dispersion of pollutants over
  two-dimensional hills: Summary report on joint Soviet-American study}.
\newblock \bibinfo{type}{Technical Report} \bibinfo{number}{EPA-600/4-81-067}.
  U.S. Environmental Protection Agency.
\bibitem[{Knoll and Keyes(2004)}]{KnollKeyes04}
\bibinfo{author}{Knoll, D.}, \bibinfo{author}{Keyes, D.}, \bibinfo{year}{2004}.
\newblock \bibinfo{title}{Jacobian-free {Newton}{\textendash}{Krylov} methods:
  a survey of approaches and applications}.
\newblock \bibinfo{journal}{J. Comput. Phys.} \bibinfo{volume}{193},
  \bibinfo{pages}{357--397}.
\newblock \DOIprefix\doi{10.1016/j.jcp.2003.08.010}.
\bibitem[{Launder and Spalding(1974)}]{LaunderSpalding74}
\bibinfo{author}{Launder, B.E.}, \bibinfo{author}{Spalding, D.},
  \bibinfo{year}{1974}.
\newblock \bibinfo{title}{The numerical computation of turbulent flows}.
\newblock \bibinfo{journal}{Comput.Methods in Appl.Mech.Eng.}
  \bibinfo{volume}{3}, \bibinfo{pages}{269--289}.
\newblock \DOIprefix\doi{10.1016/0045-7825(74)90029-2}.
\bibitem[{Liou(2006)}]{Liou06}
\bibinfo{author}{Liou, M.S.}, \bibinfo{year}{2006}.
\newblock \bibinfo{title}{A sequel to {AUSM}, part {II}: {AUSM+-up} for all
  speeds}.
\newblock \bibinfo{journal}{J. Comput. Phys.} \bibinfo{volume}{214},
  \bibinfo{pages}{137--170}.
\newblock \DOIprefix\doi{10.1016/j.jcp.2005.09.020}.
\bibitem[{Litschke and Kuttler(2008)}]{LitschkeKuttler08}
\bibinfo{author}{Litschke, T.}, \bibinfo{author}{Kuttler, W.},
  \bibinfo{year}{2008}.
\newblock \bibinfo{title}{On the reduction of urban particle concentration by
  vegetation - a~review}.
\newblock \bibinfo{journal}{Meteorol. Z.} \bibinfo{volume}{17},
  \bibinfo{pages}{229--240}.
\newblock \DOIprefix\doi{10.1127/0941-2948/2008/0284}.
\bibitem[{Liu et~al.(1996)Liu, Chen, Black and Novak}]{LiuEtAl96}
\bibinfo{author}{Liu, J.}, \bibinfo{author}{Chen, J.}, \bibinfo{author}{Black,
  T.}, \bibinfo{author}{Novak, M.}, \bibinfo{year}{1996}.
\newblock \bibinfo{title}{E-{$\varepsilon$} modelling of turbulent air flow
  downwind of a model forest edge}.
\newblock \bibinfo{journal}{Boundary Layer Meteorol.} \bibinfo{volume}{77},
  \bibinfo{pages}{21--44}.
\newblock \DOIprefix\doi{10.1007/BF00121857}.
\bibitem[{Parente et~al.(2011)Parente, Gorl{\'{e}}, van Beeck and
  Benocci}]{ParenteEtAl11}
\bibinfo{author}{Parente, A.}, \bibinfo{author}{Gorl{\'{e}}, C.},
  \bibinfo{author}{van Beeck, J.}, \bibinfo{author}{Benocci, C.},
  \bibinfo{year}{2011}.
\newblock \bibinfo{title}{Improved k{\textendash}{$\epsilon$} model and wall
  function formulation for the {RANS} simulation of {ABL} flows}.
\newblock \bibinfo{journal}{J. Wind Eng. Ind. Aerodyn.} \bibinfo{volume}{99},
  \bibinfo{pages}{267--278}.
\newblock \DOIprefix\doi{10.1016/j.jweia.2010.12.017}.
\bibitem[{Petroff et~al.(2008a)Petroff, Mailliat, Amielh and
  Anselmet}]{PetroffEtAl08a}
\bibinfo{author}{Petroff, A.}, \bibinfo{author}{Mailliat, A.},
  \bibinfo{author}{Amielh, M.}, \bibinfo{author}{Anselmet, F.},
  \bibinfo{year}{2008}a.
\newblock \bibinfo{title}{Aerosol dry deposition on vegetative canopies. {Part
  I}: {Review} of present knowledge}.
\newblock \bibinfo{journal}{Atmos. Environ.} \bibinfo{volume}{42},
  \bibinfo{pages}{3625--3653}.
\newblock \DOIprefix\doi{10.1016/j.atmosenv.2007.09.043}.
\bibitem[{Petroff et~al.(2008b)Petroff, Mailliat, Amielh and
  Anselmet}]{PetroffEtAl08b}
\bibinfo{author}{Petroff, A.}, \bibinfo{author}{Mailliat, A.},
  \bibinfo{author}{Amielh, M.}, \bibinfo{author}{Anselmet, F.},
  \bibinfo{year}{2008}b.
\newblock \bibinfo{title}{Aerosol dry deposition on vegetative canopies. {Part
  II}: A new modelling approach and applications}.
\newblock \bibinfo{journal}{Atmos. Environ.} \bibinfo{volume}{42},
  \bibinfo{pages}{3654--3683}.
\newblock \DOIprefix\doi{10.1016/j.atmosenv.2007.12.060}.
\bibitem[{Petroff et~al.(2009)Petroff, Zhang, Pryor and Belot}]{PetroffEtAl09}
\bibinfo{author}{Petroff, A.}, \bibinfo{author}{Zhang, L.},
  \bibinfo{author}{Pryor, S.}, \bibinfo{author}{Belot, Y.},
  \bibinfo{year}{2009}.
\newblock \bibinfo{title}{An extended dry deposition model for aerosols onto
  broadleaf canopies}.
\newblock \bibinfo{journal}{J. Aerosol Sci.} \bibinfo{volume}{40},
  \bibinfo{pages}{218--240}.
\newblock \DOIprefix\doi{10.1016/j.jaerosci.2008.11.006}.
\bibitem[{Richards and Hoxey(1993)}]{RichardsHoxey93}
\bibinfo{author}{Richards, P.}, \bibinfo{author}{Hoxey, R.},
  \bibinfo{year}{1993}.
\newblock \bibinfo{title}{Appropriate boundary conditions for computational
  wind engineering models using the k-{$\epsilon$} turbulence model}.
\newblock \bibinfo{journal}{J. Wind Eng. Ind. Aerodyn.} \bibinfo{volume}{46 \&
  47}, \bibinfo{pages}{145--153}.
\newblock \DOIprefix\doi{10.1016/0167-6105(93)90124-7}.
\bibitem[{Robert(1993)}]{Robert93}
\bibinfo{author}{Robert, A.}, \bibinfo{year}{1993}.
\newblock \bibinfo{title}{Bubble convection experiments with a semi-implicit
  formulation of the euler equations}.
\newblock \bibinfo{journal}{J. Atmos. Sci.} \bibinfo{volume}{50},
  \bibinfo{pages}{1865--1873}.
\newblock \DOIprefix\doi{10.1175/1520-0469(1993)050<1865:BCEWAS>2.0.CO;2}.
\bibitem[{Saad and Schultz(1986)}]{SaadSchultz86}
\bibinfo{author}{Saad, Y.}, \bibinfo{author}{Schultz, M.H.},
  \bibinfo{year}{1986}.
\newblock \bibinfo{title}{{GMRES}: A generalized minimal residual algorithm for
  solving nonsymmetric linear systems}.
\newblock \bibinfo{journal}{SIAM Journal on scientific and statistical
  computing} \bibinfo{volume}{7}, \bibinfo{pages}{856--869}.
\newblock \DOIprefix\doi{10.1137/0907058}.
\bibitem[{Seinfeld and Pandis(2006)}]{SeinfeldPandis06}
\bibinfo{author}{Seinfeld, J.}, \bibinfo{author}{Pandis, S.},
  \bibinfo{year}{2006}.
\newblock \bibinfo{title}{Atmospheric Chemistry and Physics: From Air Pollution
  to Climate Change}.
\newblock A Wiley-Interscience publication. \bibinfo{edition}{2nd} ed.,
  \bibinfo{publisher}{Wiley}.
\bibitem[{{\v{S}}{\'{i}}p and Bene{\v{s}}(2016)}]{SipBenes16-enumath}
\bibinfo{author}{{\v{S}}{\'{i}}p, V.}, \bibinfo{author}{Bene{\v{s}}, L.},
  \bibinfo{year}{2016}.
\newblock \bibinfo{title}{{CFD} optimization of a vegetation barrier}, in:
  \bibinfo{editor}{Karasözen, B.}, \bibinfo{editor}{Manguoglu, M.},
  \bibinfo{editor}{Tezer-Sezgin, M.}, \bibinfo{editor}{Göktepe, S.},
  \bibinfo{editor}{Ömür Ugur} (Eds.), \bibinfo{booktitle}{Numerical
  Mathematics and Advanced Applications - {ENUMATH }2015},
  \bibinfo{publisher}{Springer International Publishing},
  \bibinfo{address}{Cham}.
\bibitem[{Steffens et~al.(2012)Steffens, Wang and Zhang}]{SteffensEtAl12}
\bibinfo{author}{Steffens, J.}, \bibinfo{author}{Wang, Y.},
  \bibinfo{author}{Zhang, K.}, \bibinfo{year}{2012}.
\newblock \bibinfo{title}{Exploration of effects of a vegetation barrier on
  particle size distributions in a near-road environment}.
\newblock \bibinfo{journal}{Atmos. Environ.} \bibinfo{volume}{50},
  \bibinfo{pages}{120--128}.
\newblock \DOIprefix\doi{10.1016/j.atmosenv.2011.12.051}.
\bibitem[{Svensson and H{\"a}ggkvist(1990)}]{SvenssonHaeggkvist90}
\bibinfo{author}{Svensson, U.}, \bibinfo{author}{H{\"a}ggkvist, K.},
  \bibinfo{year}{1990}.
\newblock \bibinfo{title}{A two-equation turbulence model for canopy flows}.
\newblock \bibinfo{journal}{J. Wind Eng. Ind. Aerodyn.} \bibinfo{volume}{35},
  \bibinfo{pages}{201--211}.
\newblock \DOIprefix\doi{10.1016/0167-6105(90)90216-Y}.
\bibitem[{Tiwary et~al.(2005)Tiwary, Morvanb and Colls}]{TiwaryEtAl05}
\bibinfo{author}{Tiwary, A.}, \bibinfo{author}{Morvanb, H.},
  \bibinfo{author}{Colls, J.}, \bibinfo{year}{2005}.
\newblock \bibinfo{title}{Modelling the size-dependent collection efficiency of
  hedgerows for ambient aerosols}.
\newblock \bibinfo{journal}{J. Aerosol Sci.} \bibinfo{volume}{37},
  \bibinfo{pages}{990--1015}.
\newblock \DOIprefix\doi{10.1016/j.jaerosci.2005.07.004}.
\bibitem[{Turkel(1985)}]{Turkel85}
\bibinfo{author}{Turkel, E.}, \bibinfo{year}{1985}.
\newblock \bibinfo{title}{Algorithms for the {Euler} and {Navier}-{Stokes}
  equations for supercomputers}, in: \bibinfo{booktitle}{Progress and
  Supercomputing in Computational Fluid Dynamics},
  \bibinfo{organization}{Springer}. pp. \bibinfo{pages}{155--172}.
\bibitem[{Venkatakrishnan(1995)}]{Venkatakrishnan95}
\bibinfo{author}{Venkatakrishnan, V.}, \bibinfo{year}{1995}.
\newblock \bibinfo{title}{Convergence to steady state solutions of the {Euler}
  equations on unstructured grids with limiters}.
\newblock \bibinfo{journal}{J. Comput. Phys.} \bibinfo{volume}{118},
  \bibinfo{pages}{120--130}.
\newblock \DOIprefix\doi{10.1006/jcph.1995.1084}.
\bibitem[{Vranckx et~al.(2015)Vranckx, Vos, Maiheu and Janssen}]{VranckxEtAl15}
\bibinfo{author}{Vranckx, S.}, \bibinfo{author}{Vos, P.},
  \bibinfo{author}{Maiheu, B.}, \bibinfo{author}{Janssen, S.},
  \bibinfo{year}{2015}.
\newblock \bibinfo{title}{Impact of trees on pollutant dispersion in street
  canyons: A numerical study of the annual average effects in {Antwerp},
  {Belgium}}.
\newblock \bibinfo{journal}{Sci. Total Environ.} \bibinfo{volume}{532},
  \bibinfo{pages}{474--483}.
\newblock \DOIprefix\doi{10.1016/j.scitotenv.2015.06.032}.
\bibitem[{Wilson and Shaw(1977)}]{WilsonShaw77}
\bibinfo{author}{Wilson, N.R.}, \bibinfo{author}{Shaw, R.H.},
  \bibinfo{year}{1977}.
\newblock \bibinfo{title}{A higher order closure model for canopy flow}.
\newblock \bibinfo{journal}{J. Appl. Meteorol.} \bibinfo{volume}{16},
  \bibinfo{pages}{1197--1205}.
\newblock \DOIprefix\doi{10.1175/1520-0450(1977)016<1197:AHOCMF>2.0.CO;2}.

\end{thebibliography}
\end{document}